\documentclass[11pt,fleqn,a4paper]{article}

\usepackage{titling}
\usepackage{preliminary}
\usepackage{hyphenat}
\usepackage[authoryear]{natbib}
\usepackage{amssymb}
\usepackage{amsthm}
\usepackage{epstopdf}
\usepackage{amsmath}

\usepackage{scrlayer-scrpage}
\usepackage{subfigure}
\usepackage{setspace}
\usepackage[bottom]{footmisc}
\usepackage{endnotes}
\usepackage{graphicx}
\usepackage{textcomp}
\usepackage{natbib}
\usepackage[utf8]{inputenc}
\usepackage{verbatim}
\usepackage{lscape}
\usepackage{array}
\usepackage{fltpoint}
\usepackage{dcolumn}
\usepackage{booktabs}
\usepackage{rotating}
\usepackage[norounding,point]{rccol}
\usepackage{hyperref}
\usepackage{placeins}
\usepackage{multirow}
\usepackage{paralist}
\usepackage{enumitem}
\usepackage{textgreek}
\usepackage{float}
\usepackage{graphicx}
\graphicspath{ {./Images/} } 
\usepackage{tablefootnote}
\usepackage{eurosym}

\usepackage{multirow}

\usepackage{tikz-cd}
\usepackage{ctable}

\usepackage{tikz}
\usetikzlibrary{positioning}
\usetikzlibrary{snakes}

\usepackage[toc,page]{appendix}

\usepackage{multicol}
\usepackage{caption}

\usepackage{subfigure}

\usepackage{color}

\usepackage{amsmath}
\usepackage{url}

\usepackage{geometry}
\numberwithin{equation}{section}

\hypersetup{
colorlinks=true,
urlcolor=blue,
linkcolor=red,
citecolor=blue
}

\setheadwidth{textwithmarginpar}
\usepackage{setspace}
\usepackage{xspace}

\numberwithin{theorem}{section}
\numberwithin{assumption}{section}

\newcommand{\auslassen}[1]{}

\usepackage{graphicx}

\usepackage{chngcntr}

\usepackage[nottoc]{tocbibind}

\usepackage{blindtext}
\usepackage{threeparttable}
 \usepackage{threeparttablex}

\usepackage{authblk}

\usepackage{atbegshi}
\AtBeginDocument{\AtBeginShipoutNext{\AtBeginShipoutDiscard}}

\begin{document}
\begin{center}

\title{From homemakers to breadwinners?\\ How mandatory kindergarten affects maternal labour market outcomes}

\author[1]{\large Selina Gangl}
\author[2]{\large Martin Huber}

\affil[1]{\footnotesize Friedrich-Alexander-Universit\"at Erlangen-N\"urnberg (Germany), Department of Economics}
\affil[2]{\footnotesize University of Fribourg (Switzerland), Department of Economics}

\date{March 14, 2025}

\maketitle
\thispagestyle{empty}
\end{center}

\begin{abstract}
The majority of Swiss children attend mandatory and cost-free kindergarten at four. We examine the effect of this policy on maternal labour market outcomes. Using administrative data, we exploit the birthday cut-off for  kindergarten entry in the same or in the following year and apply a non-parametric regression discontinuity design (RDD). We find that mandatory kindergarten has a statistically significant positive effect on the labour market attachment of previously non-employed mothers, increasing their employment probability by 4 percentage points. In contrast, there are no significant effects on other groups or in the total sample of mothers.\\[0.2cm]
{\small  \textbf{Keywords:} Mandatory kindergarten,  maternal employment, regression discontinuity design}\\[0.1cm]
{\small  \textbf{JEL Classification:} H40, J13, J18, J21, J22  \quad }
\end{abstract}

\vfill

\begin{singlespace}

\smallskip {\scriptsize 
We thank the editor, Kompal Sinha, and three anonymous referees for helpful comments and suggestions. Furthermore, we thank Maryna Ivets, Laura Ravazzini, Regina T. Riphahn, Jürg Schweri, and Andreas Steinmayr for their  helpful suggestions on a prior version of this paper.  We have benefited from comments by seminar participants in Fribourg, the Ski \& Labour Seminar in Engelberg-Titlis, the 9th ifo Dresden Workshop on Labour Economics and Social Policy, the 10th International Conference of Panel Data Users in Switzerland in Lausanne, the Swiss Society of Economics and Statistics Annual Meeting in Geneva, the EEA in Manchester, the 3rd International BFH Conference on Discrimination in the Labor Market in Bern, the Verein f\"ur Socialpolitik - Annual Conference 2019, the 1st and 4th Workshop of the  Swiss Network on Public Economics (SNOPE), and the Lunch-time Meeting at the University of Innsbruck 2021, COMPIE 2022, European Associaton of Labour Economists (EALE) 2024, 37th Annual Conference of the European Society for Population Economics (ESPE), Workshop on Labour Economics 2024, Trier, Germany  and the Pre- and Postdoc Seminar at the Friedrich-Alexander-University Erlangen-Nürnberg. The authors declare that they have no conflict of interest. This paper uses confidential data from the Federal Statistical Office (FSO) and Old-Age and Survivor's Insurance (OASI) in Switzerland. The data can be obtained by filing a request directly with the FSO (https://www.bfs.admin.ch/bfs/de/home/dienstleistungen/datenverknuepfungen/fuer-dritte.html).  
\\}
\end{singlespace}

{\small \renewcommand{\thefootnote}{\arabic{footnote}} %
\setcounter{footnote}{0}  \pagebreak \setcounter{footnote}{0} \pagebreak %
\setcounter{page}{1} }
\pagestyle{plain}

\pagenumbering{arabic}

\newgeometry{top=3cm,bottom=3cm,right=2.5cm,left=2.5cm}

\renewcommand\thetable{\arabic{table}}
\renewcommand\thefigure{\arabic{figure}}

\section{Introduction}\label{s:intro}

The labour force attachment of mothers has been receiving much attention in scientific and public discussions, given the large employment drops observed in many countries after giving birth, see for instance \cite{Kleven2019}. In  OECD countries, the  average maternal employment rate is 71\% in 2019, with Iceland, Slovenia, and Sweden at the top with rates exceeding 85\%.\footnote{Reported by OECD. \url{https://www.oecd.org/els/family/database.htm},  retrieved 2021-08-24.} In Switzerland, which is the country considered in this study, mothers' employment rate
declines after childbirth \citep{MutterArbeitsm}, but it remains high at almost 78\% on average.\footnote{Reported by OECD. \url{https://www.oecd.org/els/family/database.htm},  retrieved 2021-08-24.} Although this figure is higher than the OECD average, most of it is due to part-time work (63\%), rather than full-time work (15\%).\footnote{Reported by OECD. \url{https://www.oecd.org/els/family/database.htm},  retrieved 2021-08-24.}

A political tool to encourage mothers' labour market outcomes  is the provision of childcare. For example, \cite{Krapf2020} show that the opportunity of placing the first child in childcare increases the maternal labour force participation in the canton Bern by 5.1 percentage points in the long term. In addition to the extensive margin, the birth of a child also affects the intensive margin of labour supply: mothers are more likely to work part-time than women without children  \citep{MutterArbeitsm}. The provision of childcare influences also mothers' stint positively: An expansion of childcare facilities raises  maternal part-time hours \citep{Ravazzini2018}, while after-school care increases mothers'  full-time employment \citep{Felfe2016}. Childcare availability may also increase maternal earnings, but not the earnings of the total household \citep{Krapf2020}. An unanswered question in this context is how mothers' labour market responses react when childcare is mandatory. While kindergarten for four-year-olds has always been cost-free, it has recently become mandatory in most Swiss cantons.

This paper examines the causal effect of mandatory  kindergarten for four-year-old children on  maternal labour behaviour and earnings in Switzerland. The aim of the policy, with regard to mothers, is to promote the compatibility of family and work. We exploit this novel setting and use  the birthday cut-off  to identify the causal effect of the mandatory kindergarten. 

The paper uses a large administrative dataset about the population of  mothers with four-year-old children living in Switzerland. We link  the Population and Households Statistics (STATPOP) with data about the Old-Age and Survivors' Insurance (OASI). Moreover, we have collected data about the  implementation year of the policy, the birthday cut-off dates per year and the obligation to offer kindergarten years. These data stem from  the cantonal departments of education, the Swiss Conference of Cantonal Departments of Education (EDK), and the cantonal laws. The State Secretariat for Economic Affairs (SECO) provides information about the cantonal unemployment rate on their homepage. Again we link this cantonal information to the administrative data. We restrict the sample to mothers whose youngest child is four-year-old and living in a canton with mandatory kindergarten.\footnote{See section 4 Data for details.} Methodologically, we exploit the random assignment at the birthday cut-off for entering the kindergarten in the same year (= treatment)
versus the following year (= control) and apply a Regression Discontinuity Design (RDD). Since parents may postpone the kindergarten entry of their child by one year and we cannot observe the actual kindergarten entry, we rather assess the intention-to-treat (ITT) effect than the average treatment effect (ATE) at the threshold. 

We find that mandatory kindergarten boosts the labour market outcomes of previously non-employed mothers.  In this subgroup, the probability of maternal employment increases by 4 percentage points on average across our outcome periods 2010 to 2017 (statistically significant at the 5\% level). Likewise, the  annual income from dependent employment increases by 1,030 CHF  on average (statistically significant at the 10\% level) and the probability of earning the income from dependent employment increases by 4 percentage points (statistically significant at the 5\% level). 
Our results suggest that mandatory kindergarten has no significant effect on the labour market outcomes in the total sample of mothers. We argue that the high participation of mothers in the labour force, the half-day care in mandatory kindergarten, and the support of grandparents for childcare are likely reasons for this finding.

Our paper fits into the literature on school policies and its impact on the family, see e.g. \cite{Landerso}. In particular, it belongs to a group of studies focusing on maternal labour market outcomes.  For example, \cite{Gelbach2002} investigates the effect of school enrolment  and \cite{BARUA2014} analyses the variation in school entry laws on maternal labour supply in the U.S. \cite{Finseraas2017} examine a decrease in the school starting age from seven to six years on maternal labour supply in Norway. The common result of these studies is that the school entry of the child raises mothers' labour supply. In line with that, \cite{Zhu2015} finds that the birth of a child shortly after the birthday cut-off leads to a  negative impact  on mothers' labour supply.  

Whereas the aforementioned studies focus on children who are starting school at age six or seven, our setting is  unique because all four-year-olds are required to enter school, or in our case, mandatory kindergarten. Therefore, this paper contributes to the existing literature by examining the impetus for maternal reintegration into the workforce \textit{two years earlier} due to mandatory kindergarten attendance of their child. This is particularly relevant because industrialized countries are currently facing labour shortages which are expected to increase due to an ageing population  and  might be  mitigated by working mothers \citep{EuropeanComm2023}. Moreover, an earlier reintegration in the labour force might decrease the risk of poverty in old age that mothers face due to interrupted working lives \citep{UN2022}.


This paper relates also to the family policy literature and its impact  on maternal labour market behaviour, see e.g. \cite{Kleven2020}. Childcare, specifically, may facilitate mothers' participation in the workforce. Hence, one  strand of literature examines whether the introduction of subsidised childcare like in Austria \citep{Kleven2020}, Canada \citep{Lefebvre2008},  Norway (\citealp{Hardoy2015}), Sweden (\citealp{Lundin2008}), and the Netherlands (\citealp{Bettendorf2015}) impacts maternal employment. Further research analyses the expansion of childcare places
as done in Spain (\citealp{NOLLENBERGER2015}), Italy (\citealp{CARTA2018}), Germany (\citealp{GEYER2015}), and Norway (\citealp{Kunze19}; \citealp{ECKHOFFANDRESEN2019}), or the provision of a legal entitlement to childcare, as in Germany \citep{BAUERNSCHUSTER2015}.
Evaluations of these policies suggest a significantly positive effect on mothers' labour force participation, but only under the condition of a low labour force participation rate and a small extent of childcare prior to the reform. However, the availability of (subsidised) childcare can lead to a crowding out of private childcare, which may offset any increase in overall childcare provision and, as a result, fail to improve maternal labour market outcomes \citep{HAVNES2011}. For this reason, \cite{GOUX} find only an effect of childcare availability on maternal labour supply for the subgroup of single mothers in France.

While this literature analyses the impact of childcare on the labour market behaviour of a self-selected group of mothers who choose to enrol their children in childcare, it does not encompass the effect on \textit{all mothers}. Our paper addresses this inquiry by examining whether mandatory kindergarten incentivizes  all mothers to participate in the labour market and the resulting impact on their wages. This information seems to be crucial for policy makers seeking to promote maternal labour force participation.

The findings of our study align with those of \cite{HAVNES2011}, \cite{GOUX}, and \cite{Lundin2008}, indicating hardly no impact of childcare on maternal  labour market outcomes. In comparison, \cite{Berlinski2007} and  \cite{Lefebvre2008} report significant positive effects. These divergent conclusions may partly stem from variations in the socio-economic  settings of the studied countries. For instance, a high rate of maternal employment, such as that observed in Sweden could elucidate the near negligible effect of childcare policies \citep{Lundin2008}. Similarly, in Switzerland, the focus of our study, where maternal employment rates hover around 80\%, the presence of mandatory kindergarten shows only discernible effects on labour market outcomes of previously non-employed mothers. Another feature of the Swiss setting is that  working parents mainly use the support of grandparents for childcare in Switzerland  \citep{JacobsFoundation2018}. The insignificant  impact of mandatory kindergarten on  labour market response in the total sample of mothers could potentially be attributed to  a crowding out effect of informal childcare, as suggested in \cite{HAVNES2011}. Different policy designs may also explain the mixed results in the studies. For example, \cite{Lefebvre2008} analyses the effect of full-time childcare and find significant and large effects. While the mandatory kindergarten is  half-day care in most Swiss cantons such that mothers are often faced with the challenge of balancing their work around the limited kindergarten hours or finding another (costly) care option. This circumstance could be a possible explanation for the fact that mandatory kindergarten does not have a far-reaching impact on labour market outcomes in the total sample of mothers.

The remainder of this paper is organised as follows: In the next section, we provide information about the preschool policy in Switzerland. Thereafter, we present the data source, descriptive statistics, and define the sub-sample. Then, we discuss the empirical strategy. In the following section, we report the results and the robustness checks. Finally, we draw a conclusion from the empirical analysis.

\section{Institutional Background}
\label{s:inst}

 This section provides an overview of the mandatory and cost-free kindergarten for four year old children in Switzerland. Education policy in Switzerland is decentralized, with each of the 26 cantons having autonomy over the quality and structure of education. This decentralization leads to variations in aspects such as the school curriculum, the age of entry into kindergarten, and the number of school years \citep{Abstimmung2006}. Hence, each canton decides about the implementation of the mandatory kindergarten. Table \ref{tab: Cut-off dates} in the Online Appendix shows that Basel-Stadt was the first canton to introduce mandatory kindergarten for four-year-olds in 2005.  In 2006 a national referendum about an educational reform took place, which was accepted by 86\% of the voters. One year later, an inter-cantonal  "HarmoS" concordat was established, 15 cantons entered into the concordat, while seven cantons rejected it and four cantons are still indecisive (as of 2019). The main goal of the concordat is to harmonise the mandatory school education among the cantons. This agreement also includes an age decrease for mandatory school attendance. All children turning four before the first of August must enter kindergarten \citep{EDK2007}. Therefore, other cantons followed Basel-Stadt and introduced mandatory kindergartens. But also in this case, neither the implementation year nor the shift of the birthday cut-off date is regulated on a central level, see Table \ref{tab: Cut-off dates} in the Online Appendix for an overview.
 
 In 2017, almost two-thirds of the cantons had enforced mandatory kindergarten attendance for four-year-olds.\footnote{Aargau, Basel-Land, Basel-Stadt, Bern, Fribourg, Geneva, Glarus, Jura, Neuch\^atel, St Gallen, Schaffhausen, Solothurn, Thurgau, Ticino, Vaud, Valais, and Zurich.} In these cantons, children who turn four before or on the cutoff date must enter kindergarten in the same year, while four-year-olds born after the cutoff date do not enter kindergarten until the following year. Table \ref{tab: Cut-off dates} in the Online Appendix shows that these birthday cut-off dates vary from year to year and canton to canton and settle on 31.7. The reason for this shift is the joint decision by the cantons to determine a uniform cut-off date. Most cantons have decided to gradually shift the cut-off date so that the rush to kindergartens remains roughly the same during the transition period and does not increase particularly sharply in one year and overload the infrastructure.

The twofold aim of the mandatory kindergarten is to promote the development of children, but also to encourage the compatibility of family and work \citep{EDK2014}. In this paper, we focus on the latter and examine whether the obligation to attend kindergarten at the age of four affects mothers' work behaviour. Hence, the length of the mandatory kindergarten may be crucial for the maternal work decision. The kindergarten hours differ from  canton to canton: At the lower bound is the canton Valais, where four-year-old children are supposed to  attend 12 lessons per week,\footnote{EDK Kantonsumfrage 2016/17} each lesson is 45 minutes long, i.e. the children spend in sum 9 hours per week in the kindergarten. At the upper bound of kindergarten hours is the canton Ticino, where children attend the kindergarten on four days from the morning till the afternoon and an additional half-day. The mandatory kindergarten is part of early childhood education, which takes place prior to primary education \citep{OECD2015} and corresponds to level 0 in the International Standard Classification of Education (ISCED) \citep{EuropeanComm16/17}.  In the German- and Italian-speaking parts of Switzerland, kindergarten is rather a separate institution, whereas in the French-speaking part, kindergarten is represented as two additional classes in primary school \citep{EDK2016}.

 Red-shirting behaviour of the parents, i.e. deferring a four-year-old child from kindergarten, is possible, but is associated with hurdles that vary from canton to canton. In some cantons, an application by the parents is sufficient, in other cantons a medical certificate is required which proves that the admission of a child to the kindergarten would not be reasonable even with other educational assistance.\footnote{https://www.swissmom.ch/de/kind/kindergarten-und-schule/mit-4-jahren-schon-in-den-kindergarten-19063.}

\section{Hypotheses}
\label{s:hypotheses}

As the literature review reveals, childcare reforms as well as a lower school starting age affect mothers' work behaviour. In the former case, a subsidy either reduces the financial burden or provides more childcare places. In the latter case, the effect comes from the obligation to attend school at a younger age. In contrast, we evaluate the impact of the mandatory kindergarten on mothers' work behaviour. Therefore, the analysis combines features from both literature strands: The kindergarten is  an obligation and is free of charge like in the school starting age literature. On the other side, we evaluate a policy that affects children in the childcare age. This chapter presents the hypothesis and the possible channels.

We evaluate the effect of the mandatory kindergarten on maternal labour market behaviour within a specific age window: We compare mothers with slightly older children who were born prior to a specific birthday cut-off and are supposed to enter kindergarten in the same year (= treated) with mothers whose children are slightly younger, born after a specific birthday cut-off and who are supposed to enter kindergarten in the following year (= control).\footnote{One exception is the canton of Ticino that offers kindergarten for 3-year-old children.}
Mothers with a slightly younger child may spend their time by taking care of their child or must organize a (costly) childcare to allocate their time to work. Contrarily, mothers, whose child is supposed to enter kindergarten, can spend their time on other activities, for example, labour market participation. Consequently, a mother  who would take care of her child in the absence of the reform can rearrange her time: Following the model of the allocation of time \cite{Becker1965}, the mother can spend more time in market work and/or leisure activities. \citet*{SwartBergeWiel2017} find that mothers increase their labour supply after the school enrolment of their youngest child, given that they take care of their children prior to the school start. Therefore, we expect in our setting that only one subgroup of mothers increases their labour supply: Mothers whose children are supposed to enter kindergarten, but who would take care of them if no policy is implemented. All in all, we hypothesize a positive, but small  effect because the effect is zero for mothers who would send their child to kindergarten anyway.

\section{Data}\label{s:data}

Our evaluation strategy is a non-parametric regression discontinuity design (RDD) to examine the effect of the reform at the  birthday cut-off for entering the kindergarten in the same versus in the following year. It relies on a large administrative dataset including the universe of mothers with four-year-old children living in Switzerland, the so-called Population and Households Statistics (STATPOP). We link the latter with data coming from the old-age and survivors' insurance (OASI) in Switzerland by a unique identifier. STATPOP provides information on the stock and structure of the Swiss resident population from 2010-2018.  This enables us to examine the population of mothers whose youngest child is four-year-old and living in a canton with mandatory kindergarten. Furthermore, the data contain personal characteristics from the mother, their children, and the father of the four-year-old child. Two variables are central for our regression discontinuity approach (RDD): Firstly, the canton of residence to determine whether the four-year-old lives in a canton with mandatory kindergarten in the current year.  Secondly, the exact date of the birth of the four-year-old determines whether a child was born prior (= treated) or after (= control) a specific birthday cut-off date. Additionally, other personal characteristics such as the marital status, country of birth, resident permit of the mother, father, and the children are also contained in the dataset.

The OASI data contain information about the income, unemployment benefits, disability indemnities for every insured person in Switzerland from 2010 - 2017. Based on these data,  we calculated the "Income from work" as the total income minus transfer payments (for instance, unemployment insurance and welfare benefits) and minus other indemnities (like the compensation for military service)  for the following subgroups: i) employees whose employer resides in Switzerland, ii) voluntary insured employees whose employer is not liable for contributions in Switzerland, and iii) self-employed persons.\footnote{excluding agriculture.} Subsequently, we generated the variable "In labour force", which is 1 if a person earns income from work or if she receives unemployment benefits and 0 otherwise. We calculated the dummy "Out of labour force" as 1 minus the value of "In labour force". Furthermore, we constructed the dummy variable "Employed", which turns 1 if the person earns an income from work and participates in the labour market and is 0 otherwise. Vice versa, a person is indicated as "Unemployed" if she participates in the labour force but is not employed. The variable "Income from dependent employment" gives the sum of the income for mothers working in dependent employment, whereas the variable "Income from dependent employment (binary)" indicates whether a mother works in dependent employment.
In the last step, we matched this database with cantonal information about the unemployment rates, the yearly birthday cut-off date of the mandatory kindergarten, and an indicator of whether a canton joined the "HarmoS" concordat. The "HarmoS" concordat is an association of some cantons to harmonize mandatory education like the uniform school entry age.\footnote{More information about the institutional background can be found in Online Appendix B.2.1}

The database contains a total of 687,749 mothers with at least one four-year-old child between 2010 and 2018. Since the last period in the OASI data is 2017, we drop the STATPOP observations from 2018 such that 606,434 observations remain. Another data restriction is based on the empirical literature, because childcare seems to not affect the mothers' labour market outcomes when there is a younger sibling of the four-year-old child in the household (see, for example, \cite{Berlinski2007}, \cite{fitzpatrick2012}, and \cite{Gelbach2002}). We overcome this concern by restricting the subsample to mothers whose youngest child is four years old, which decreases the sample size to 360,234 mothers.
Three cantons with mandatory kindergarten did not determine a birthday cut-off date, hence we exclude observations from Aargau, Fribourg prior to 2013, and all the municipalities of the canton Bern except the federal city of Bern, after that 289,728 observations are in the sample. Furthermore, we drop observations with negative income entries\footnote{for example due to adjustment entries.} in the OASI data and keep 289,631 observations.  We exclude observations, which live in cantons without a mandatory kindergarten for four-years-old. Since two sources\footnote{see "Regierungsratsbeschluss betreffend Stichtag f\"ur die Einschulung f\"ur die Schuljahre 2011/12 bis 2015/16 (\S 56 Abs. 1 Schulgesetz)" and "EDK/IDES-Kantonsumfrage" for the years 2011 to 2016.} reported two consecutive days as birthday cut-off dates in Basel-Stadt, we took the later cut-off date and excluded observations on the previous day from our analysis.\footnote{to run a Donut-RDD for this canton.} Finally, 201,993 observations remain in our evaluation dataset.

\begin{figure}[ht!]
 \caption{Timeline of measured variables}
 \label{Kiga:timeline}
\centering
\begin{threeparttable}

\begin{tikzpicture}[snake=zigzag, line before snake = 5mm, line after snake = 5mm]
    \draw[ ->]  (-1,0) -- (9,0);

    \foreach \x in {-1, 3, 8.8}
      \draw (\x cm,3pt) -- (\x cm,-3pt);

      \draw (-1,0) node[below=3pt] {$ t=-1 $} node[above=3pt] {Time-dependent covariates };
    \draw (3,0) node[below=3pt] {$ t=0 $}    node[above=3pt] { Treatment };
     \draw (6,0)  node[below=3pt]  {$ 0 \leq \ t \leq \ 5$}   node[above=3pt] { Outcome periods  };
\draw (8.8,0) node[below=3pt] {$ t=5 $};
  \end{tikzpicture}

\begin{tablenotes}[flushleft]
\footnotesize
\item Note:  The timeline shows the time points at which the variables were measured relative to the treatment.
        \end{tablenotes}
\end{threeparttable}
\end{figure}
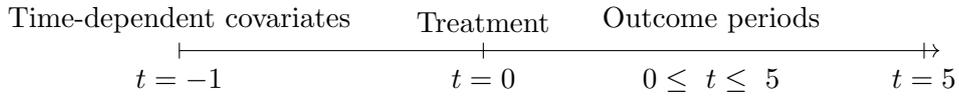

Figure \ref{Kiga:timeline}  provides a timeline for the measurement of the key variables in our analysis, with t denoting a specific year. The running variable will be henceforth denoted by D and depicts whether the child turns four prior (D = 1) or after the cut-off date (D = 0), measured in the baseline period (t = 0). The outcome periods start in the year the child turns four (t = 0) and end five years later (t = 5). Personal characteristics (e.g. number of children), are measured one year before the child turns four (t = -1).

We check the balance of labour market relevant covariates as suggested in \cite{LEE2008}, by running several RDD estimations where each covariate serves as an outcome. Table \ref{descrRDD} reports the sample mean, the balance checks, and the relative number of missings for each variable at the threshold. Mothers', fathers' and cantonal covariates are generally well balanced, which points to a quasi-random assignment of mandatory kindergarten at the threshold. We also see from the table that mothers close to the threshold have on average 1.87 children, are predominately born in Switzerland and live in a relationship. Furthermore, 73\% of these mothers are employed, earn on average 32,760 Swiss francs (CHF) per year, which mainly stem from dependent employment. Fathers labour market characteristics show with 90\% a very high employment rate, an annual mean gross income of 94,342 CHF, and 86\% of this income comes from dependent employment. The cantonal characteristics reveal a rather low mean unemployment rate (3.6\%) and a high probability of participating in the HarmoS concordat (94\%).

\begin{table}[!h]
\center
\footnotesize
\caption{Balance check: Covariates}
\label{descrRDD}
\centering
\begin{threeparttable}

 \resizebox{\linewidth}{!}{

\begin{tabular}{lccccp{1.5cm}}
\hline\hline
 & Sample mean & Coefficient &Standard error & P-value & Relative number of missings in \% \\
  \hline
\textit{Mother's characteristics} & & & & &\\
[0.15cm]
Number Children & 1.87 & 0.00 & 0.02 & 0.93 & 0.00 \\
[0.15cm]
  Born in Switzerland & 0.52 & 0.00 & 0.01 & 0.99 & 2.76 \\
  [0.15cm]
\textit{Nationality} & & & & & \\
  Resident permit B & 0.11 & -0.00 & 0.01 & 0.88 & 6.70 \\
  Resident permit C & 0.26 & 0.01 & 0.01 & 0.59 & 6.70 \\
  Other resident & 0.63 & -0.00 & 0.01 & 0.99 & 6.70 \\
  [0.15cm]
\textit{Age} & & & & &\\
  Age & 35.62 & 0.01 & 0.14 & 0.92 & 6.70 \\
  [0.15cm]
\textit{Relationship} & & & & &\\
  In relationship & 0.82 & -0.01 & 0.01 & 0.24 & 6.70 \\
   Not in relationship & 0.14 & 0.01 & 0.01 & 0.21 & 6.70 \\
   Terminated relationship & 0.05 & 0.00 & 0.01 & 0.92 & 6.70 \\
    [0.15cm]
\textit{Mother's labour market characteristics} & & & & &\\
 Out of labour force (binary) & 0.26 & 0.01 & 0.01 & 0.67 & 5.90 \\
  Employed (binary) & 0.73 & 0.00 & 0.01 & 0.72 & 5.90 \\
  Unemployed (binary) & 0.01 & 0.00 & 0.00 & 0.75 & 5.90 \\
  Total income from work (in CHF) & 32,760.09 & -1,345.75 & 944.46 & 0.15 & 5.90 \\
  Income from dependent employment (binary) & 0.70 & -0.01 & 0.01 & 0.53 & 5.90 \\
   Income from dependent employment (in CHF) & 31,204.52 &  447.98 & 1,066.99 & 0.68 & 5.90 \\
  [0.3cm]
\textit{Father's labour market characteristics} & & & & &\\
  Out of labour force (binary) & 0.09 & 0.00 & 0.01 & 0.86 & 5.90 \\
  Employed (binary)& 0.90 & 0.00 & 0.01 & 0.84 & 5.90 \\
  Unemployed (binary) & 0.01 & 0.00 & 0.00 & 0.97 & 5.90\\
  Total income from work (in CHF)  & 94,342.30 & -440.71 & 2,505.84 & 0.86 & 5.90 \\
   Income from dependent employment (binary) & 0.86 & 0.01 & 0.01 & 0.49 & 5.90 \\
 Income from dependent employment (in CHF) & 88,951.52 & -35.00 & 2,584.53 & 0.99 & 5.90\\
 [0.3cm]
\textit{Cantonal characteristics\textit} & & & & &\\
HarmoS member & 0.94 & 0.00 & 0.01 & 0.98 & 0.00 \\
Unemployment rate & 3.55 & 0.04 & 0.03 & 0.25 & 0.00 \\
   \hline
\end{tabular}
}
\begin{tablenotes}[flushleft]
\footnotesize
\item Note:  The table presents the balance check of the labour market relevant covariates as suggested in \cite{LEE2008}. The data comes from STATPOP (2010 - 2017) and OASI (2010 - 2017), the calculations are done by ourselves. Income is deflated, the base year is 2011. The official currency in Switzerland is the Swiss Franc (CHF), which had an average exchange rate of 1.04 USD/CHF in the last decade. Residence permit B allows foreign nationals in Switzerland for specific purposes with a five-year validity, while Residence permit C grants unrestricted residency after five or ten years of lawful stay, with specific agreements for citizens of certain EU/EFTA countries.
 \end{tablenotes}
\end{threeparttable}
\end{table}

Table \ref{descr2}  reports the balance checks for cantonal dummies at the threshold. Due to the different birthday cut-off dates in  the German- and the French-speaking part of the canton Valais, we analyse Upper Valais and Central/Lower Valais separately. In most cases, our tests do not reject the null hypothesis that the cantonal dummies are balanced around the birthday cut-off, with the two exceptions of Basel-Land and Ticino.\footnote{Ticino is the only canton implementing mandatory kindergarten offer for 3-year-old children.}  For visualization, Figures \ref{fig:baseline_cov_mother} to  \ref{fig:baseline_cov_cantons} in the Online Appendix provide plots of the conditional mean of each of the baseline covariates against the running variable.

\begin{table}[!h]
\center
\footnotesize
\caption{Balance check: Cantonal dummies}
\label{descr2}
\centering
\begin{threeparttable}

\begin{tabular}{lcccc}
 \hline\hline
 & Sample mean & Coefficient & Standard error & P-value \\
  \hline
Basel-Stadt & 0.04 & -0.00 & 0.01 & 0.60 \\
  St Gallen & 0.10 & -0.01 & 0.01 & 0.59 \\
  Thurgau & 0.06 & -0.00 & 0.01 & 0.98 \\
  Zurich & 0.32 & 0.00 & 0.02 & 0.81 \\
  Fribourg & 0.05 & -0.00 & 0.01 & 0.80 \\
  Geneva & 0.10 & 0.01 & 0.01 & 0.16 \\
  Glarus & 0.01 & -0.00 & 0.00 & 0.60 \\
  Neuch\^atel & 0.04 & 0.00 & 0.01 & 0.76 \\
  Basel-Land & 0.05 & -0.02 & 0.01 & 0.01 \\
  Jura & 0.01 & -0.00 & 0.00 & 0.40 \\
  Solothurn & 0.04 & -0.01 & 0.01 & 0.30 \\
  Bern & 0.02 & 0.00 & 0.00 & 0.87 \\
  Vaud & 0.12 & 0.00& 0.01 & 0.67 \\
  Schaffhausen & 0.01 & 0.00 & 0.00 & 0.11 \\
  Ticino & 0.03 & 0.01 & 0.00 & 0.00 \\
  Upper Valais & 0.01 & 0.00 & 0.00& 0.91 \\
  Central and Lower Valais & 0.02 & -0.01 & 0.01 & 0.37 \\
   \hline
\end{tabular}

\begin{tablenotes}[flushleft]
\footnotesize
\item Note:  The table presents the balance check of the cantonal dummies as suggested in \cite{LEE2008}. The data comes from STATPOP (2010 - 2017) and OASI (2010 - 2017), the calculations are done by ourselves.
 \end{tablenotes}
\end{threeparttable}
\end{table}

Table \ref{meanout} provides the mean outcomes per period for the whole sample and reveals that the outcome means are rather stable over time. In period 0, on average 76\% of mothers work in the labour market and almost all of them are employed. They earn an average  gross annual income of 33,902 CHF, with the largest part (32,353 CHF) stemming from dependent employment. Figure \ref{fig:outcomeplots} in the Online Appendix plots the conditional mean of each outcome, which is pooled over the outcome periods, against the running variable (difference between the cut-off date and the birth date of the child) when controlling for baseline covariates.

\begin{table}[h!]
\center
\footnotesize
\caption{Mean outcomes per period}
\label{meanout}
\begin{center}
 \resizebox{\linewidth}{!}{
\begin{tabular}{lcccccc}
 \hline\hline
& \multicolumn{6}{c}{Mean} \\
\hline
 & Period 0 & Period  1 & Period  2 & Period  3 & Period  4 & Period  5 \\
  \hline
  Out of labour force (binary)& 0.24 & 0.23 & 0.23 & 0.22 & 0.22 & 0.23 \\
  Employed (binary) & 0.75 & 0.76 & 0.76 & 0.77 & 0.77 & 0.76 \\
  Unemployed (binary) & 0.01 & 0.01 & 0.01 & 0.01 & 0.01 & 0.01 \\
  Annual work income (in CHF) & 33,902.41 & 34,490.92 & 35,035.26 & 35,681.94 & 36,080.12 & 35,992.51 \\
   Income from dependent employment (binary)  & 0.72 & 0.73 & 0.73 & 0.74 & 0.74 & 0.74 \\
  Income from dependent employment (in CHF)  & 32,353.25 & 32,879.31 & 33,407.43 & 34,080.92 & 34,553.32 & 34,750.57 \\
   \hline
\end{tabular}
}\\
 \vskip0.3cm
 {\footnotesize Note: The table presents the mean outcomes from period 0 to 5. The data comes from STATPOP (2010 - 2017) and OASI (2010 - 2017), the calculations are done by ourselves. Income is deflated, the base year is 2011. \par}
\end{center}
\par

\end{table}

	\clearpage

\section{Econometric approach}\label{s:metrics}

We use an RDD for evaluating the effect of mandatory kindergarten on mothers' labour market outcomes. The setting of our  paper implies an RDD, because of the following rule of the treatment (D): All children who turn four prior to or exactly at the cut-off are supposed to enter kindergarten in the current year (D=1), whereas children who turn four after the cut-off are supposed to enter kindergarten in the following year (D=0). In this case, we identify the average treatment effect (ATE) at the threshold (see \cite{LeeandLemieux2010}), i.e. among mothers, whose children are born at the cut-off.  Due to this rule, it is not possible to observe children born on the same day, simultaneously in the treatment and control group. To make the treatment and control group as comparable as possible in terms of covariates, we analyse only those observations within a specific bandwidth around the cut-off date.

One concern of identifying the ATE at the cut-off in our setting is the so-called "red shirting behaviour", i.e.  parents postpone the kindergarten entrance of their children by one year.  Since there is no micro-data about kindergarten attendance in Switzerland, we can neither evaluate whether red shirting takes place nor instrument the kindergarten attendance (see, for example, \cite{Hahn2001}) if this is the case. Therefore, we can only identify the intention-to-treat-effect (ITT) in our setting.

The identification of the effect relies on the assumption that the potential outcomes must be continuous around the cut-off \citep{Hahn2001}. This assumption implies the comparability of mothers in their characteristics on both sides of the cut-off. In our setting, the assumption holds if the difference between the cut-off date and children's birth date is  as good as randomly assigned at the cut-off. In other words, the continuity assumption is fulfilled if parents cannot precisely sort their children in the treatment or the control group and is violated otherwise.

In our case, the running variable specifies the difference between the birth date of the child and the cut-off date and is therefore a discrete value. A standard test for checking the plausibility of the continuity of the potential outcomes at the cut-off is the McCrary test which analyses the continuity of density of the running variable at the cut-off. A discontinuity of the density at the cut-off would point to selective sorting (also known as bunching) and, therefore, likely violate the continuity of potential outcomes. This test works well if the running variable is continuous, yet it is in general inconsistent when
the running variable is discrete. For this reason, we use Frandsen's test instead, which delivers a consistent result even in the presence of a discrete running variable. Frandsen's test uses only data points, which are exactly at or adjoining to the threshold, whereas the McCrary Test extrapolates in areas away from the threshold, which leads to inconsistency if the running variable is discrete \citep{Frandsen2017}.

\begin{figure}[!h]
\centering
\caption{Density plot of running variable and Frandsen's manipulation test\\
}
\label{tab:Frandsenstest}
\begin{threeparttable}
{ \includegraphics[scale=0.9]{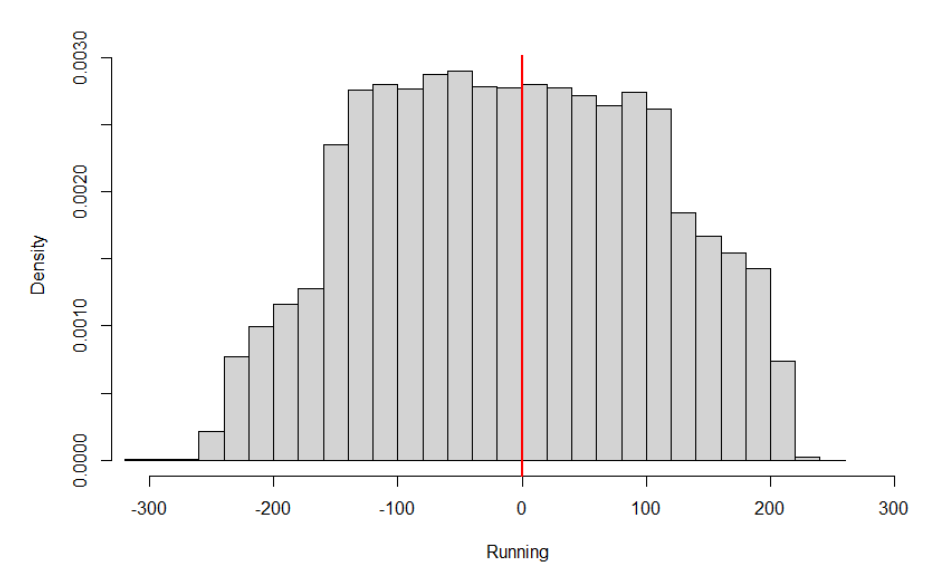}}
\centering

 Manipulation test (k = 0) p-value = 0.34\\
 Manipulation test (k = 0.1) p-value = 1.00\\
 Manipulation test (k = 0.2) p-value =  1.00  \\
 \vskip0.3cm
\begin{tablenotes}[flushleft]
\item Note:  The figure presents the density plot of the running variable and the Frandsen's test for the non-sorting assumption at the birthday cut-off. The running variable is defined as the difference between the birthday cut-off date and children's birth date. K corresponds to the self-defined maximum of the probability mass function (pmf) curve which is still considered as "non-sorting," see \cite{Frandsen2017}. Data stems from STATPOP (2010 - 2017) and OASI (2010 - 2017), calculations are done by ourselves.
 \end{tablenotes}
\end{threeparttable}
\end{figure}

The upper part of Figure \ref{tab:Frandsenstest} depicts the density plot of the running variable. Since the density below and above the birthday cut-off (depicted as red line) is very similar, it does not point to sorting behaviour. To analyse manipulation more precisely, the lower part of Figure \ref{tab:Frandsenstest}  reports the results of Frandsen's test, where K corresponds to our self-defined maximum of the probability mass function (pmf) curve which is still considered as "non-sorting," see \cite{Frandsen2017}. K must at least equal zero, meaning that a non-linearity is not allowed in the pmf \citep{Frandsen2017}, which represents our first specification. Furthermore, to allow for a non-linearity, K equals 0.1 and 0.2, as robustness tests in the following two specifications. The associated p-value indicates whether there is a statistical difference in the pmf of both sides of the threshold. Figure \ref{tab:Frandsenstest} indicates  no significant deviation from the expected density of the running variable, regardless of the specification of K. In short, we do not find any sorting behaviour around the birthday cut-off, pointing to a as good as randomized  treatment.

\vskip0.5cm



If the continuity assumption holds, the ATE (or in our case the ITT) at the threshold is non-parametrically identified given the bandwidth approaches zero, see \cite{Hahn2001}. Equation \eqref{KigaRDD: eqn} presents the identification result, where $\gamma$ gives the parameter of interest, $Y_{d}$ denotes the potential outcome (e.g. hypothetical employment) under treatment status $d$ $\in$ $\{1,0\}$,  $R$ the running variable, and $r$ the cut-off value $(r = 0$).

\vspace{-.75cm}
\begin{eqnarray}
\gamma = E [Y_{1} - Y_{0} |R=r]
\label{KigaRDD: eqn}
\end{eqnarray}

We use the following local linear regression to estimate the parameter of interest within a specific data window around the cut-off of the running variable:

\vspace{-.75cm}
\begin{eqnarray}
Y_{i}\ =\ \alpha\ +\beta_0D_{i}\ +\ \beta_{1}R_{i}\ +\ \beta_2R_{i}*D_{i}\ +\ \beta_3Z_{i}+\epsilon_{i}
\label{eqn2}
\end{eqnarray}

Y$_{i}$ denotes the outcome (e.g. labour force status), D$_{i}$ represents the treatment status, and R$_{i}$ the running variable centered around the cut-off (=0)  of individual i. $\beta_0$ is the average treatment effect of the mandatory kindergarten on mother's labour market behaviour at the threshold and $\beta_{1\ }$ and $\beta_{2\ }$ give the average effect of the running variable on the outcome. Z$_{i}$ is a vector of additional covariates for individual i.\footnote{We include all covariates listed in Table~\ref{descrRDD} and ~\ref{descr2}.} We control for covariates to get lower standard errors, which is possible because the covariates (e.g. maternal total income one year earlier) are good predictors of the outcomes (e.g. maternal total income one year later). For the estimation, we use the "rdrobust" command from the eponymous package  for the statistical software R \citep{Calonico2021}, with a uniform kernel function and a heteroskedasticity-robust plug-in residuals variance estimator without weights. As recommended by \cite{GelmanandImbens2019}, we estimate a local linear regression (p = 1) and use a local quadratic regression (q = 2) to perform bias-correction. The bandwidth h is computed in a data-driven way by using a MSE-optimal bandwidth (companion command "rdbwselect").

\section{Results}\label{s:results}

This chapter provides the estimated results as well as the robustness checks. At first, we present the ITT when pooling the outcome periods, then the effects over the years, and finally the heterogeneous effects for different subgroups. Since we use an RDD approach, all effects are measured at the cut-off.

\subsection{Average effects}

 Table \ref{RDD: Empirical results with covariates} reports the ITT estimates at the cut-off for pooled outcome periods when including covariates and corresponding missing dummies. It suggests the sample mean of the outcome (not only at the threshold), the effects along with clustered standard errors at the individual level and p-values, the bandwidth size, the number of observations in the bandwidth. We do not find statistically significant effects of mandatory kindergarten on maternal labour market outcomes in the total sample.\footnote{With only a 1.5\% rate of inter-cantonal movement in 2022 \citep{BFS2022}, the minimal migration between cantons suggests a negligible impact on the RDD estimates.}
 
 One potential explanation for the lack of impact on maternal labour supply might be that mandatory kindergarten crowds out informal childcare arrangements. In Switzerland, it is common for working parents to rely on the support of grandparents for the care of their children 
 \citep{JacobsFoundation2018}. The mandatory kindergarten may have reduced the necessity of informal childcare options, but this did not affect the overall impact on mothers' labour supply. The effect of mandatory kindergarten might be also diminished by the presence of formal childcare services. However, daycare in Switzerland is known to be expensive compared to other countries \citep{SternFelfe2015}, and there is a substantial demand surplus for childcare places \citep{BundesamtfurSozialversicherungen}. These factors contribute to a lower daycare attendance rate, with only 36\% of children under the age of four currently attending daycare \citep{JacobsFoundation2018}. Finally, the labour force supply of mothers in Switzerland is already high, which might also explain  the lack of a sensible effect of mandatory kindergarten on maternal labour market participation in the total sample of mothers. This finding is in line with other studies that have found either no effect or only a modest effect of childcare on maternal employment in the context of high maternal labor force participation (see, e.g. \cite{Lundin2008} and \cite{HAVNES2011}).
 
An important factor to consider when examining the lack of impact on maternal income is the limited flexibility provided by half-day mandatory kindergarten. With the majority of cantons offering half-day care, mothers are often faced with the challenge of balancing their work schedules around the limited hours of kindergarten. This constraint can significantly hinder their ability to pursue full-time employment opportunities. Furthermore, the reliance on alternative childcare arrangements for the rest of the day, such as private daycare, can come at an additional cost to mothers. These expenses, combined with the logistical challenges of coordinating multiple care providers, can create financial and logistical burdens that further discourage mothers from pursuing full-time employment. Therefore, the lack of access to full-day care following mandatory half-day kindergarten not only restricts the hours available for maternal employment but also introduces financial and logistical barriers that impede the potential increase in maternal income.


We check the robustness of our ITT effects by estimating the RDD approach without covariates (Table \ref{result1oKov} in the Online Appendix) and find rather similar point estimates, yet the precision decreases. Another check consists of varying the optimal bandwidth, we multiplied this by 1.5 and 2/3 (Tables  \ref{robust1} and \ref{robust2} in the Online Appendix). Our findings indicate that the effect sizes remain relatively stable, with the exception of the effects on "Annual income from dependent employment," as shown in Tables \ref{robust1} and \ref{robust2} in the Online Appendix, and "Annual work income" as presented in Table \ref{robust2} in the same Appendix.

\begin{landscape}
\begin{table}[!h]
\center
\footnotesize
\caption{RDD: Empirical results with covariates}
\label{RDD: Empirical results with covariates}
\centering
\begin{threeparttable}
 \renewcommand\TPTminimum{\textwidth} 
\begin{tabular}{lcccccp{2cm}}
 \hline\hline

  & Sample mean & Coefficient & Standard error & P-value & Bandwidth  & Observations within bandwidth \\
  \hline
Out of labour force (binary) & 0.23 & -0.00 & 0.01 & 0.60 & 72.11 & 297,628 \\ 
  Employed (binary) & 0.76 & 0.01 & 0.01 & 0.45 & 70.52 & 289,613\\ 
  Unemployed (binary) & 0.01 & -0.00 & 0.00 & 0.16 & 52.99 & 214,832\\ 
  Annual work income (in CHF) & 34,860.92 & 591.87 & 756.10 & 0.43 & 52.11 & 214,832\\ 
  Income dependent employment (binary) & 0.73 & 0.01 & 0.01 & 0.29 & 67.27 & 277,425\\ 
  Annual income from dependent employment (in CHF) & 33,075.03 & -52.90 & 594.99 & 0.93 & 69.82 & 285,555\\ 
  \hline
\end{tabular}
\begin{tablenotes}[flushleft]
\footnotesize
\item Note:  This table reports the local linear estimates of  equation \ref{eqn2}. The following control variables are included: Number of children, born in Switzerland, resident permit, age, relationship status, labour market characteristics of the mother and the father of the four-year old child, cantonal characteristics and dummies. Bandwidth shows the MSE-optimal bandwidth chosen by \citep{Calonico2021}. The standard errors are clustered at the individual level. Data stems from STATPOP (2010 - 2017) and OASI (2010 - 2017), calculations are done by ourselves.  Income is deflated, the base year is 2011. The official currency in Switzerland is the Swiss Franc (CHF), which had an average exchange rate of 1.04 USD/CHF in the last decade. Number of observations in total: 735,520. The number of observations within the specified bandwidth is displayed in the corresponding column.
   \end{tablenotes}
\end{threeparttable}
\end{table}

\end{landscape}

\subsection{Effects over years}

In a next step, we analyse the effects in separate and consecutive outcome periods, relative to the year the child turns four. Figures \ref{fig:over_years_m_cov} show the results of the different outcomes, starting in period 0 (i.e. the year the child turns four) and ending in period 5. The dots depict the ITT at the cut-off and the bands represent the point-wise 95\% confidence intervals based on robust standard errors. We do not find any statistically significant effect for the outcomes in comparison to the pooled estimates. This might be caused by lower power due to a decrease in the number of observations. Moreover, we find unstable point estimates, both the sign and the effect size change over periods. The plots reveal further that the effects are, especially in the later periods, imprecisely estimated because of the decrease in the number of observations. Figure \ref{fig:over_years_o_cov} in the Online Appendix depicts the plots without covariates and shows similar results in terms of significance, point estimates, and precision.

\subsection{Heterogeneous effects}

From a policy perspective, it is interesting to know whether the effect of mandatory kindergarten differs for specific subgroups. Non-employed mothers may react differently to the kindergarten policy than employed mothers. Therefore,  we split the sample according to the maternal labour force status one year before the youngest child turns four. 

First, we start with the presentation of the tests for the RDD assumptions. The Tables \ref{descrRDDE} and \ref{descrRDDN} in the Online Appendix show, similar as for the entire sample in Table \ref{descrRDD}, that the labour market relevant covariates are balanced for both sub-samples. Analogous to the balance checks for cantonal dummies of the entire sample in Table \ref{descr2}, we conclude from Tables \ref{descr2emp} and \ref{descr2non} that in most cases, our tests do not reject the null hypothesis that the cantonal dummies are balanced around the birthday cut-off.  In contrast to result of the entire sample in Figure \ref{tab:Frandsenstest}, the results of the Frandsen's test in the Figures \ref{tab:Frandsenstestempl} and \ref{tab:Frandsenstestnempl} in the Online Appendix suggests  evidence for manipulation at the cut-off if we assume that the distribution around the cut-off must be precisely linear (k = 0), but does not point to manipulation if we allow for non-linearities (k = 0.1 and k = 0.2). We follow the argumentation of \cite{OOSTERBEEK2021}, which states that the RDD becomes invalidated by manipulation only if there exist systematic disparities between mothers on either side of the birthday cut-off. However, Tables \ref{descrRDDE} and \ref{descrRDDN} in the Online Appendix reveal no discernible association between being born slightly after versus before the birthday cut-off and the predetermined covariates.

Second, we report the estimation results.  Table \ref{empl} represents the results for employed mothers,  whereas Table \ref{nempl} reports the effects for non-employed mothers. The results suggest that the effect is driven by non-employed mothers who start working when their youngest, four-year-old child is required to attend kindergarten.  Three effects in Table \ref{nempl} remain statistically significant at the 10\% level after applying the Benjamini-Hochberg procedure.\footnote{We inspect also the heterogeneous effects for older versus younger mothers as well as divorced mothers and do not find statistically significant effects.} To further validate the robustness of our findings, we conduct an RDD with a fake cut-off set 30 days earlier than the actual threshold, and the results in Table \ref{nempl_fake} in the Online Appendix indicate that the effects are insignificant. 

Additionally, we employ a Donut-RDD approach in Table \ref{donut} in the Online Appendix, which excludes observations with running variables close to the cut-off (values between -2 and 1). The results obtained under this design confirm that all estimated effects maintain the same direction as in the main analysis. However, the statistical significance of the effects is generally reduced. Specifically, only the effects on employment, out of the labour force, and income-dependent employment exhibit p-values close to 10\%, while the remaining effects show p-values that are considerably beyond any conventional levels of statistical significance.

Third, we provide numbers on the effect of the kindergarten policy on reducing the labour shortage. As shown in Table \ref{nempl}, the effect of mandatory kindergarten on employment is an increase of 4 percentage points for the subgroup of previously non-employed mothers. If we extrapolate this local effect to all 179,895 non-employed mothers with four-year-olds in our sample, mandatory kindergarten results in 7,196 additional employed mothers. When this number is compared to the total number of employees in Switzerland in the first quarter of 2010\footnote{4,530,917; Federal Statistical Office: https://www.bfs.admin.ch/bfs/en/home/statistics/industry-services/businesses-employment/jobs-statistics/jobs.assetdetail.32008419.html}, it corresponds to an increase in aggregate employment of 0.002 percentage points. Comparing these 7,196 additional employed mothers to the job vacancies in Switzerland in the first quarter of 2010\footnote{51,162; Federal Statistical Office: https://www.bfs.admin.ch/bfs/en/home/statistics/industry-services/businesses-employment/jobs-statistics/vacancies.assetdetail.32008426.html}, we see a reduction in vacant positions of 0.14 percentage points. We assume that the effects can be extrapolated. Therefore, it should be noted that this is only an approximation subject to errors. It's also worth noting that the administrative dataset lacks information on working hours or whether mothers work full- or part-time, making it impossible to calculate the employment increase in full-time equivalents.\\

Additionally, we exploit the spatial variation and the staggered implementation of the mandatory kindergarten in a Difference-in-Differences Approach. The description of the data, method and the results are presented in the Online Appendix. We find that labour market participation, employment and the part-time rate of mothers increase because of mandatory kindergarten. However, the precision of these results is low due to a small sample size.

\begin{figure}[h!]
\centering
   \caption{RDD: Effects over years with covariates}
\label{fig:over_years_m_cov}
\begin{threeparttable}
        \subfigure[Out of labour force (binary)]{%
      \includegraphics[width=0.3\textwidth]{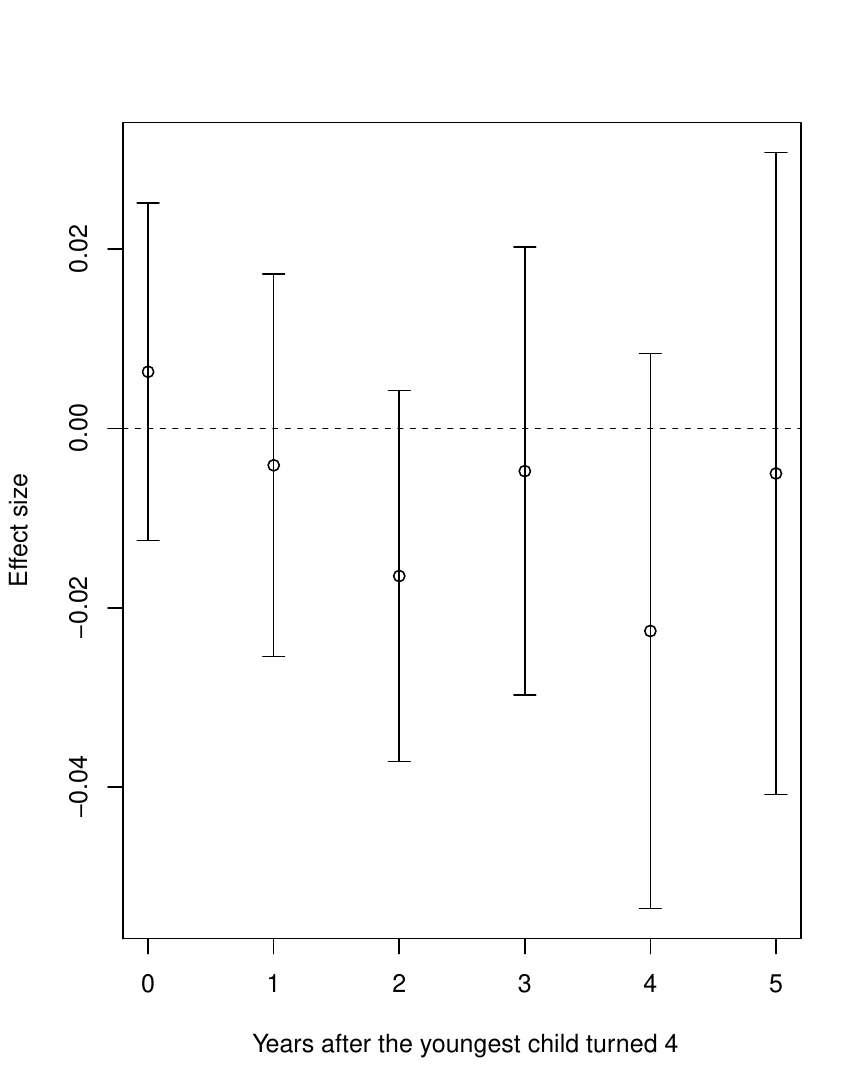}
       \label{fig:out}}
        \quad
        \subfigure[Employed (binary)]{%
      \includegraphics[width=0.3\textwidth]{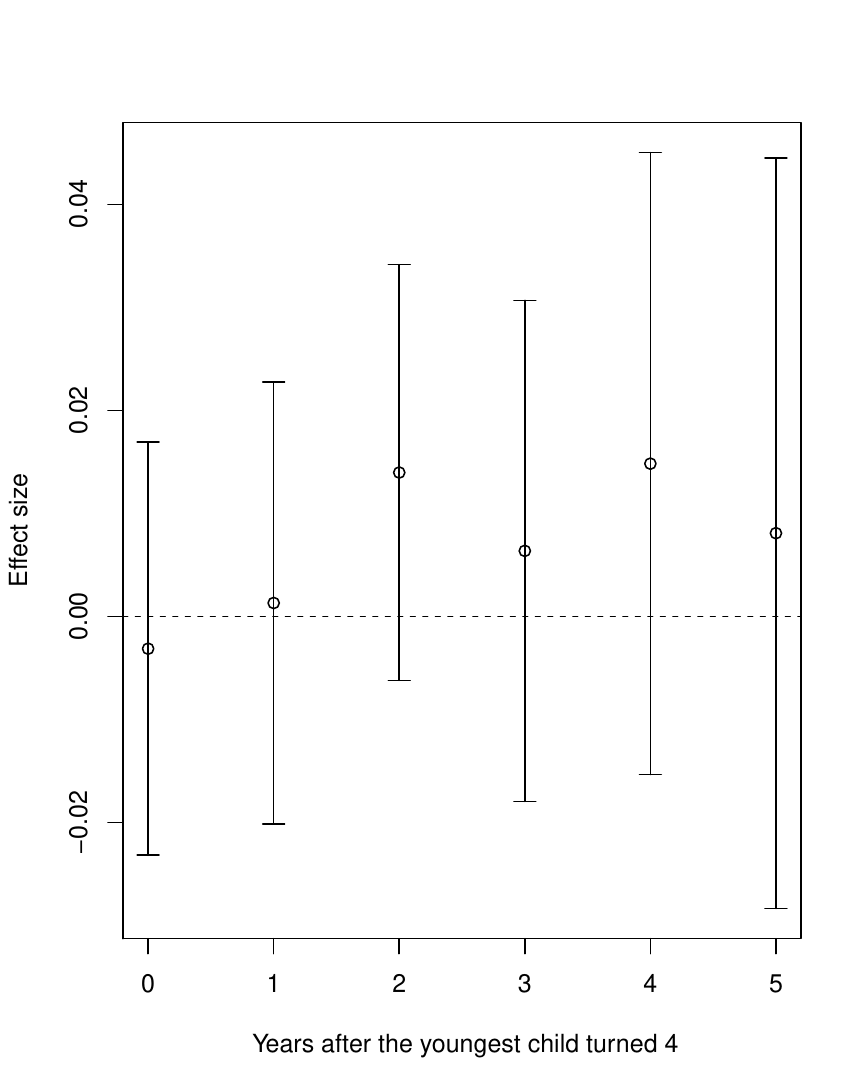}
       \label{fig:emp}}
        \subfigure[Unemployed (binary)]{%
      \includegraphics[width=0.3\textwidth]{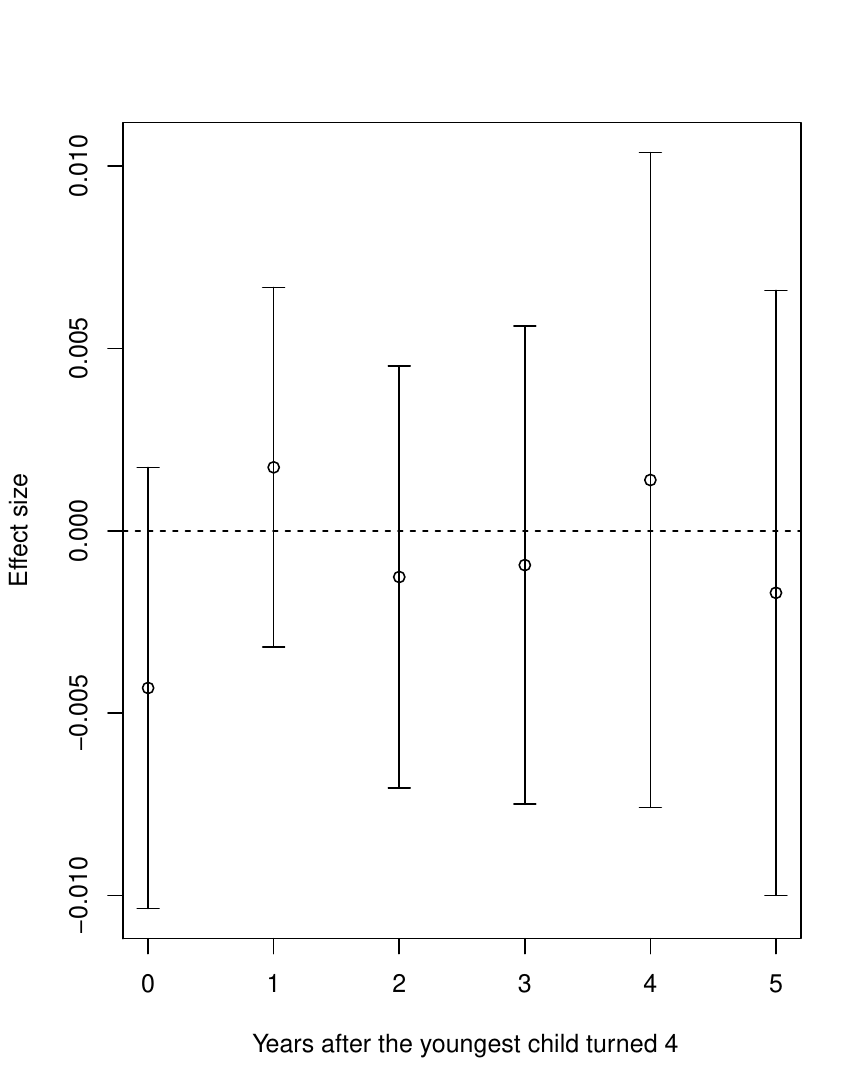}
       \label{fig:unem} }
       \quad
        \subfigure[Total annual work income(in CHF)]{%
      \includegraphics[width=0.3\textwidth]{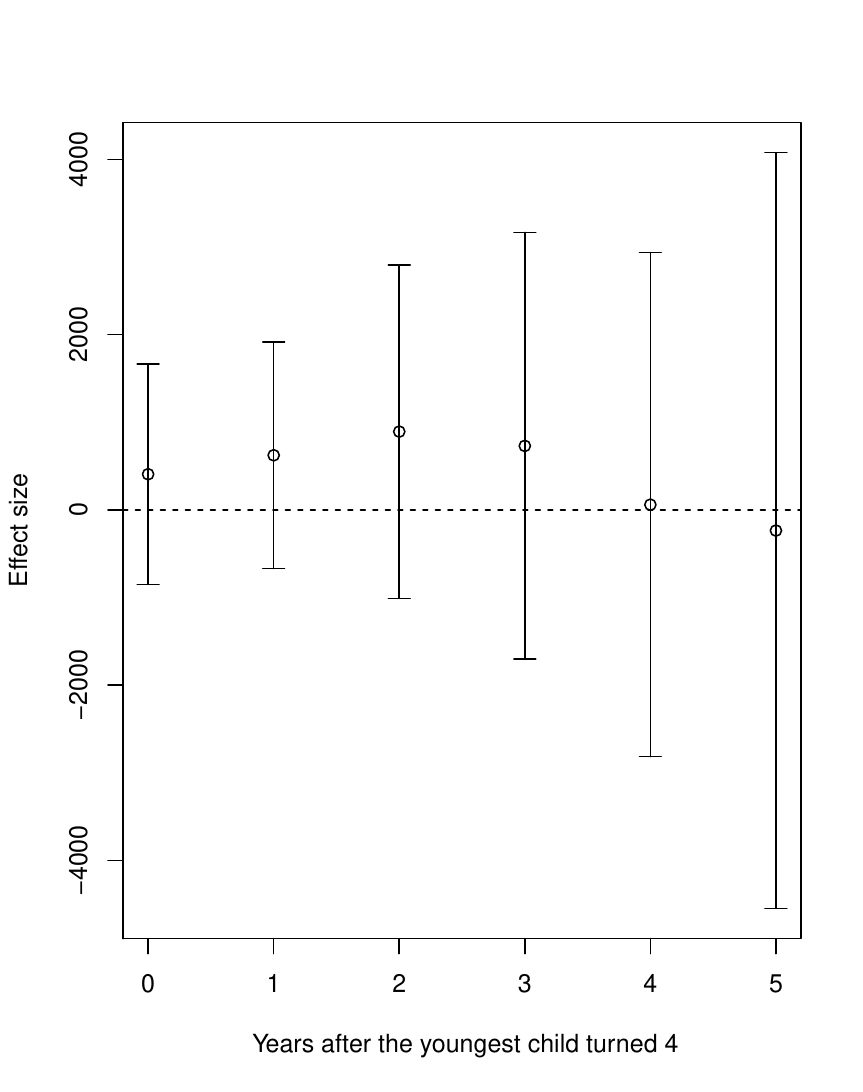}
      \label{fig:work income} }
      \subfigure[Income from dependent employment	 (binary)]{%
 \label{fig:icdum}
      \includegraphics[width=0.3\textwidth]{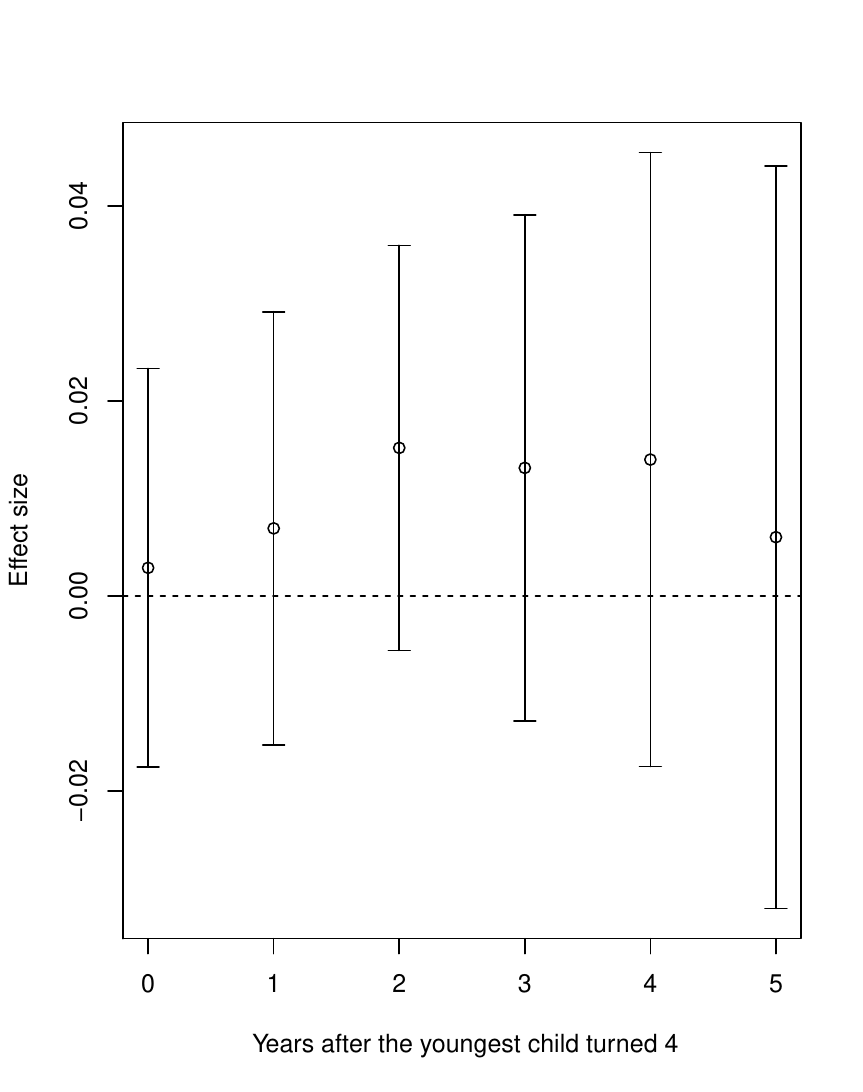}
       }
        \subfigure[Income from dependent employment (in CHF)]{%
        \label{fig:icdepe}
      \includegraphics[width=0.3\textwidth]{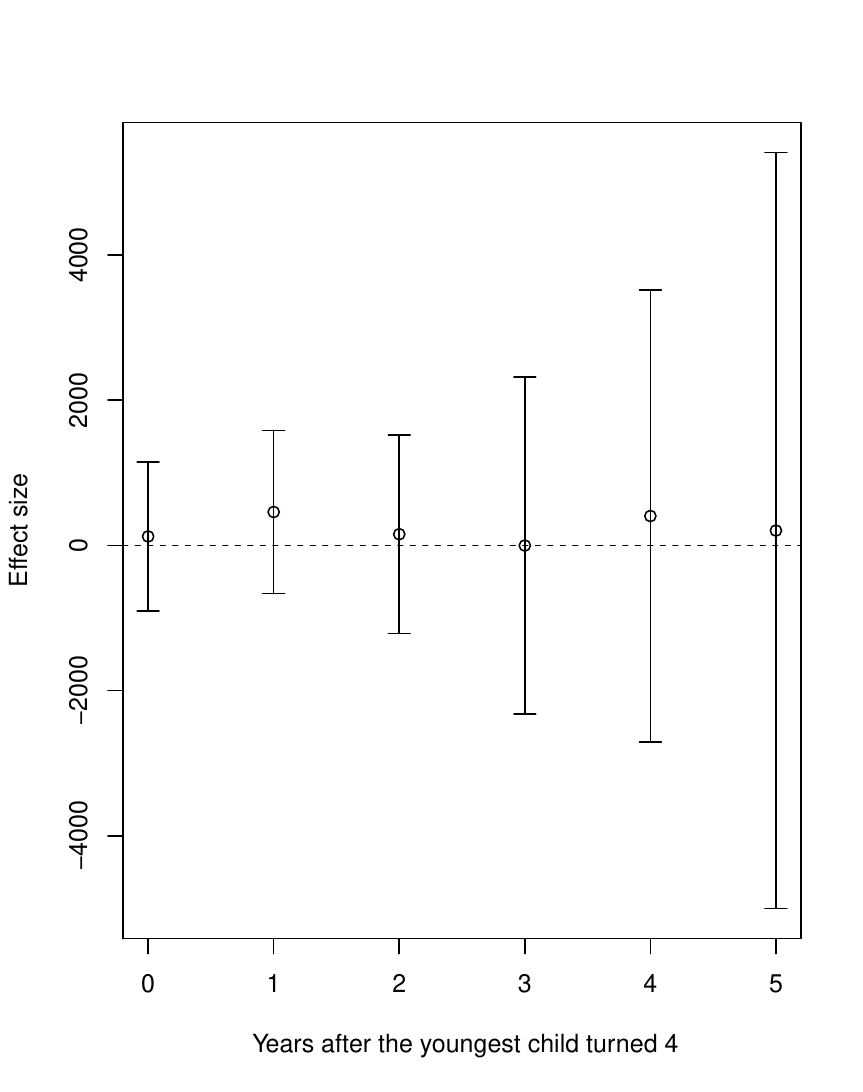}
       }

\begin{tablenotes}[flushleft]
\footnotesize
\item Note: This figure shows the effect of mandatory kindergarten on each outcome variable separately and over time. Dots represent ITTs at the threshold, bands correspond to 95\% confidence intervals.The following control variables are included: Number of children, born in Switzerland, resident permit, age, relationship status, labour market characteristics of the mother and the father of the four-year old child, cantonal characteristics and dummies. Data stems from STATPOP (2010 - 2017) and OASI (2010 - 2017), calculations are done by ourselves.  Income is deflated, the base year is 2011. The official currency in Switzerland is the Swiss Franc (CHF), which had an average exchange rate of 1.04 USD/CHF in the last decade.
\end{tablenotes}
\end{threeparttable}

 \label{fig:over_years_m_cov}

\end{figure}

	\clearpage

	\begin{landscape}
	\begin{table}[!h]
\center
\footnotesize
\caption{RDD: Heterogenous effects: Employed mothers}
\label{empl}
\centering
\begin{threeparttable}
 \renewcommand\TPTminimum{\textwidth} 
\begin{tabular}{lcccccp{2cm}}
 \hline\hline

  & Sample mean & Coefficient & Standard error & P-value & Bandwidth  & Observations within bandwidth \\
  \hline
Out of labour force (binary) & 0.06 & 0.00 & 0.01 & 0.58 & 67.03 & 183,521\\ 
  Employed (binary) & 0.93 & -0.00 & 0.01 & 0.72 & 67.66 & 183,521\\ 
  Unemployed (binary) & 0.01 & -0.00 & 0.00 & 0.54 & 51.19 & 138,858\\ 
  Annual work income (in CHF) & 46,460.42 & -531.87 & 698.65 & 0.45 & 72.22 & 196,930\\ 
  Income dependent employment (binary) & 0.90 & 0.00 & 0.01 & 0.96 & 61.60 & 167,217\\ 
  Annual income from dependent employment (in CHF) & 44,347.23 & -606.55 & 650.57 & 0.35 & 62.65 & 170,025\\ 
  \hline
\end{tabular}
\begin{tablenotes}[flushleft]
\footnotesize
\item Note:  This table reports the local linear estimates of  equation \ref{eqn2} only for previously employed mothers. Bandwidth shows the MSE-optimal bandwidth chosen by \citep{Calonico2021}. The following control variables are included: Number of children, born in Switzerland, resident permit, age, relationship status, labour market characteristics of the mother and the father of the four-year old child, cantonal characteristics and dummies. The standard errors are clustered at the individual level. Data stems from STATPOP (2010 - 2017) and OASI (2010 - 2017), calculations are done by ourselves.  Income is deflated, the base year is 2011. The official currency in Switzerland is the Swiss Franc (CHF), which had an average exchange rate of 1.04 USD/CHF in the last decade. Number of observations in total: 484,153.
  \end{tablenotes}
\end{threeparttable}
\end{table}

\begin{table}[!h]
\center
\footnotesize
\caption{RDD: Heterogenous effects: Non-employed mothers}
\label{nempl}
\centering
\begin{threeparttable}
 \renewcommand\TPTminimum{\textwidth} 
\begin{tabular}{lcccccp{2cm}}
 \hline\hline

  & Sample mean & Coefficient & Standard error & P-value & Bandwidth  & Observations within bandwidth \\
  \hline
Out of labour force (binary) & 0.70 & -0.04 & 0.02 & 0.05 & 59.46 & 59,463\\ 
  Employed (binary) & 0.30 & 0.04 & 0.02 & 0.04 & 58.26 & 58,459\\ 
  Unemployed (binary) & 0.01 & -0.00 & 0.00 & 0.39 & 59.39 & 59,463\\ 
  Annual work income (in CHF) & 5,582.01 & 936.74 & 652.42 & 0.15 & 55.40 & 55,452\\ 
  Income dependent employment (binary) & 0.28 & 0.04 & 0.02 & 0.03 & 57.77 & 57,634\\ 
  Annual income from dependent employment (in CHF) & 4,789.60 & 1,030.41 & 596.34 & 0.08 & 54.96 & 54,394\\ 

  \hline
\end{tabular}
\begin{tablenotes}[flushleft]
\footnotesize
\item Note:  This table reports the local linear estimates of  equation \ref{eqn2} only for previously non-employed mothers. Bandwidth shows the MSE-optimal bandwidth chosen by \citep{Calonico2021}. The following control variables are included: Number of children, born in Switzerland, resident permit, age, relationship status, labour market characteristics of the mother and the father of the four-year old child, cantonal characteristics and dummies. The standard errors are clustered at the individual level. Data stems from STATPOP (2010 - 2017) and OASI (2010 - 2017), calculations are done by ourselves.  Income is deflated, the base year is 2011. The official currency in Switzerland is the Swiss Franc (CHF), which had an average exchange rate of 1.04 USD/CHF in the last decade. Number of observations in total: 179,895.
 \end{tablenotes}
\end{threeparttable}
\end{table}

\end{landscape}

\section{Conclusion}\label{s:concl}

This paper analyses the impact of mandatory and cost-free kindergarten for four-year-old children on maternal labour market behaviour using Swiss administrative data.
Our identification strategy exploits  birthday cut-off dates as  random assignment of the mandatory kindergarten and compares labour market responses for mothers whose children enter kindergarten in the current year versus the following year. We apply a non-parametric regression discontinuity design (RDD) and find statistically significant effects on the employment and earnings of previously non-employed mothers, but not of the total sample of mothers.



%
%
Our findings contribute to the literature on childcare and school entry by analysing the mandatory and free of charge kindergarten.
Current literature suggests that attending childcare or school  affects maternal labour force participation positively in countries with low maternal labour force participation and low childcare attendance rate (see, for example,  \cite{NOLLENBERGER2015} or \cite{BAUERNSCHUSTER2015}). In Switzerland, maternal labour force participation exceeds the OECD average. Our study’s result  that mandatory kindergarten only sensibly affects labour market decisions of previously non-employed  mothers, but not of other groups, is in line with no or only modest effects of childcare on maternal employment found in countries with a high maternal labour force participation (see, e.g. \cite{Lundin2008} and \cite{HAVNES2011}). 

Furthermore,  working parents mainly use the support of grandparents for childcare in Switzerland  \citep{JacobsFoundation2018}. The insignificant  impact of mandatory kindergarten on  labour market response in the total sample of mothers might also be explained by a crowding out effect of informal childcare (see e.g. \cite{HAVNES2011}). Since mandatory kindergarten is  half-day care in most cantons, mothers are often faced with the challenge of balancing their work around the limited kindergarten hours or finding another (costly) care option. This circumstance might potentially account for the absence of a widespread impact of mandatory kindergarten on the earnings in the total sample of mothers. For future research, it would be interesting to examine whether a mandatory full-day kindergarten impacts maternal labour force participation and earnings.

\clearpage
\newpage

\begin{appendices}

\setcounter{section}{0}

\numberwithin{equation}{section}
\numberwithin{lemma}{section}
\numberwithin{proposition}{section}
\numberwithin{figure}{section}
\numberwithin{table}{section}

\section{Acknowledgement}\label{s:ack}
The study has been realised using the data from the Federal Statistical Office (FSO) and the Central Compensation Office (CCO). We are grateful for the data provision as well as the linking and anonymizing of the data conducted by the FOS.

\newpage

\section{Online Appendix}\label{s:app}
\newgeometry{includefoot,left=2cm,right=2cm,bottom=0.5cm,top=1cm}
 \subsection{Regression Discontinuity Design}\label{s:app_main}
\renewcommand\thetable{\thesection.1.\arabic{table}}
\renewcommand\thefigure{\thesection.1.\arabic{figure}}

\begin{samepage}
\begin{table}[h!]
\caption{Overview birthday cut-off dates}
\label{tab: Cut-off dates}
\centering
\begin{threeparttable}
{\footnotesize
\begin{tabular}{ |p{4.5cm}||p{4.5cm}|p{4.5cm}||}
\hline
    Canton  & Year(s) & Cut-off date \\
    \hline
   Basel-Stadt & 2005 - 10 & 30.04.\\

               & 2011 & 16.05.\\
               & 2012 & 01.06.\\
               & 2013 & 16.06.\\
               & 2014 & 01.07. \\
               & 2015 & 16.07. \\
               & 2016 & 31.07.  \\
               & 2017 & 31.07. \\
               \hline
   St Gallen & 2008 - 17 & 31.07.\\
   \hline
   Thurgau & 2008& 31.05.\tablefootnote{Municipalities can decide whether they use the 31.7. as cut-off date from 2008 an or use this stepwise approach: \url{https://www.tg.ch/news/news-detailseite.html/485/news/2817/newsarchive/1}, last downloaded 2019/10/07.}\\
   & 2009& 30.06.\\
   & 2010-17& 31.07.\\	
   \hline	
   Zurich & 2008-13 & 30.04.\\
    & 2014 & 15.05.\\
    & 2015 & 31.05.\\
    & 2016 & 15.06.\\
    & 2017 & 30.06.\\
    \hline
   Fribourg & 2009 - 17 & 31.07.\\
   \hline
  Geneva & 2011 - 17&  31.07.\\
  \hline
  Glarus & 2011 - 17 & 31.07. \\
  \hline
  Neuch\^{a}tel & 2011  - 17 & 31.07.\\
  \hline
  Basel-Land & 2012 & 15.05.\\\
  & 2013 & 31.05.\\
  & 2014 & 15.06.\\
  & 2015 & 30.06.\\
  & 2016 & 15.07.\\
  & 2017 & 31.07.\\
  \hline
  Jura & 2012 - 17 & 31.07.\\
  \hline
   Solothurn & 2012& 31.05.\\
   & 2013 & 30.06.\\
   & 2014 - 17 & 31.07.\\
   \hline
   Bern & 2013& 31.05.\\
   & 2014& 30.06.\\
   & 2015 - 17& 31.07.\\
   \hline
   Aargau & 2013 - 17 & 30.04. - 31.07.\tablefootnote{Transition period of 6 years.}\\
   \hline
   Vaud &2013 - 17 &31.07.\\
   \hline
   Schaffhausen& 2014 & 30.06.\\
   & 2015 - 17 & 31.07.\\
   \hline
   Ticino & 2015 - 17 & 31.07.\\
   \hline
   Upper Valais & 2015 & \emph{\color{blue}28.02.}\\
   & 2016 & 30.04.\\
   & 2017 & 30.06.\\
   \hline
   Central and Lower Valais & 2015 & \emph{\color{blue}31.08.}\\
    &2016& 31.07.\\
   &2017& 31.07.\\

\hline
\end{tabular}
}
\begin{tablenotes}[flushleft]
\footnotesize
\item Note: This table reports the birthday cut-off dates for the cantons with mandatory kindergarten for four-year-old children. Dates in \emph{\color{blue}blue} represent the earliest or the latest cut-off ever.
 \smallskip
 \singlespace
\item Sources: \cite{Basel-Stadt2010}, \cite{Basel-Stadt2013}, \cite{Basel-Stadt2016}, \cite{StGallen2007}, \cite{Thurgau2007}, \cite{Kantonsrat}, \cite{Kuesnacht}, \cite{Fribourg2008}, \cite{Genf2010}, \cite{Glarus2009}, \cite{Neuchatel2011}, \cite{BaselLand2011}, \cite{Jura2011},  \cite{Solothurn2012}, \cite{Bern2013}, \cite{Aargau2010}, \cite{Schaffhausen2014}, \cite{Tessin2011}, and \cite{Wallis2014}.
\end{tablenotes}
\end{threeparttable}
\end{table}

\end{samepage}
\clearpage

\begin{figure}[h!]
 \centering
 \caption{RDD: Baseline maternal covariates}
  \label{fig:baseline_cov_mother}
       \begin{threeparttable}
        \subfigure[Number Children]{%
      \includegraphics[width=0.3\textwidth]{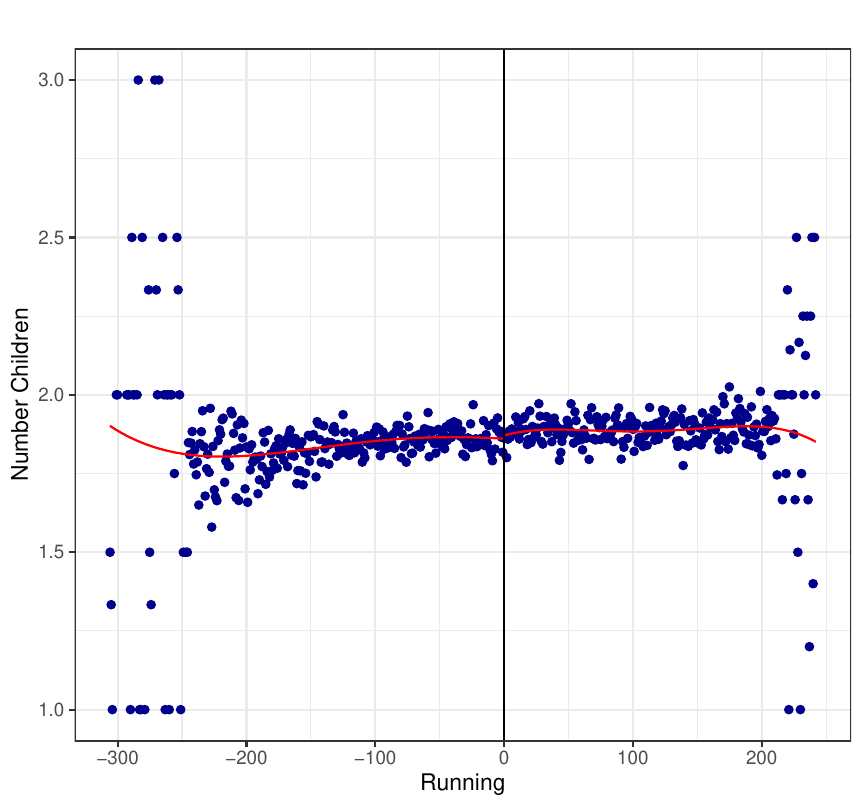}
       \label{fig:out}}
        \quad
        \subfigure[Born in Switzerland]{%
      \includegraphics[width=0.3\textwidth]{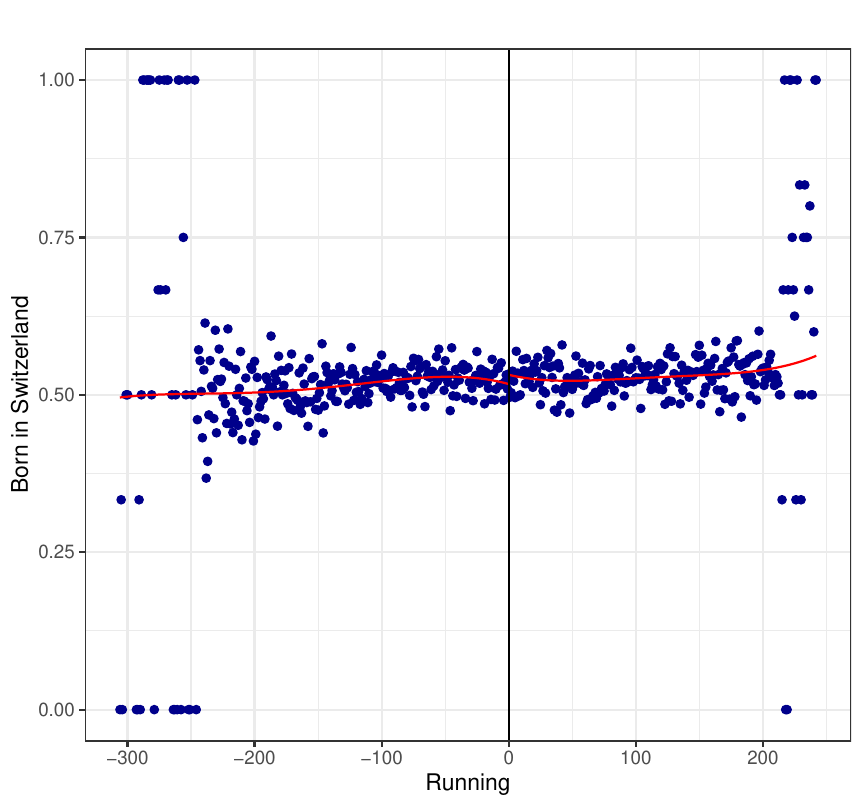}
       \label{fig:emp}}
        \subfigure[Resident permit B]{%
      \includegraphics[width=0.3\textwidth]{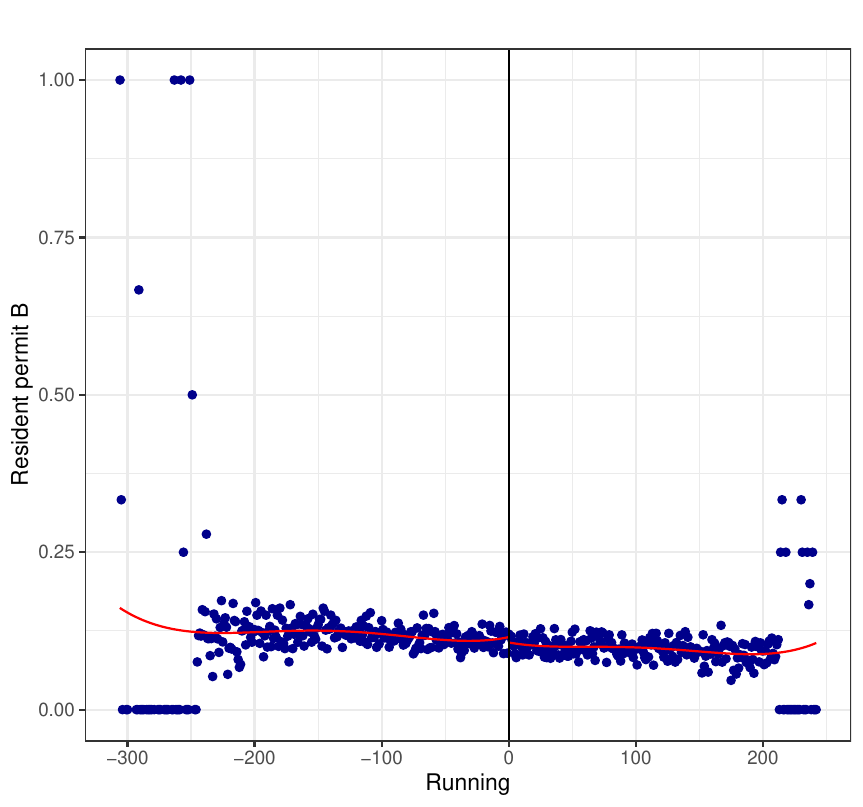}
       \label{fig:unem} }
       \quad
        \subfigure[Resident permit C]{%
      \includegraphics[width=0.3\textwidth]{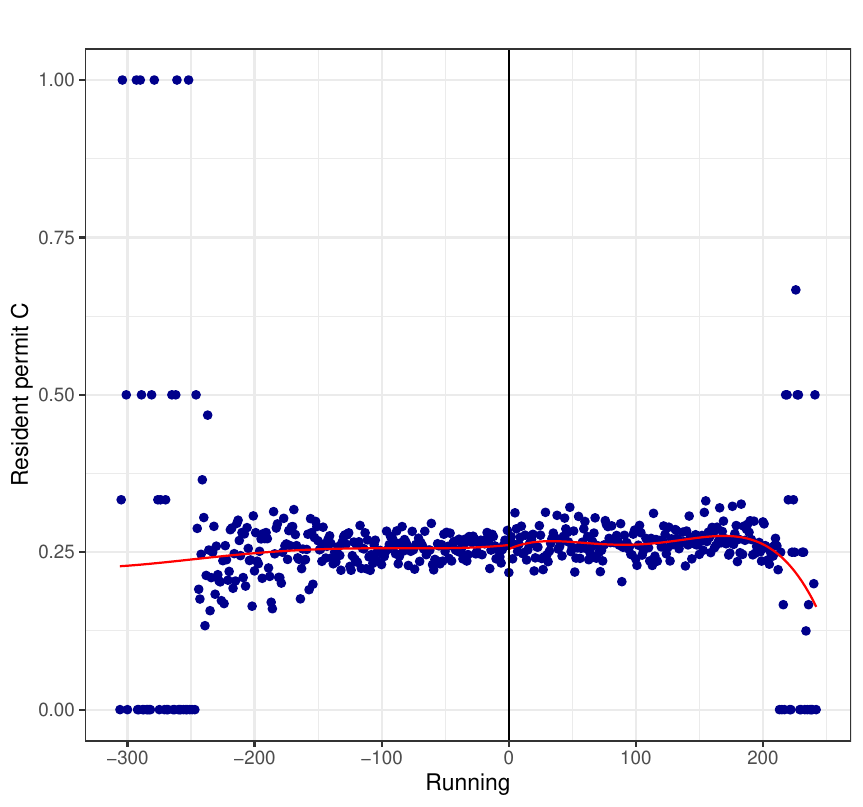}
      \label{fig:work income} }
      \subfigure[Other resident]{%
 \label{fig:icdum}
      \includegraphics[width=0.3\textwidth]{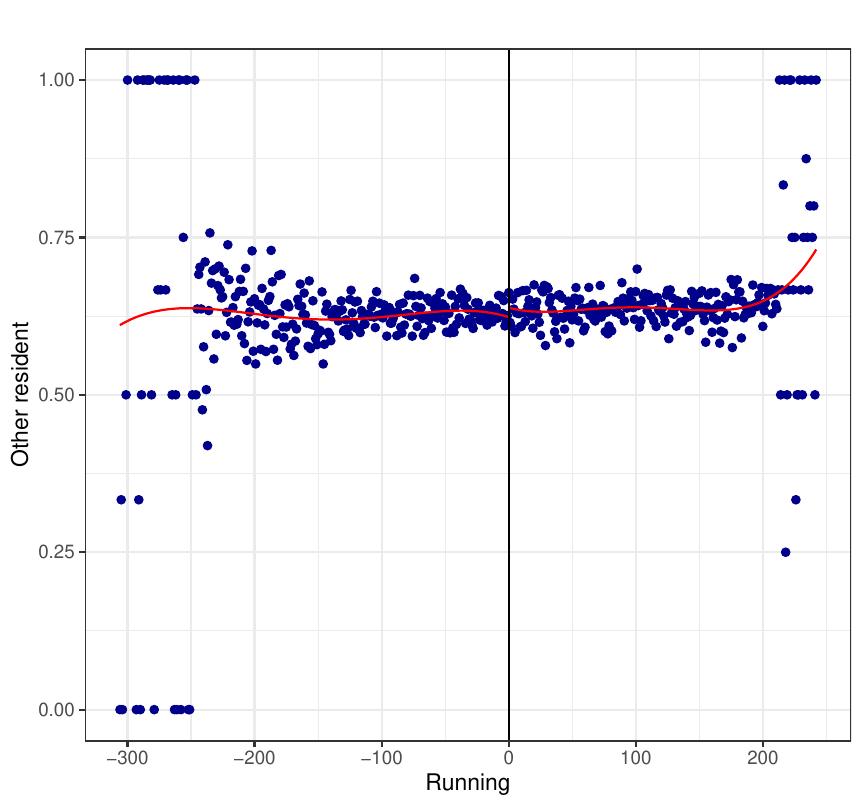}
       }
        \subfigure[Age]{%
        \label{fig:icdepe}
      \includegraphics[width=0.3\textwidth]{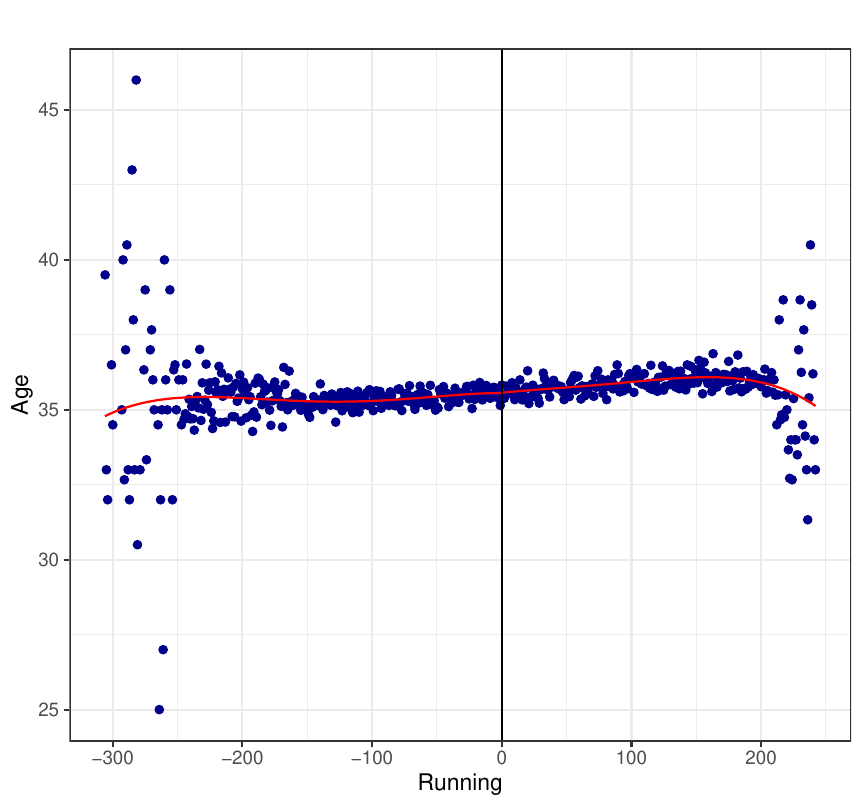}
       }
  \subfigure[In relationship]{%
      \includegraphics[width=0.3\textwidth]{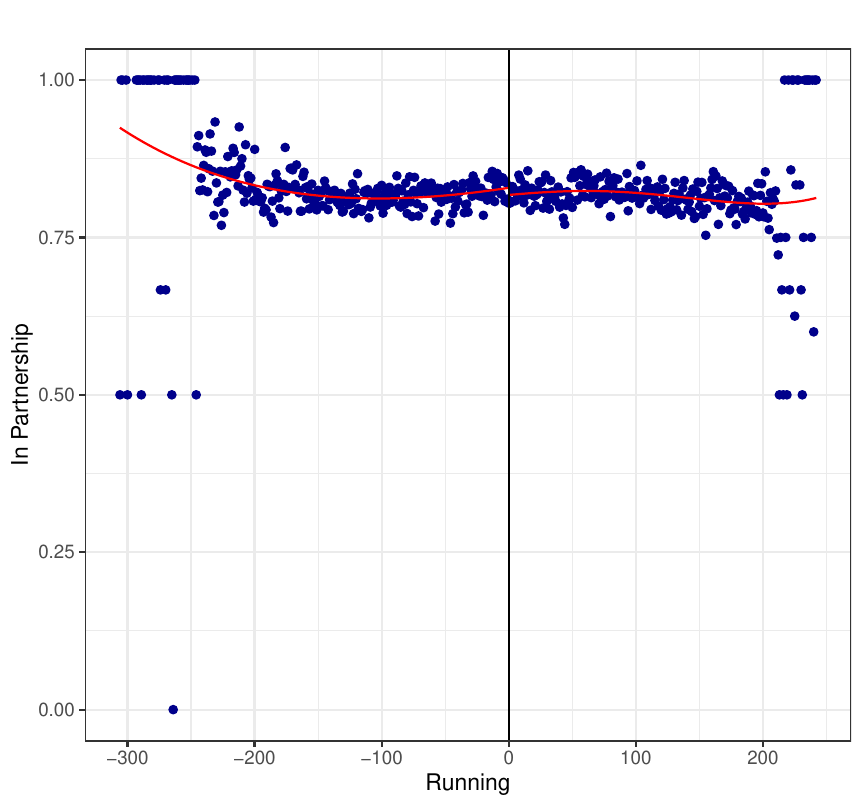}
       \label{fig:out}}
        \quad
        \subfigure[Not in relationship]{%
      \includegraphics[width=0.3\textwidth]{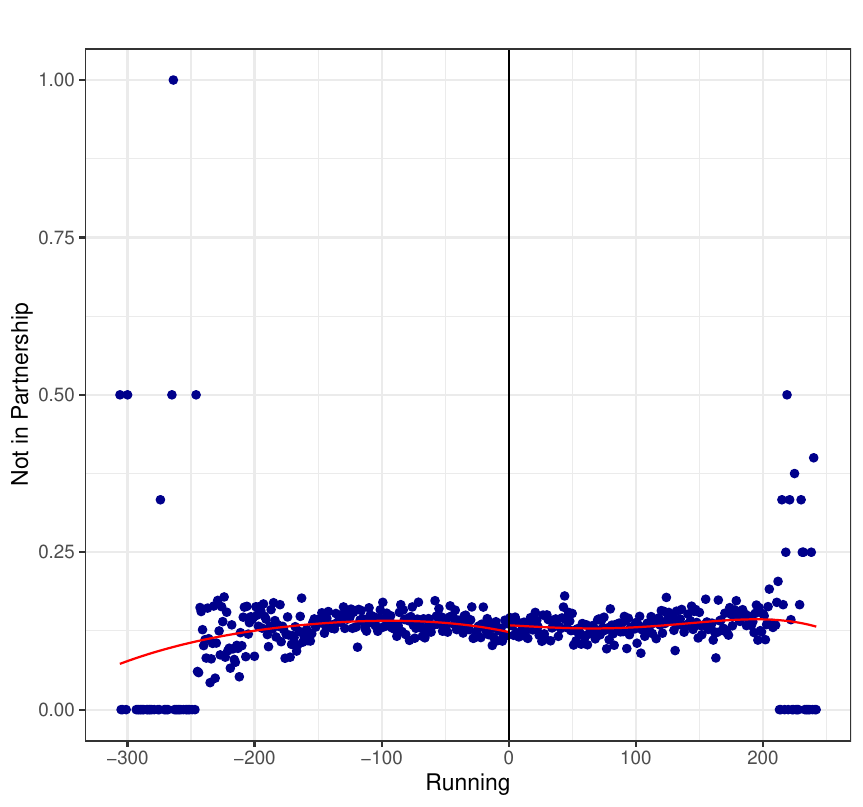}
       \label{fig:emp}}
        \subfigure[Terminated relationship]{%
      \includegraphics[width=0.3\textwidth]{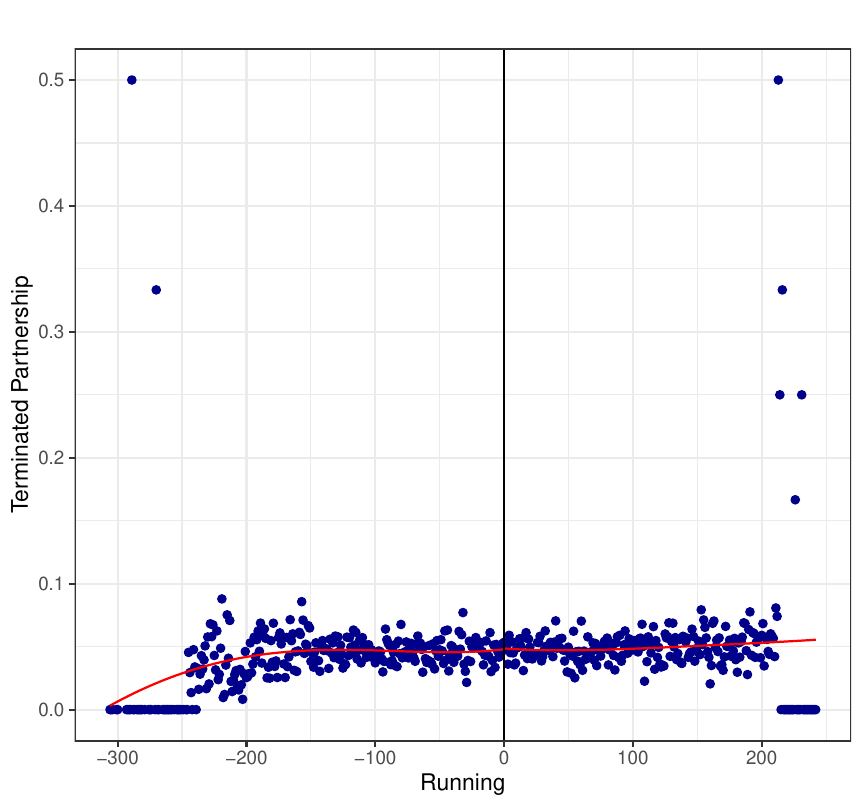}
       \label{fig:unem} }
     \begin{tablenotes}[flushleft]
\footnotesize
\item Note: The figure provides plots for baseline maternal covariates. The dots represent the conditional means of the respective covariate within bins defined upon values of the running variable and the solid line shows the quartic global polynomial fit when regressing the respective covariate on the running variable above and below the cut-off, respectively. The bin width is chosen according to the default option in the rdplot command of the rdrobust package. Data stems from STATPOP (2010 - 2017) and OASI (2010 - 2017), calculations are done by ourselves.
\end{tablenotes}
\end{threeparttable}
       \end{figure}

       \begin{figure}[ht]
       \centering
        \label{fig:baseline_cov_labour_market_mother}
       \caption{RDD: Baseline maternal labour market covariates}
       \begin{threeparttable}
         \subfigure[Out of labour force (binary)]{%
      \includegraphics[width=0.3\textwidth]{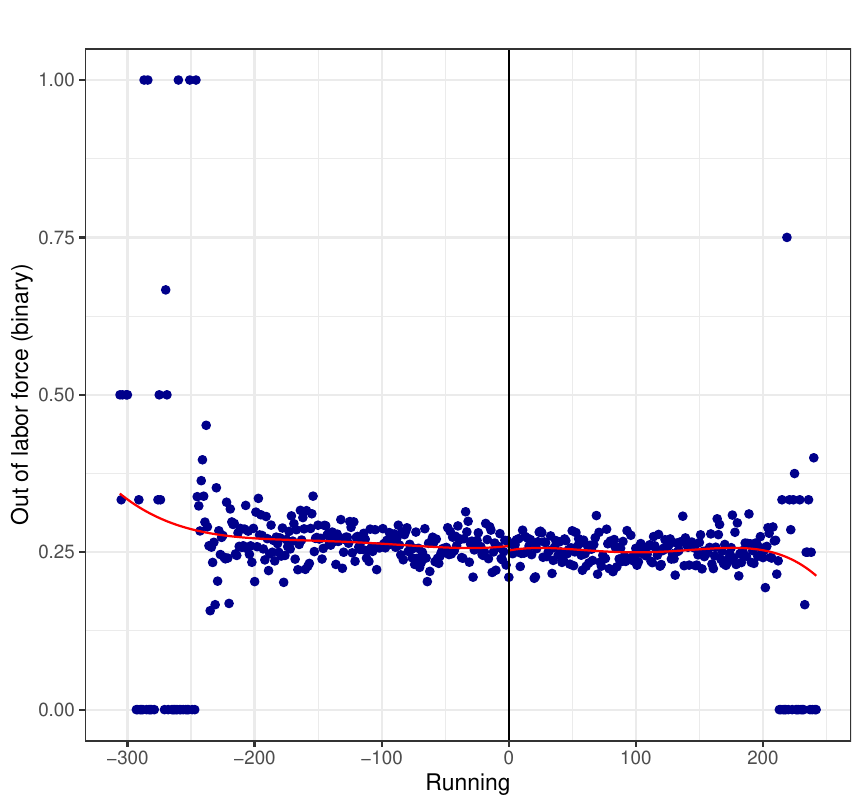}
      \label{fig:work income} }
      \subfigure[Employed (binary)]{%
 \label{fig:icdum}
      \includegraphics[width=0.3\textwidth]{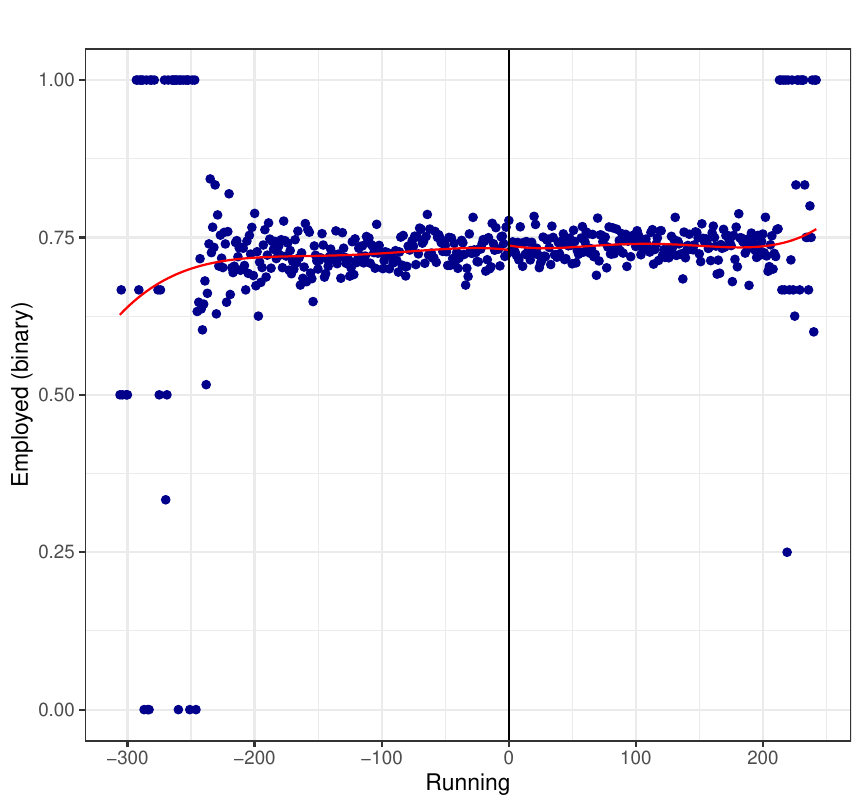}
       }
        \subfigure[Unemployed (binary)]{%
        \label{fig:icdepe}
      \includegraphics[width=0.3\textwidth]{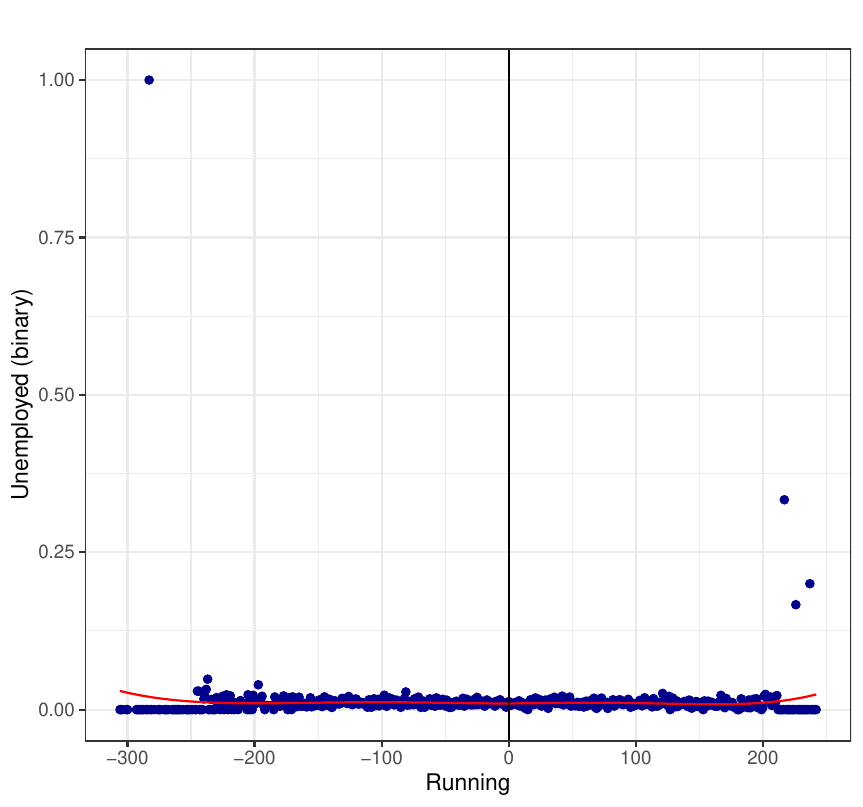}
       }
        \subfigure[Total income from work (in CHF)]{%
      \includegraphics[width=0.3\textwidth]{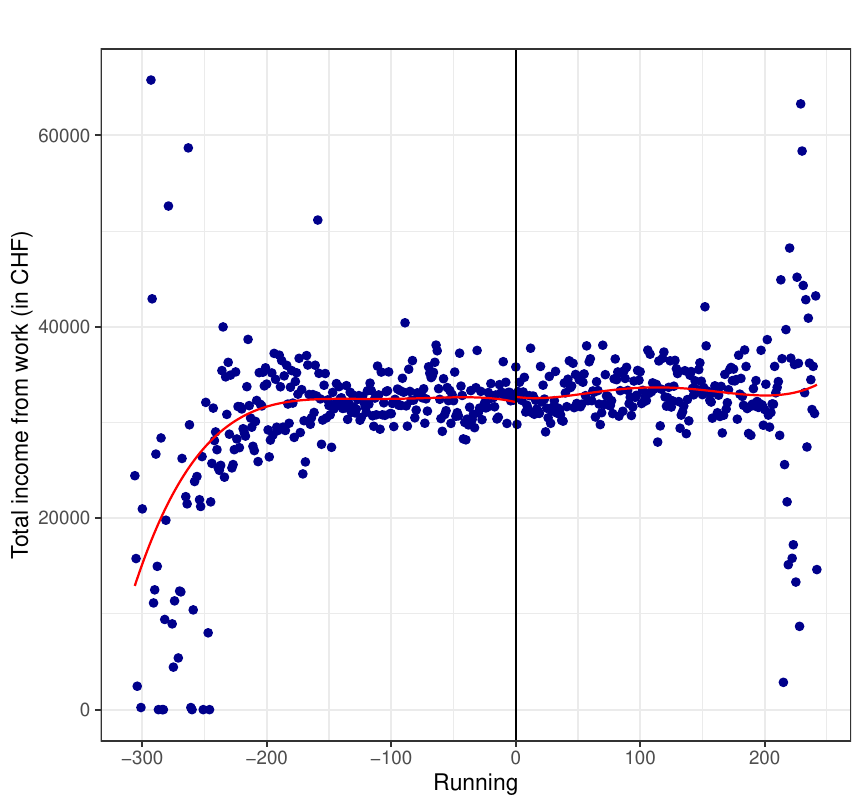}
       \label{fig:emp}}
        \subfigure[Income from dependent employment (binary)]{%
      \includegraphics[width=0.3\textwidth]{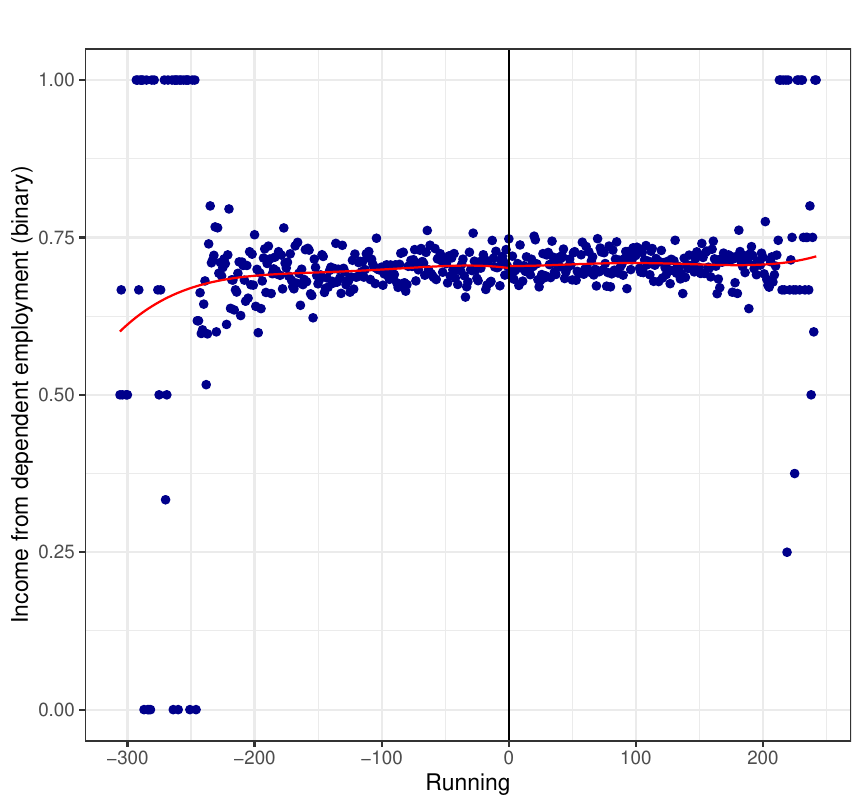}
       \label{fig:unem} }
       \quad
        \subfigure[Income from dependent employment (in CHF)]{%
      \includegraphics[width=0.3\textwidth]{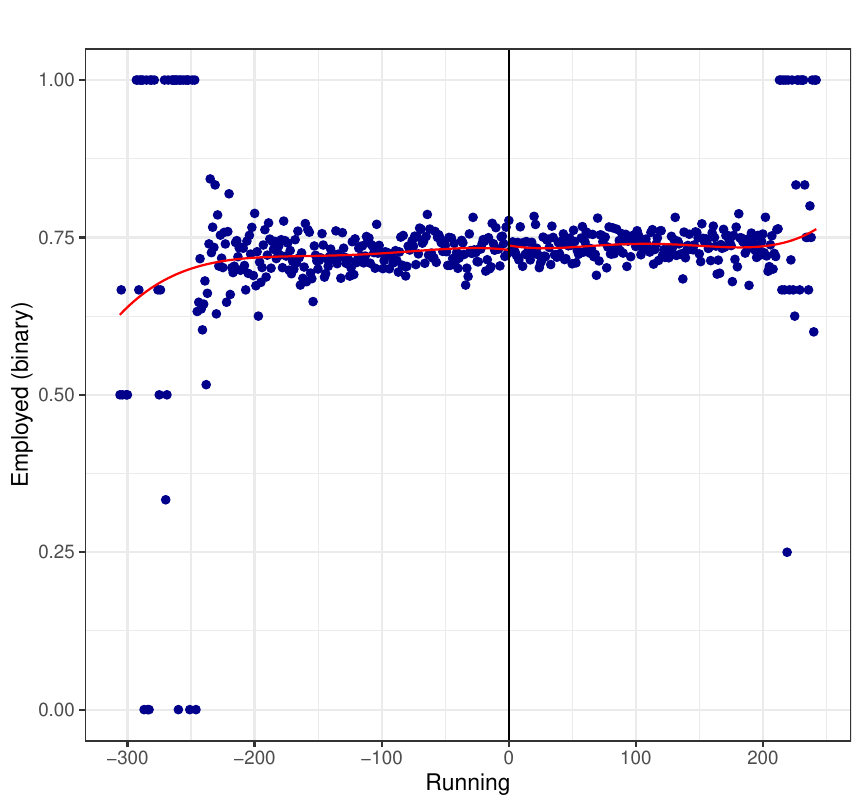}
      \label{fig:work income} }

\begin{tablenotes}[flushleft]
\footnotesize
\item Note: The figure provides plots for pre-treatment maternal labour market outcomes. The dots represent the conditional means of the respective variable within bins defined upon values of the running variable and the solid line shows the quartic global polynomial fit when regressing the respective variable on the running variable above and below the cut-off, respectively. The bin width is chosen according to the default option in the rdplot command of the rdrobust package. Data stems from STATPOP (2010 - 2017) and OASI (2010 - 2017), calculations are done by ourselves.  Income is deflated, the base year is 2011. The official currency in Switzerland is the Swiss Franc (CHF), which had an average exchange rate of 1.04 USD/CHF in the last decade.
\end{tablenotes}
\end{threeparttable}
\end{figure}

       \begin{figure}[ht]
       \centering
               \label{fig:baseline_cov_labour_market_father}
       \caption{RDD: Baseline paternal labour market covariates}
       \begin{threeparttable}
         \subfigure[Out of labour force (binary)]{%
      \includegraphics[width=0.3\textwidth]{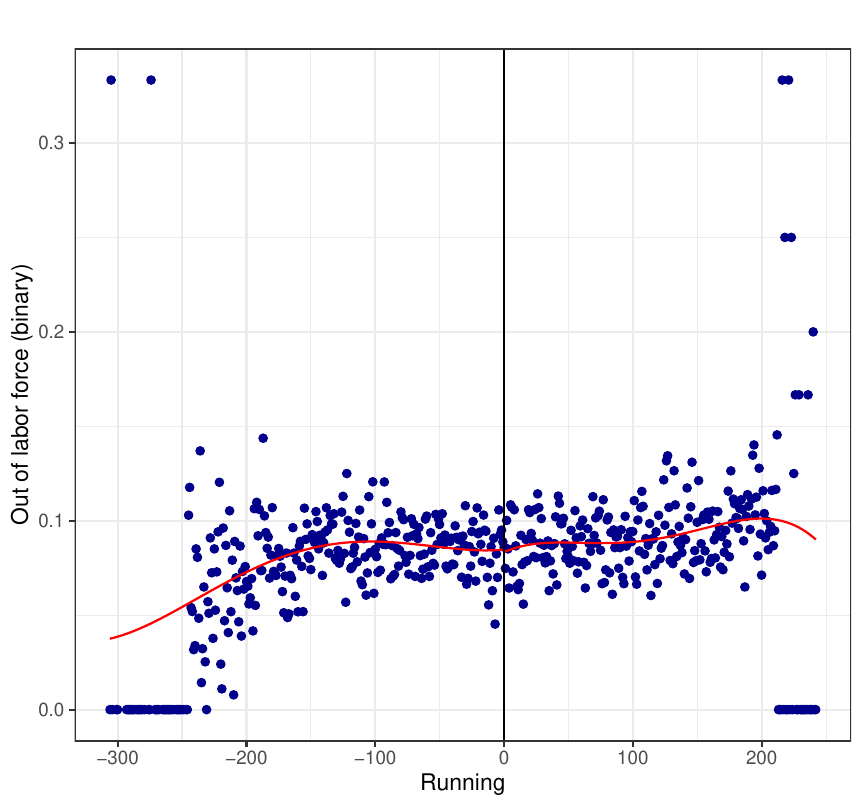}
      \label{fig:work income} }
      \subfigure[Employed (binary)]{%
 \label{fig:icdum}
      \includegraphics[width=0.3\textwidth]{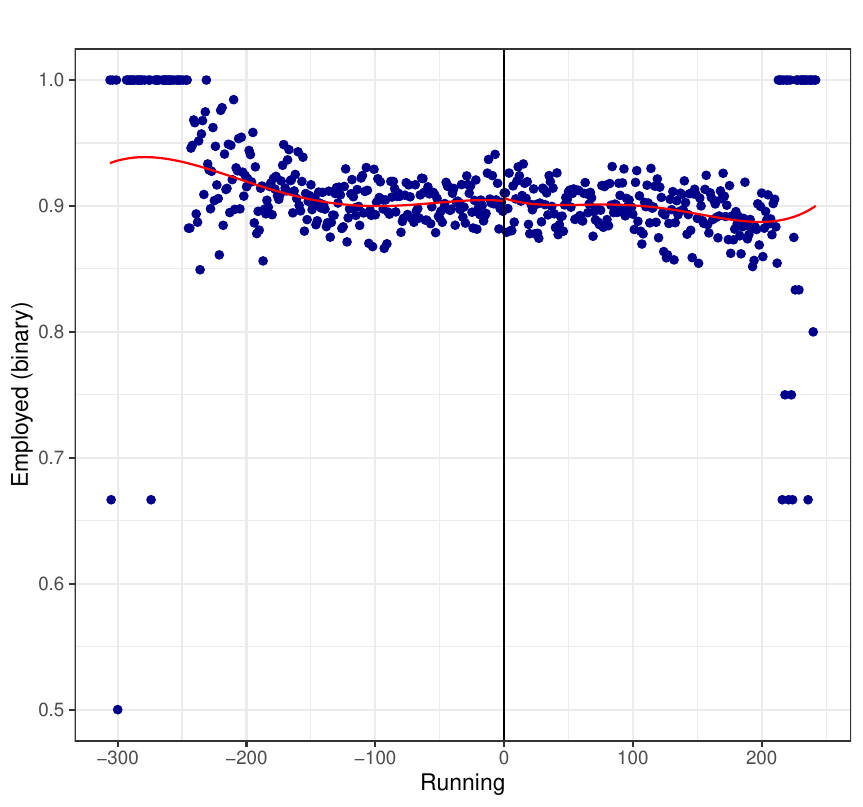}
       }
        \subfigure[Unemployed (binary)]{%
        \label{fig:icdepe}
      \includegraphics[width=0.3\textwidth]{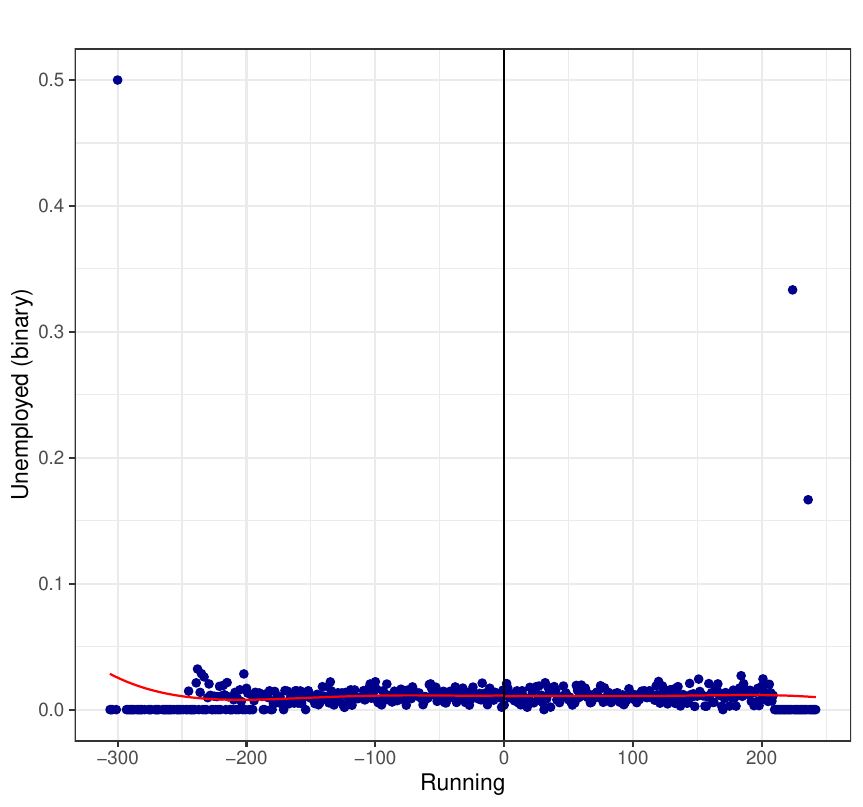}
       }
        \subfigure[Total income from work (in CHF)]{%
      \includegraphics[width=0.3\textwidth]{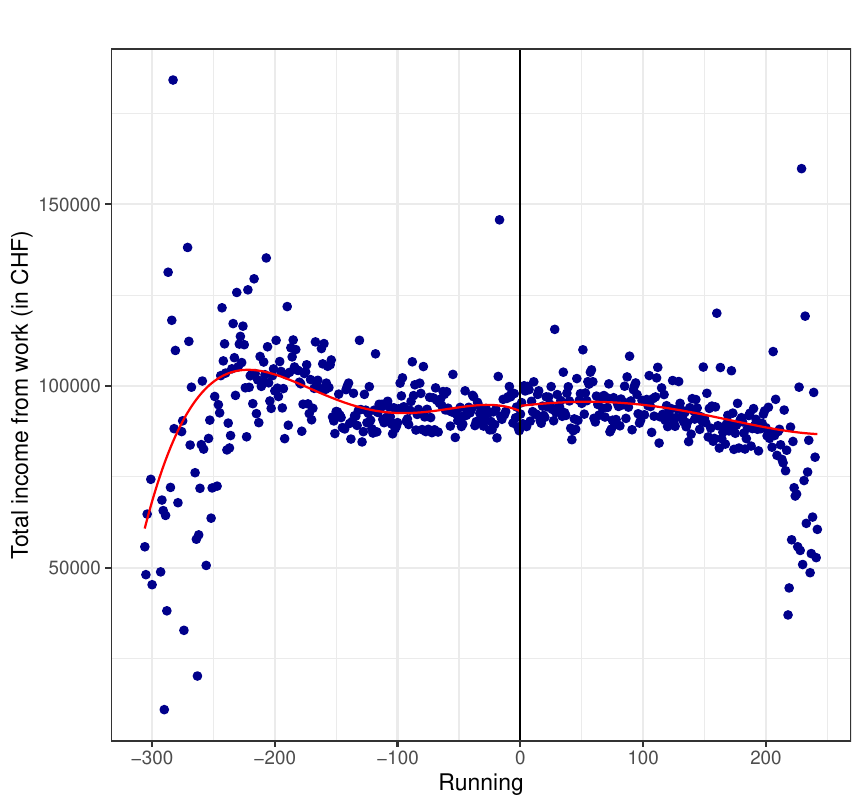}
       \label{fig:emp}}
        \subfigure[Income from dependent employment (binary)]{%
      \includegraphics[width=0.3\textwidth]{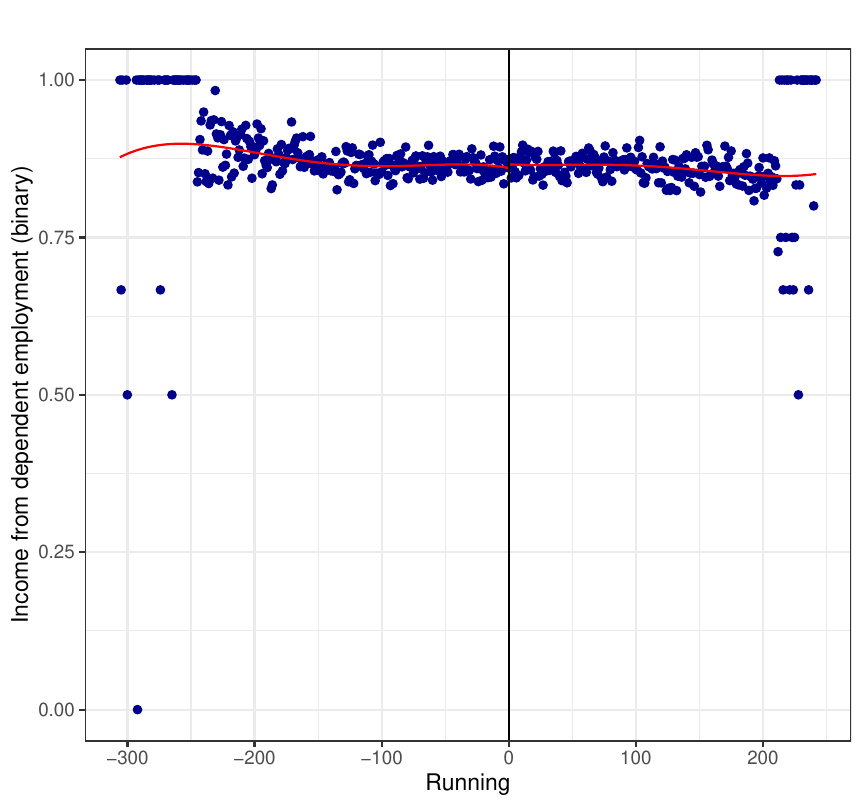}
       \label{fig:unem} }
       \quad
        \subfigure[Income from dependent employment (in CHF)]{%
      \includegraphics[width=0.3\textwidth]{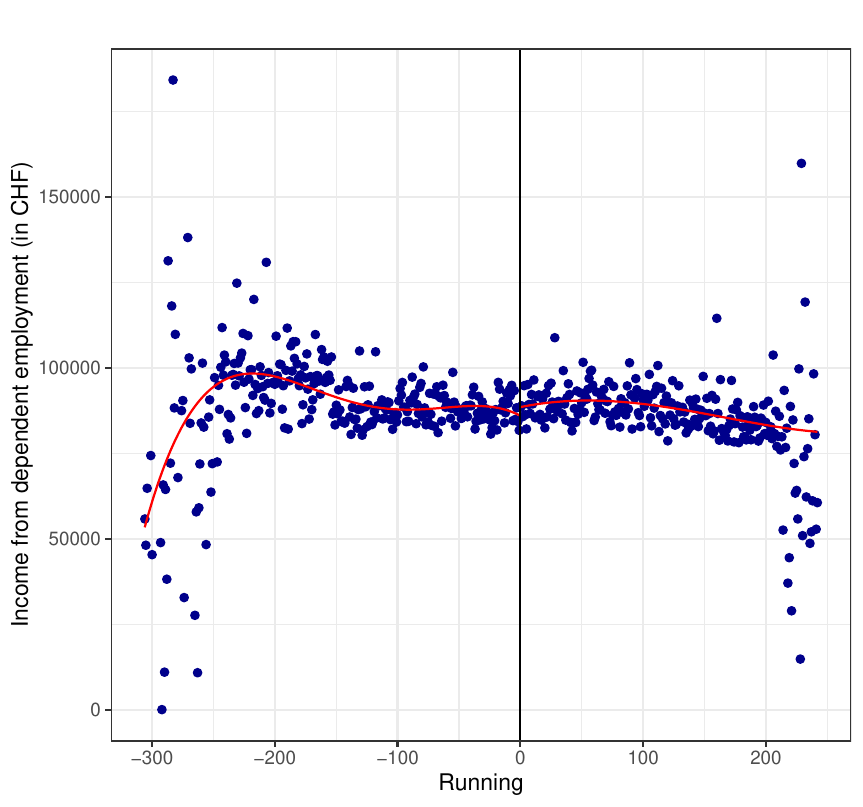}
      \label{fig:work income} }

\begin{tablenotes}[flushleft]
\footnotesize
\item Note: The figure provides plots for pre-treatment paternal labour market outcomes. The dots represent the conditional means of the respective variable within bins defined upon values of the running variable and the solid line shows the quartic global polynomial fit when regressing the respective variable on the running variable above and below the cut-off, respectively. The bin width is chosen according to the default option in the rdplot command of the rdrobust package. Data stems from STATPOP (2010 - 2017) and OASI (2010 - 2017), calculations are done by ourselves.  Income is deflated, the base year is 2011. The official currency in Switzerland is the Swiss Franc (CHF), which had an average exchange rate of 1.04 USD/CHF in the last decade.
\end{tablenotes}
\end{threeparttable}
\end{figure}

       \begin{figure}[ht]
       \centering
       \label{fig:baseline_cov_cantonal_dummies}
       \caption{RDD: Baseline cantonal covariates}
       \begin{threeparttable}
         \subfigure[HarmoS member]{%
      \includegraphics[width=0.3\textwidth]{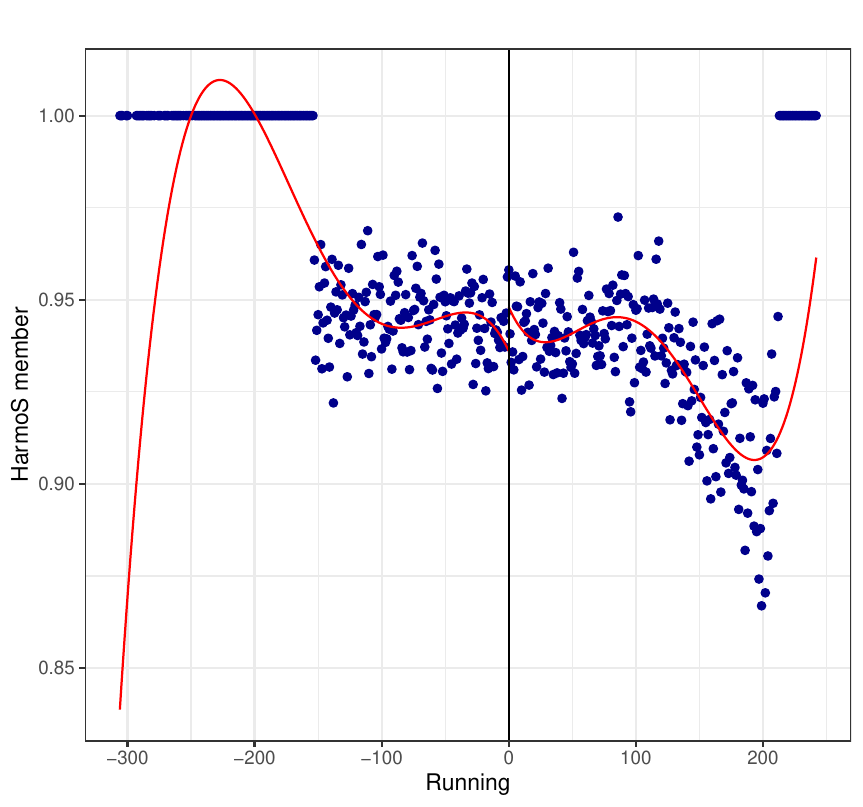}
      \label{fig:work income} }
      \subfigure[Unemployment rate]{%
 \label{fig:icdum}
      \includegraphics[width=0.3\textwidth]{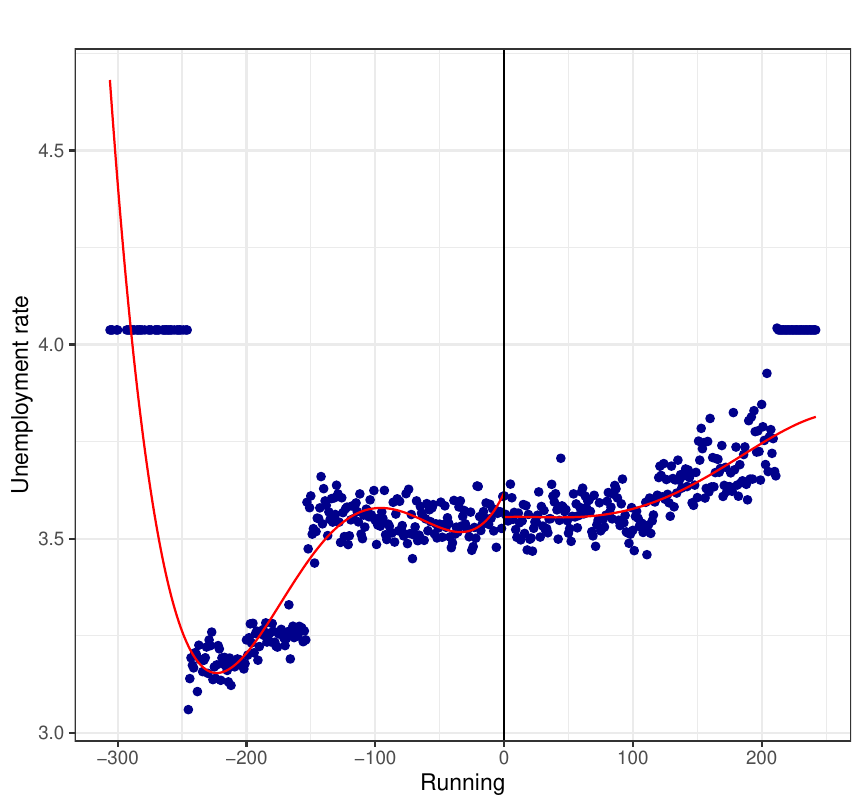}
       }

\begin{tablenotes}[flushleft]
\footnotesize
\item Note: The figure provides plots for the baseline cantonal covariates. The dots represent the conditional means of the respective covariate within bins defined upon values of the running variable and the solid line shows the quartic global polynomial fit when regressing the respective covariate on the running variable above and below the cut-off, respectively. The bin width is chosen according to the default option in the rdplot command of the rdrobust package. Data stems from STATPOP (2010 - 2017) and OASI (2010 - 2017), calculations are done by ourselves.
\end{tablenotes}
\end{threeparttable}
\end{figure}

\begin{figure}[h!]
\centering
   \caption{RDD: Baseline cantonal covariates}
\label{fig:baseline_cov_cantons}

        \subfigure[Basel-Stadt]{%
      \includegraphics[width=0.3\textwidth]{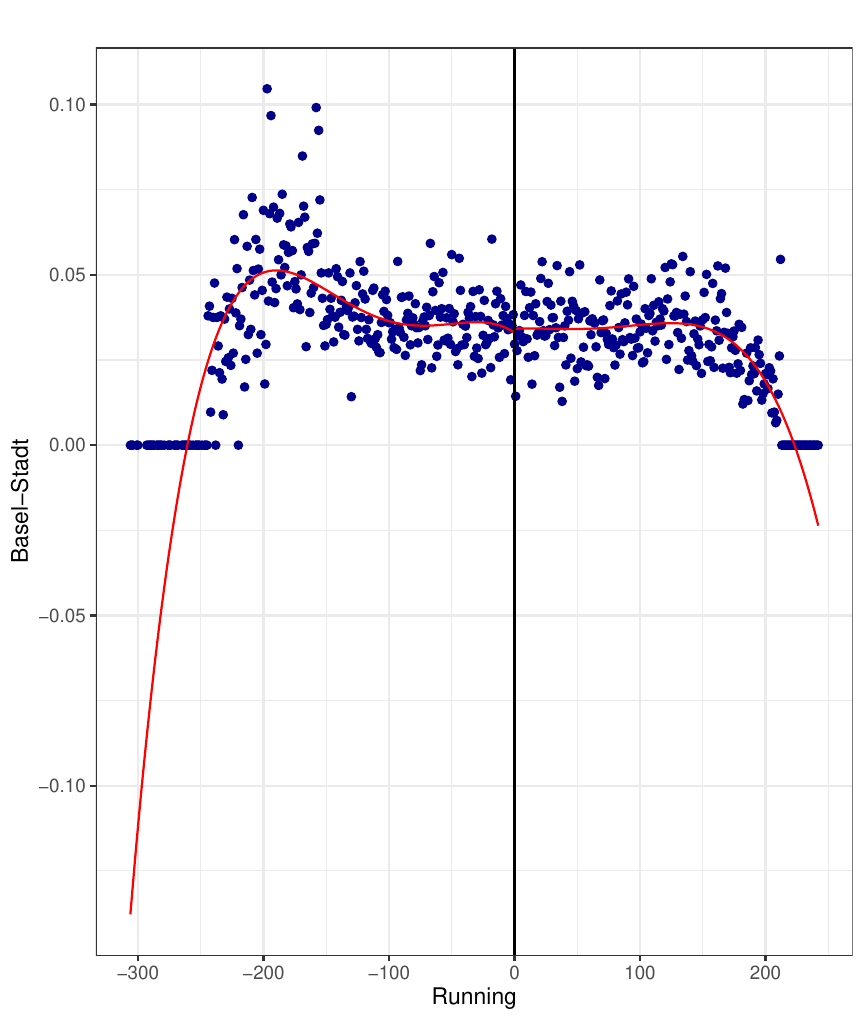}
       \label{fig:out}}
        \quad
        \subfigure[St Gallen]{%
      \includegraphics[width=0.3\textwidth]{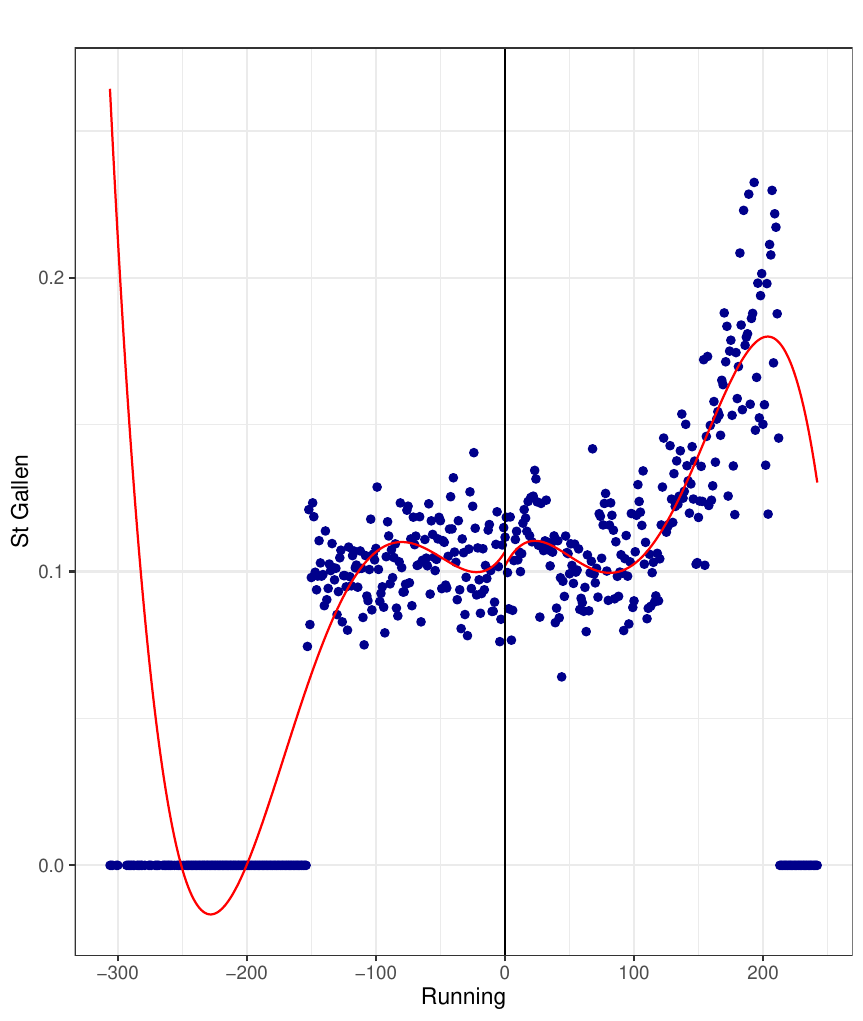}
       \label{fig:emp}}
        \subfigure[Thurgau]{%
      \includegraphics[width=0.3\textwidth]{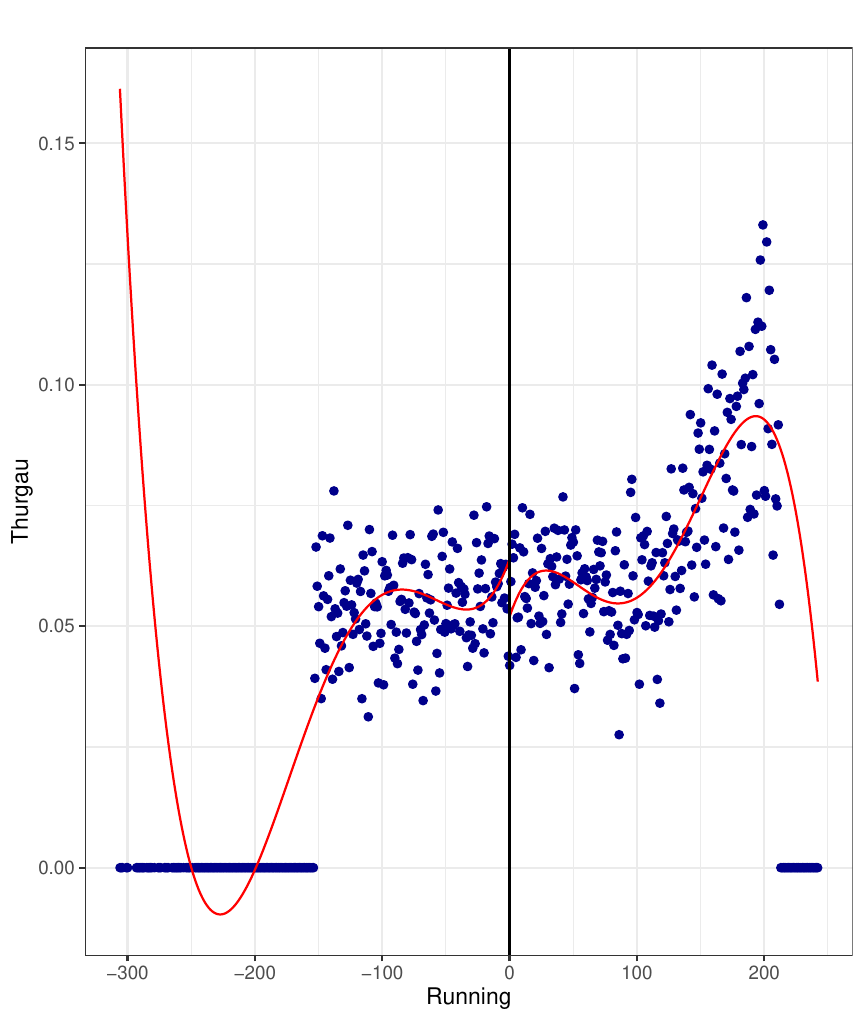}
       \label{fig:unem} }
       \quad
        \subfigure[Zurich]{%
      \includegraphics[width=0.3\textwidth]{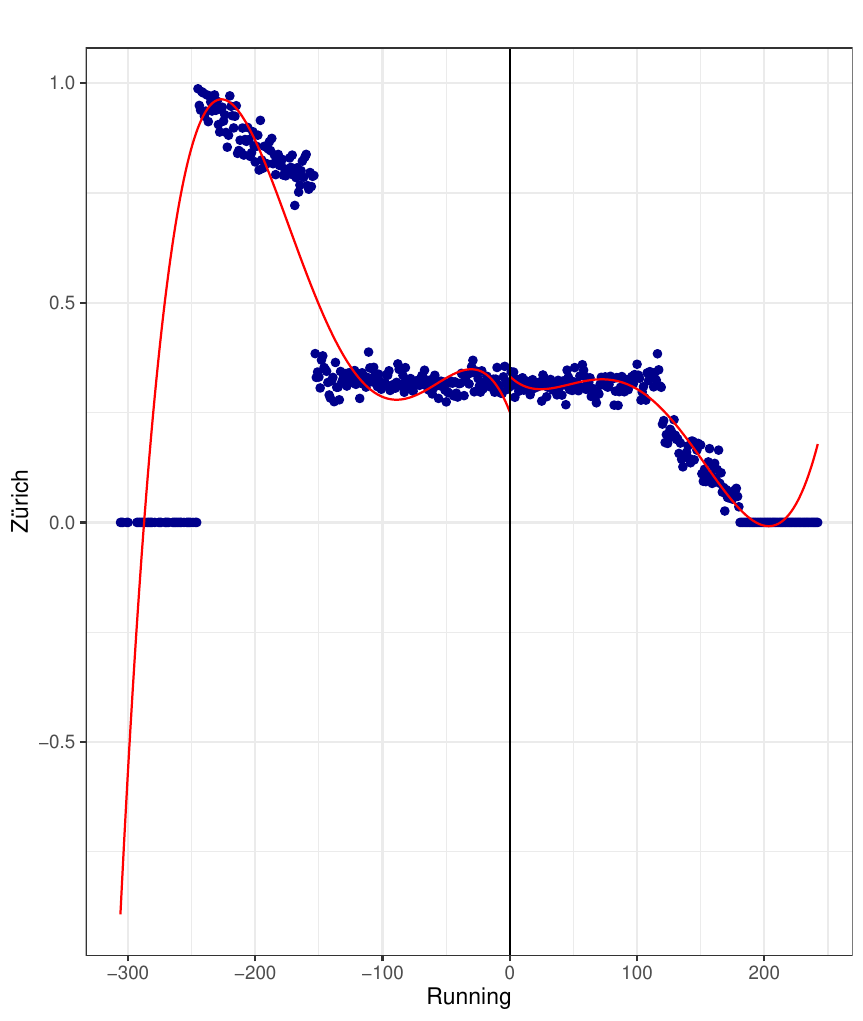}
      \label{fig:work income} }
      \subfigure[Fribourg]{%
 \label{fig:icdum}
      \includegraphics[width=0.3\textwidth]{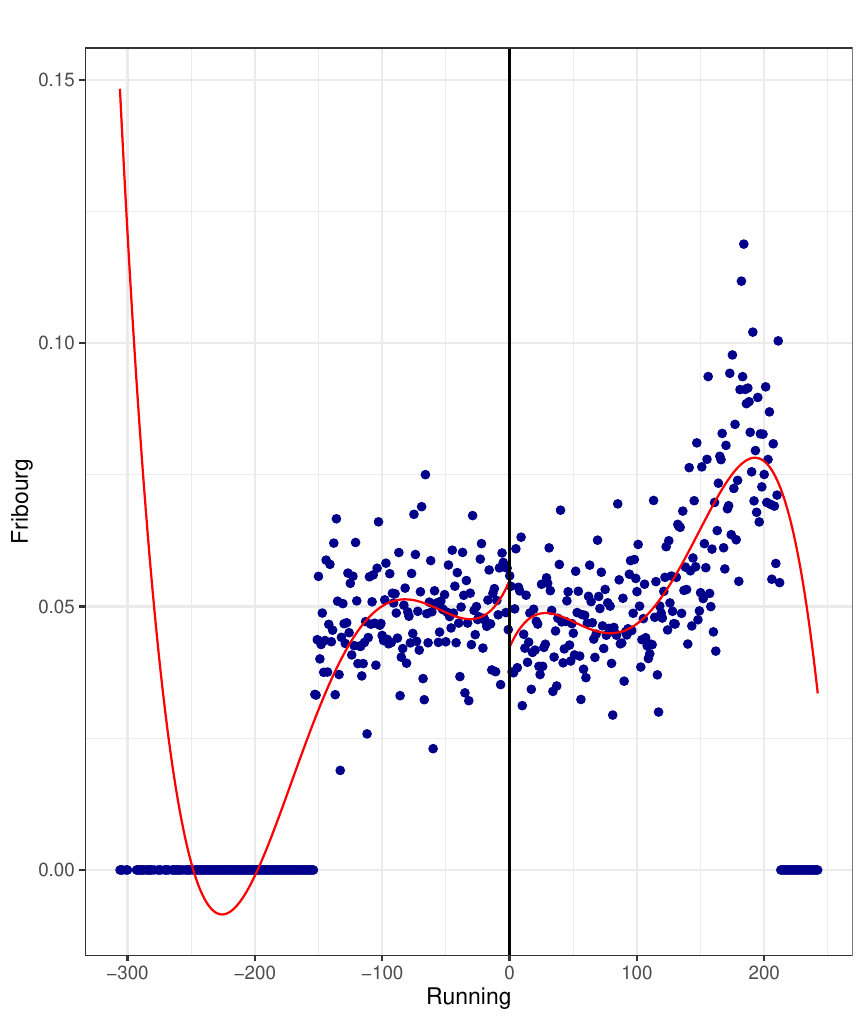}
       }
        \subfigure[Geneva]{%
        \label{fig:icdepe}
      \includegraphics[width=0.3\textwidth]{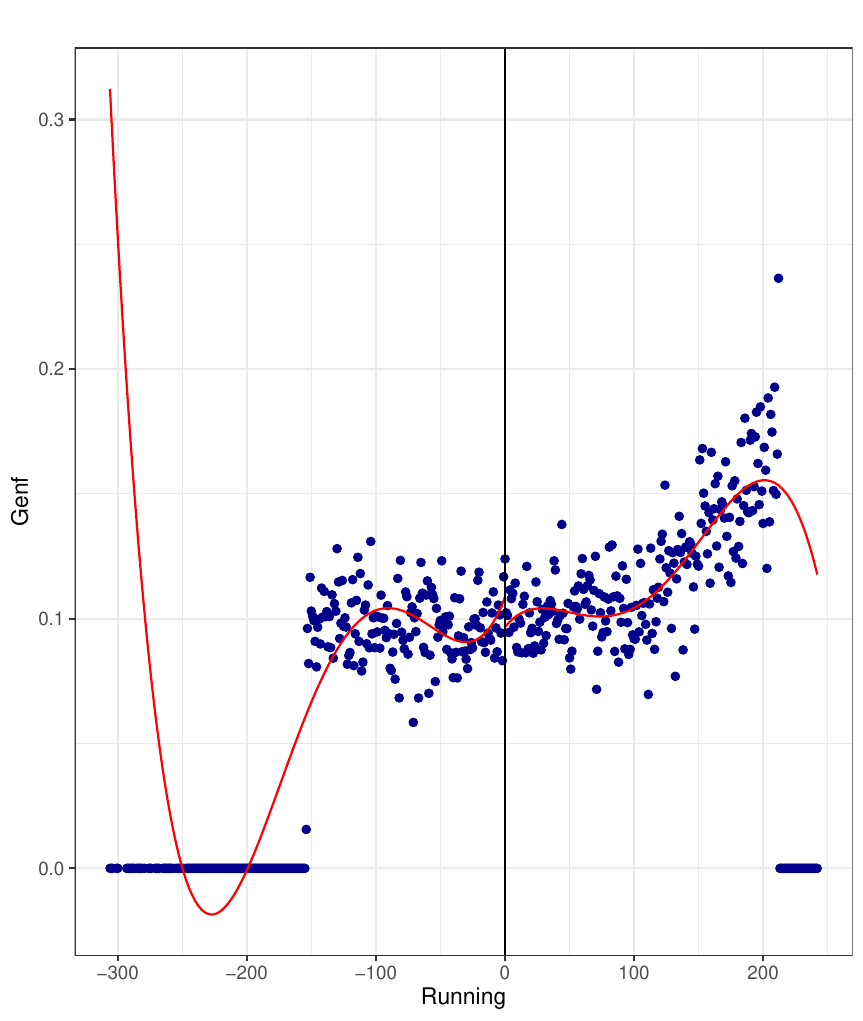}
       }
  \subfigure[Glarus]{%
      \includegraphics[width=0.3\textwidth]{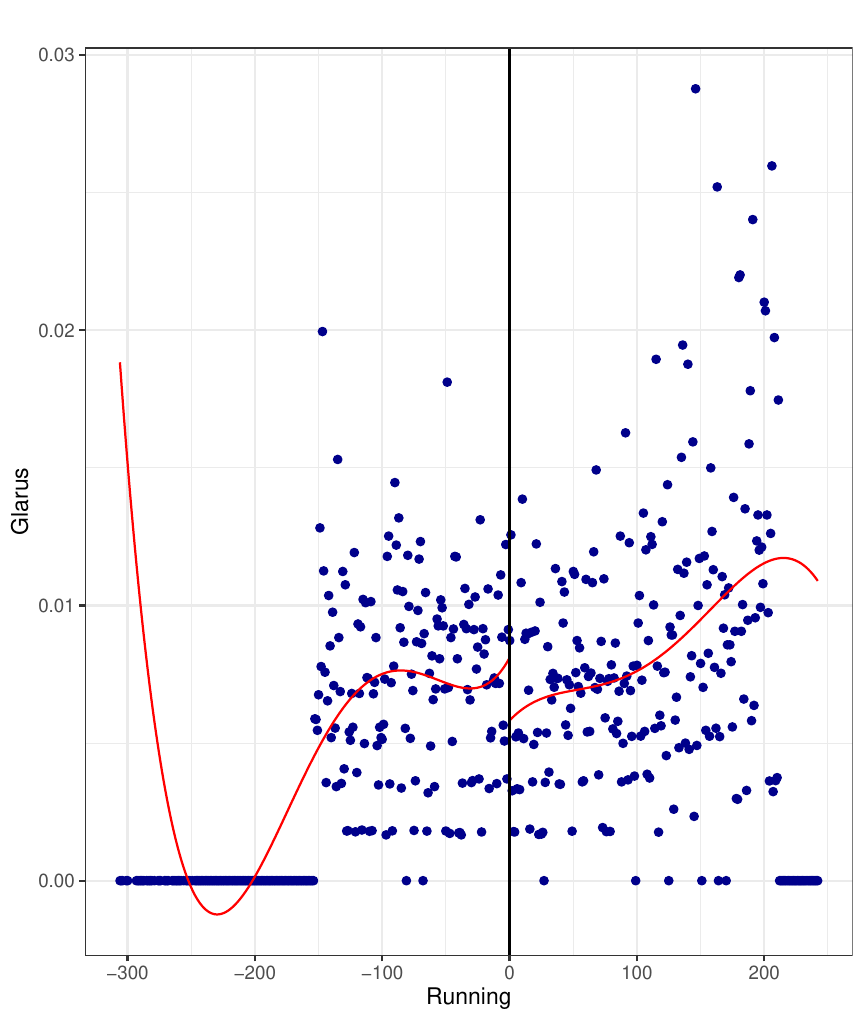}
       \label{fig:out}}
        \quad
        \subfigure[Neuchâtel]{%
      \includegraphics[width=0.3\textwidth]{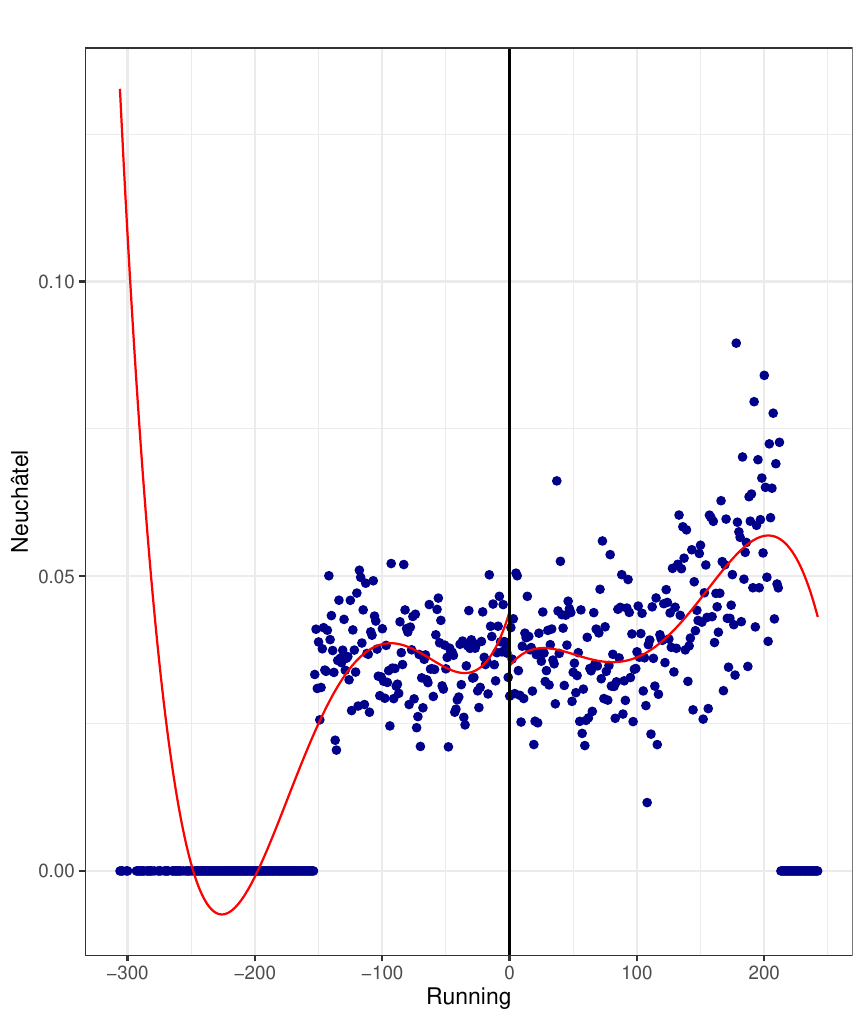}
       \label{fig:emp}}
        \subfigure[Basel-Land]{%
      \includegraphics[width=0.3\textwidth]{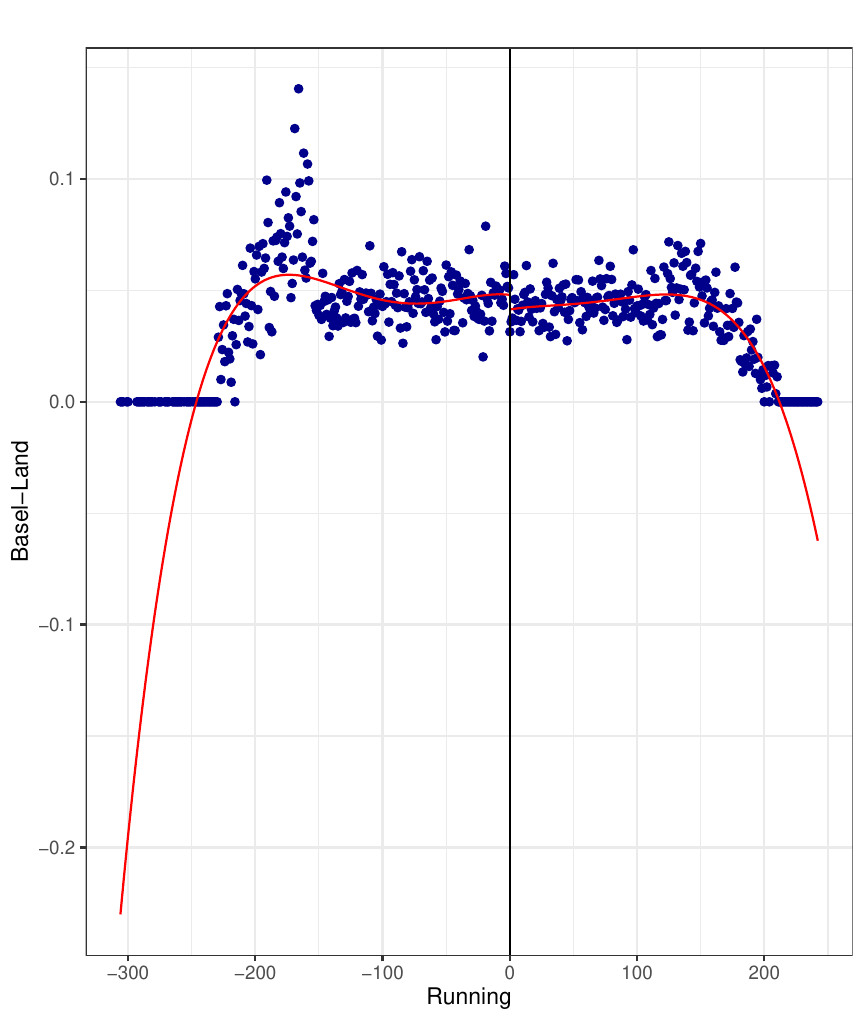}
       \label{fig:unem} }
       \end{figure}

       \begin{figure}[ht]\ContinuedFloat
       \centering
       \begin{threeparttable}
        \subfigure[Jura]{%
      \includegraphics[width=0.3\textwidth]{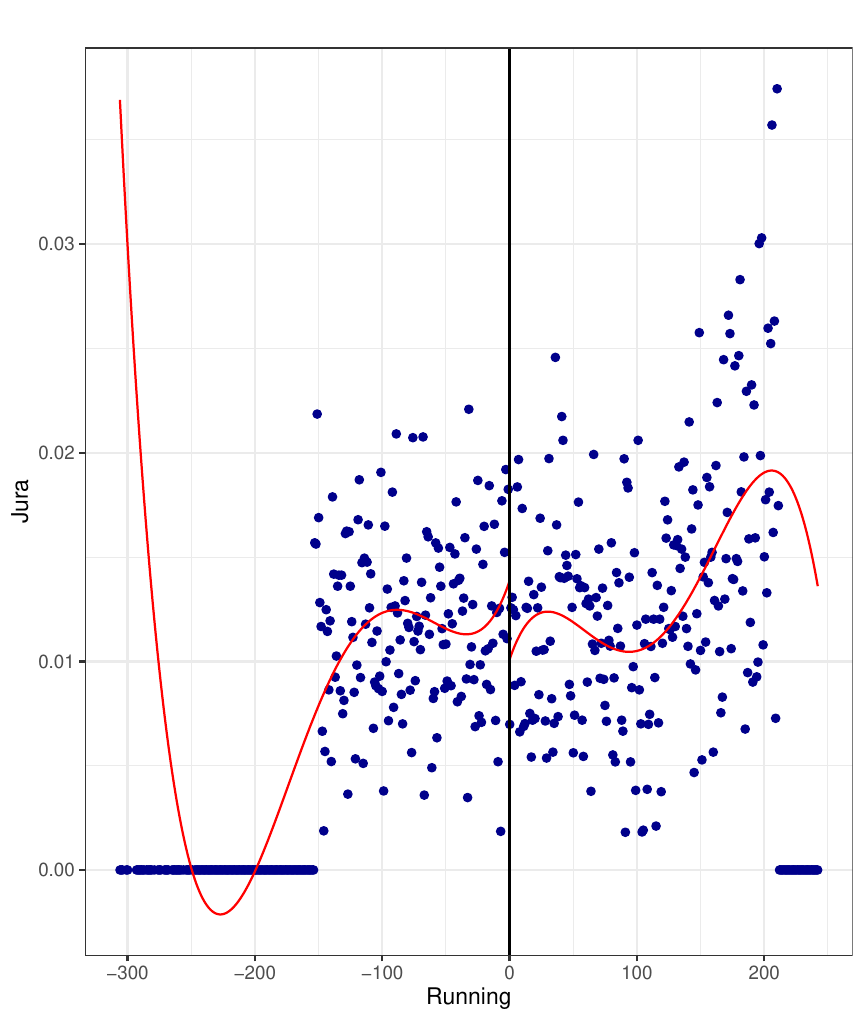}
      \label{fig:work income} }
      \subfigure[Solothurn]{%
 \label{fig:icdum}
      \includegraphics[width=0.3\textwidth]{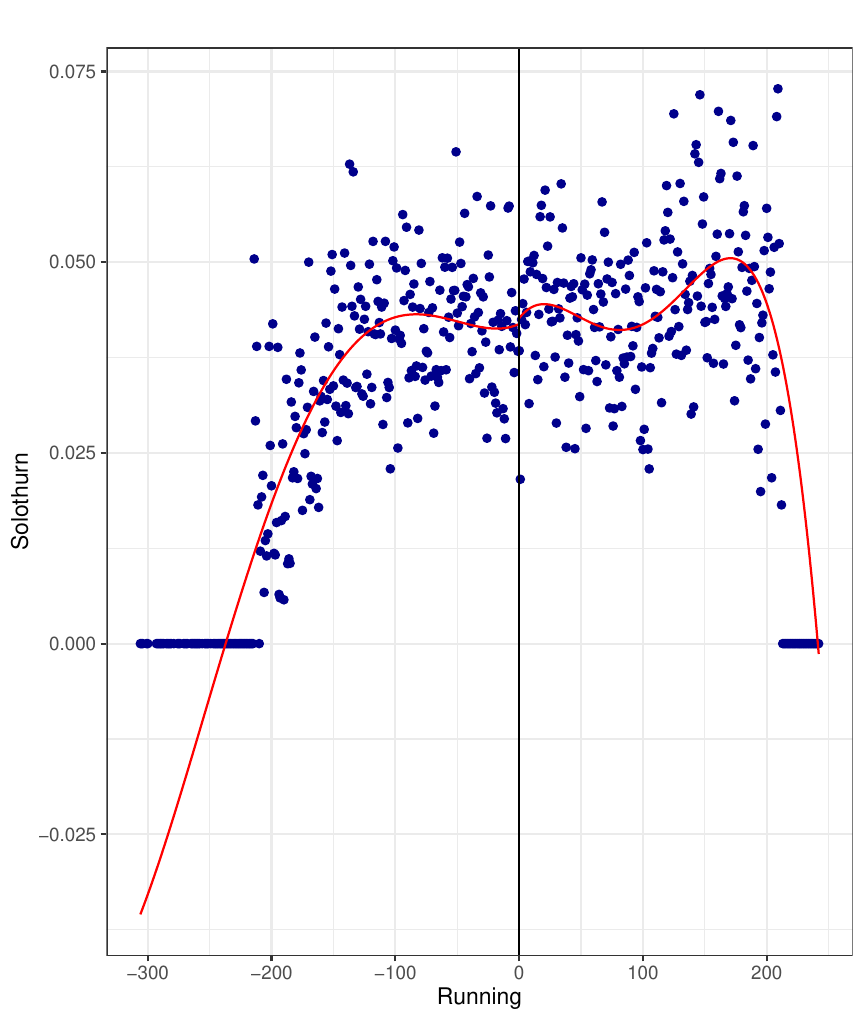}
       }
        \subfigure[Bern]{%
        \label{fig:icdepe}
      \includegraphics[width=0.3\textwidth]{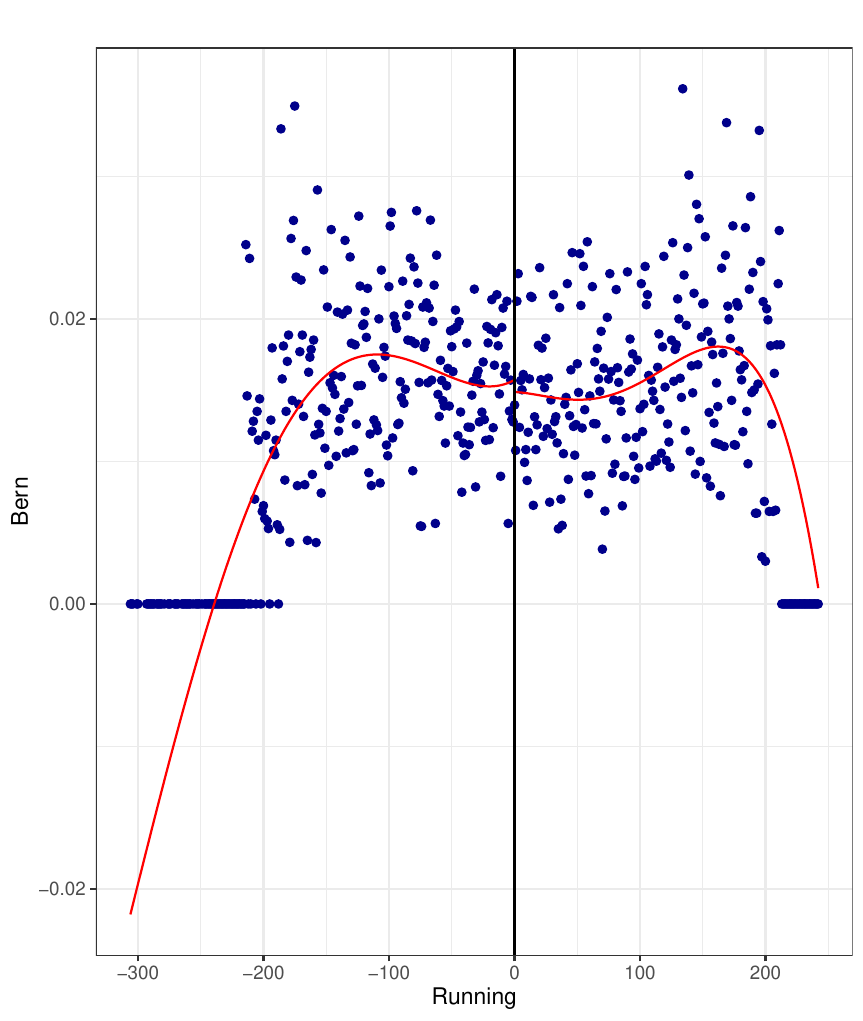}
       }
        \subfigure[Vaud]{%
      \includegraphics[width=0.3\textwidth]{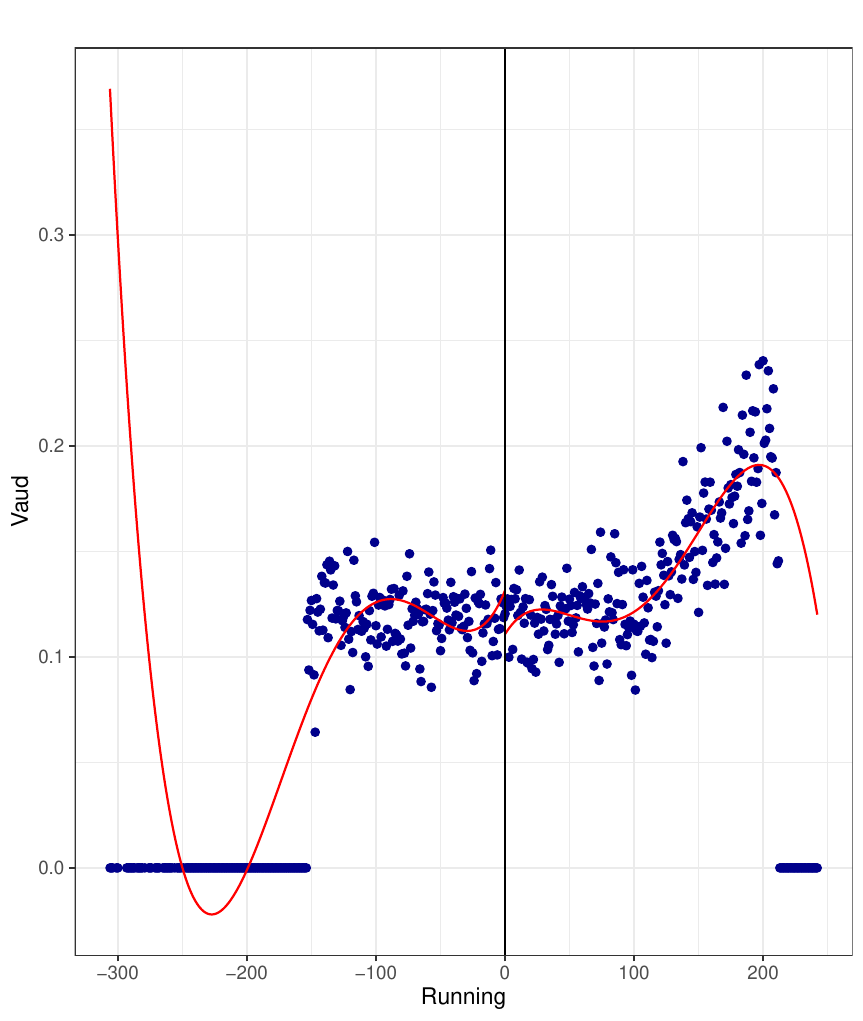}
       \label{fig:emp}}
        \subfigure[Schaffhausen]{%
      \includegraphics[width=0.3\textwidth]{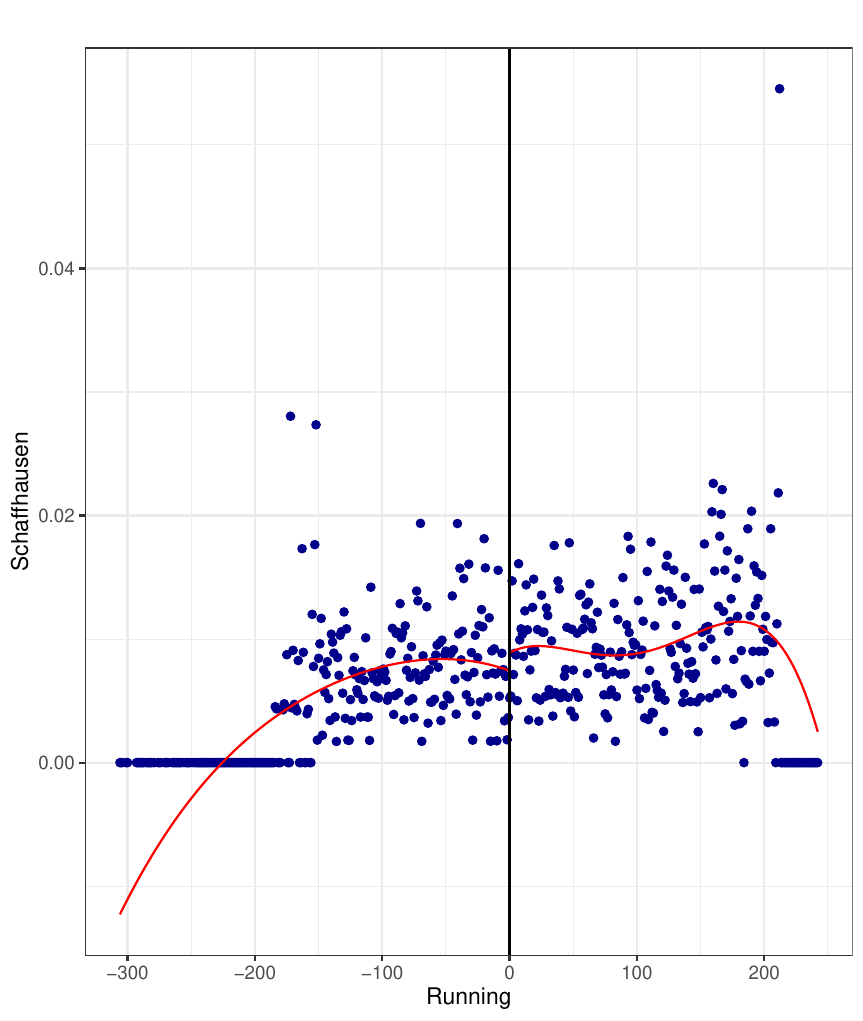}
       \label{fig:unem} }
       \quad
        \subfigure[Ticino]{%
      \includegraphics[width=0.3\textwidth]{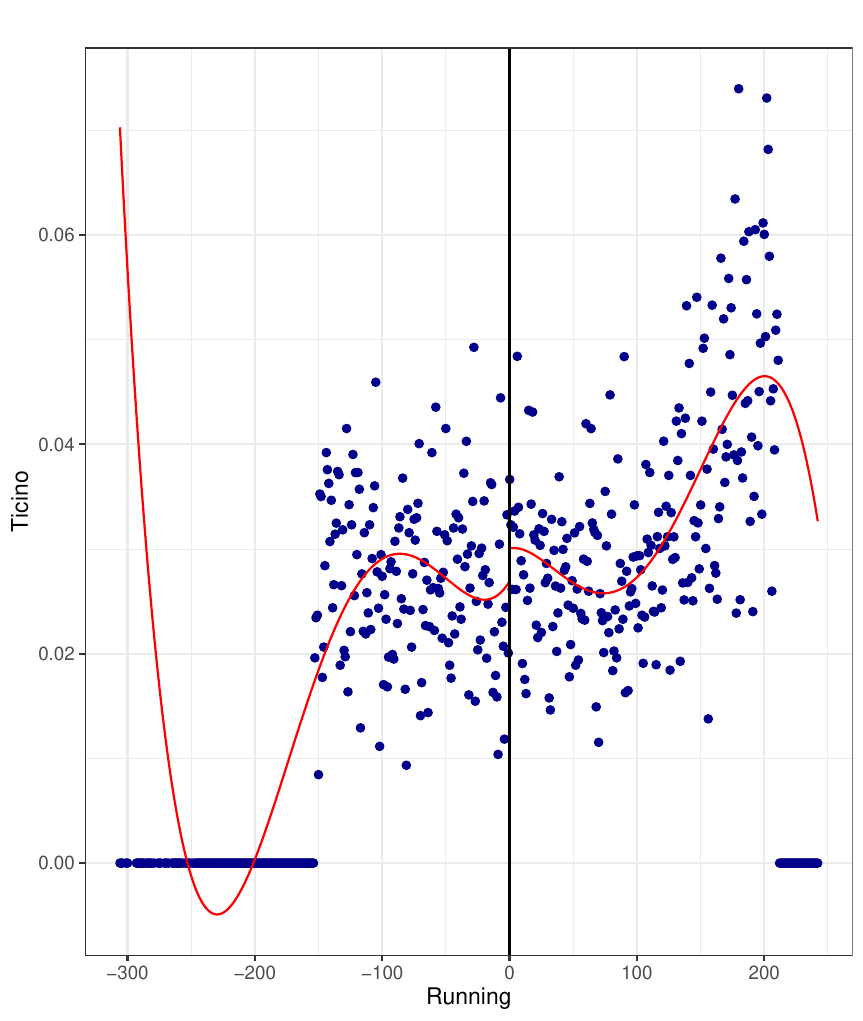}
      \label{fig:work income} }
      \subfigure[Upper Valais]{%
 \label{fig:icdum}
      \includegraphics[width=0.3\textwidth]{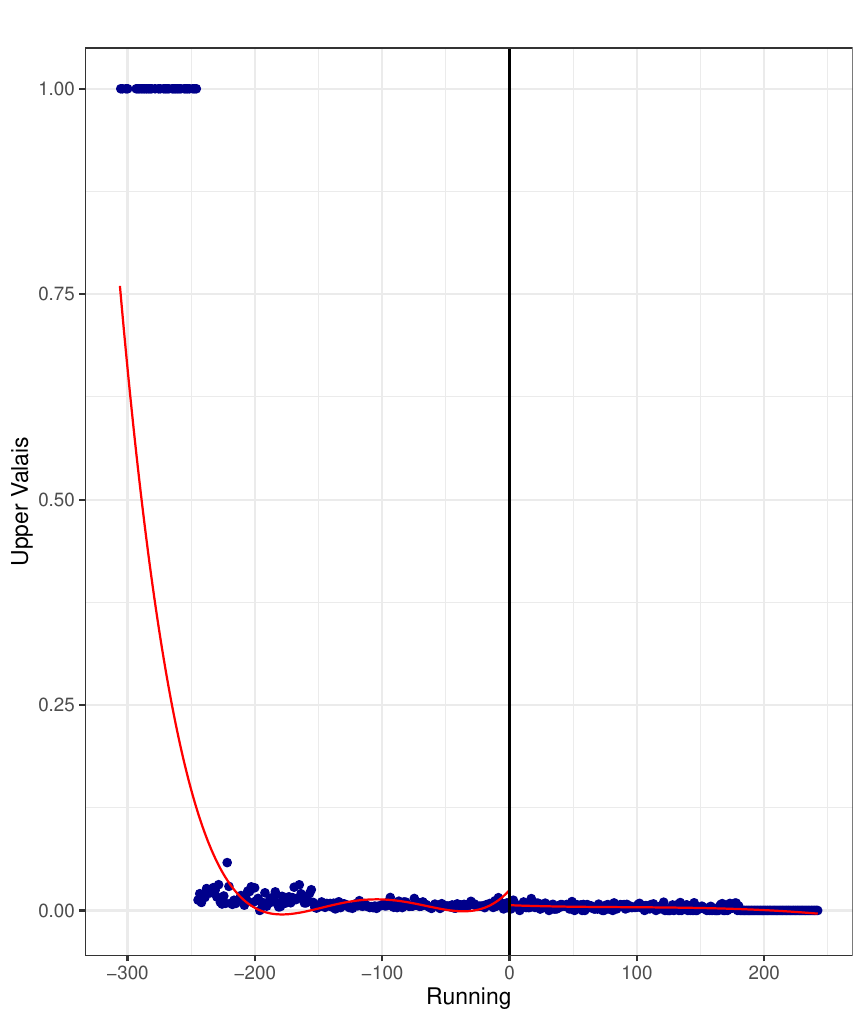}
       }
        \subfigure[Central and Lower Valais]{%
        \label{fig:icdepe}
      \includegraphics[width=0.3\textwidth]{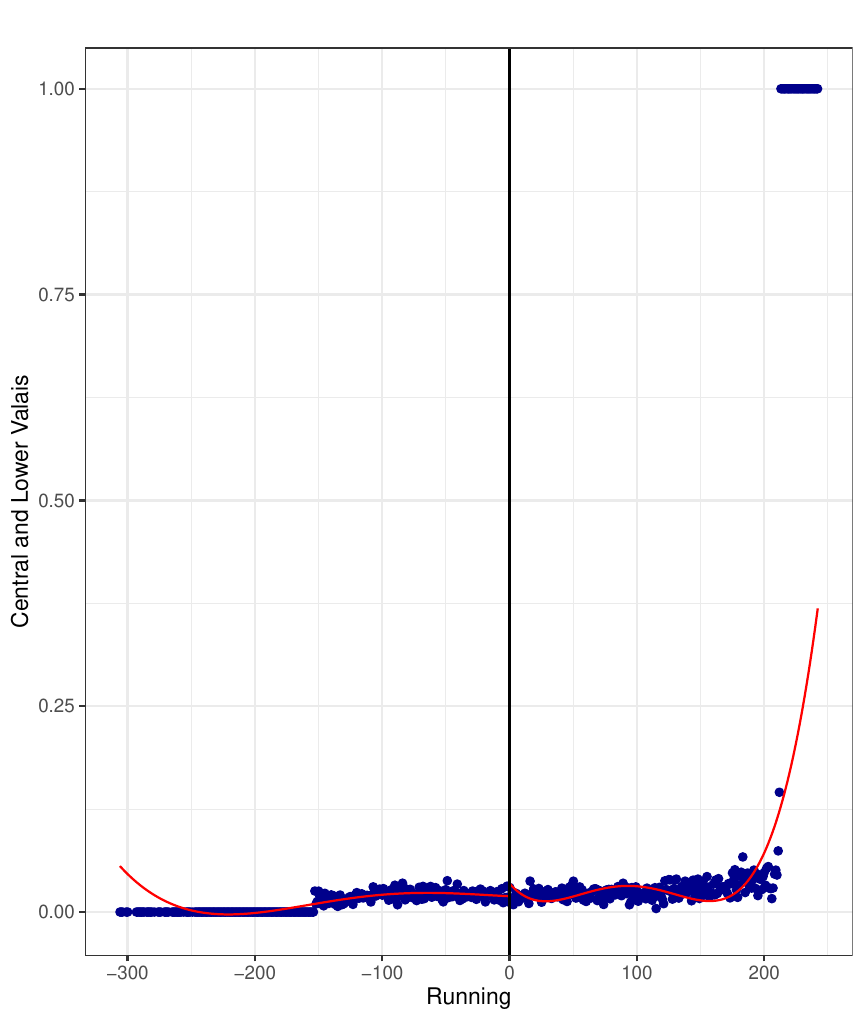}
       }

\begin{tablenotes}[flushleft]
\footnotesize
\item Note: The figure provides plots for cantonal dummies. The dots represent the conditional means of the respective covariate within bins defined upon values of the running variable and the solid line shows the quartic global polynomial fit when regressing the respective covariate on the running variable above and below the cut-off, respectively. The bin width is chosen according to the default option in the rdplot command of the rdrobust package. Data stems from STATPOP (2010 - 2017) and OASI (2010 - 2017), calculations are done by ourselves.
\end{tablenotes}
\end{threeparttable}

\end{figure}

\begin{figure}[h!]
 \centering
 \caption{RDD: Outcome plots}
  \label{fig:outcomeplots}
       \begin{threeparttable}
        \subfigure[Out of labour force (binary)]{%
      \includegraphics[width=0.3\textwidth]{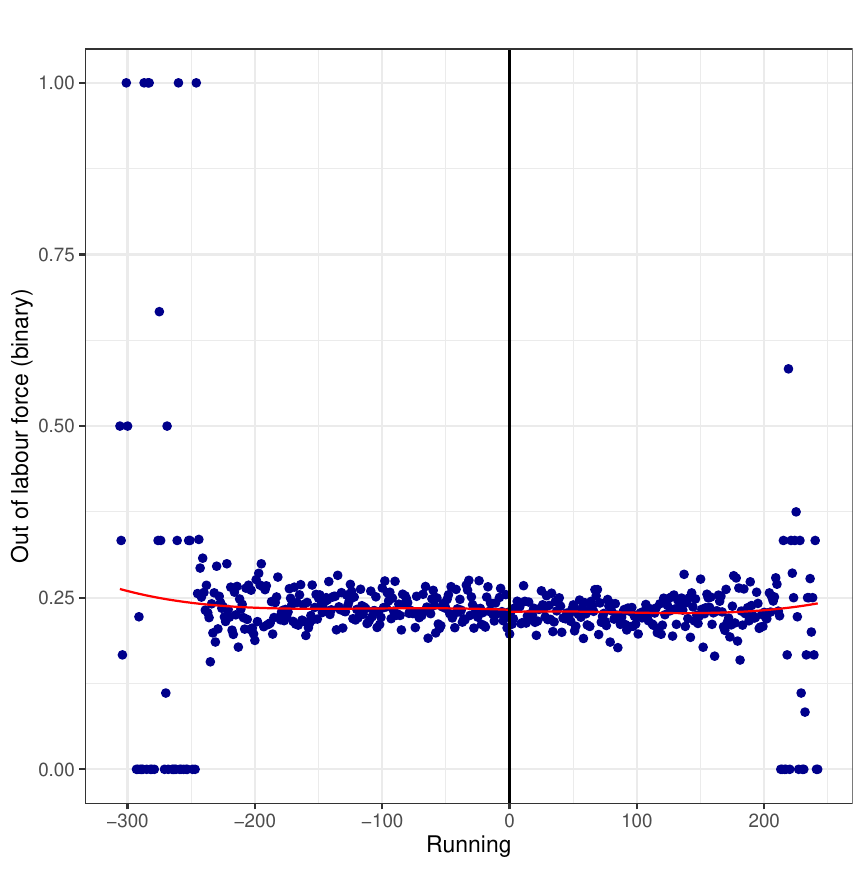}
       \label{fig:out}}
        \quad
        \subfigure[Employed (binary)]{%
      \includegraphics[width=0.3\textwidth]{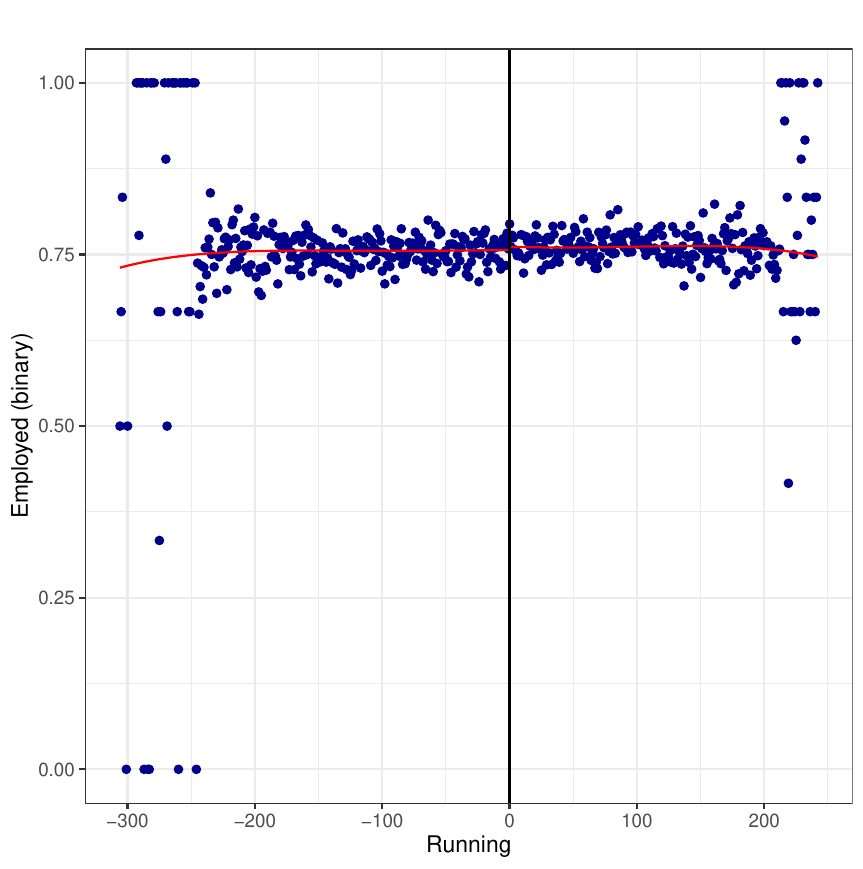}
       \label{fig:emp}}
        \subfigure[Unemployed (binary)]{%
      \includegraphics[width=0.3\textwidth]{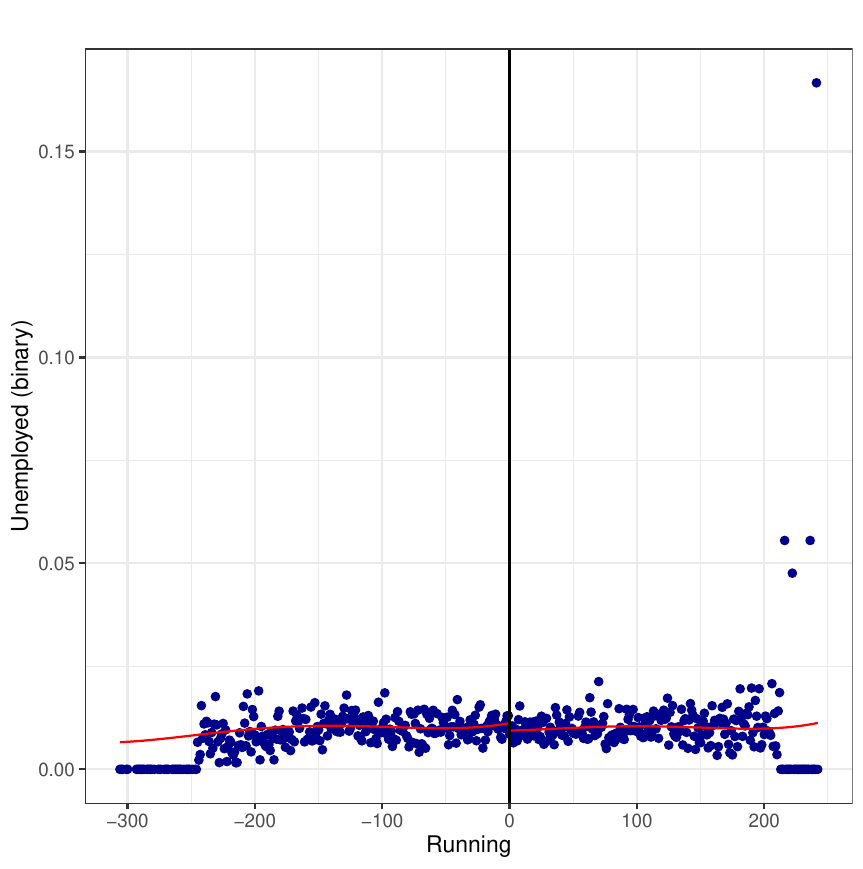}
       \label{fig:unem} }
       \quad
        \subfigure[Total annual work income (in CHF)]{%
      \includegraphics[width=0.3\textwidth]{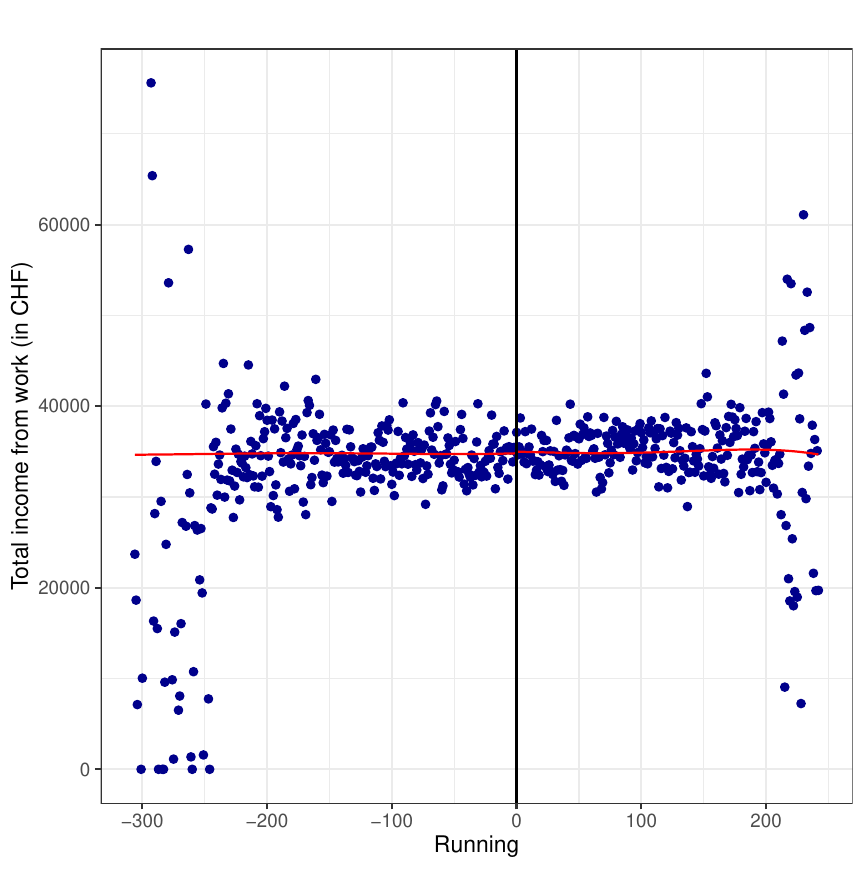}
      \label{fig:work income} }
      \subfigure[Income from dependent employment (binary)]{%
 \label{fig:icdum}
      \includegraphics[width=0.3\textwidth]{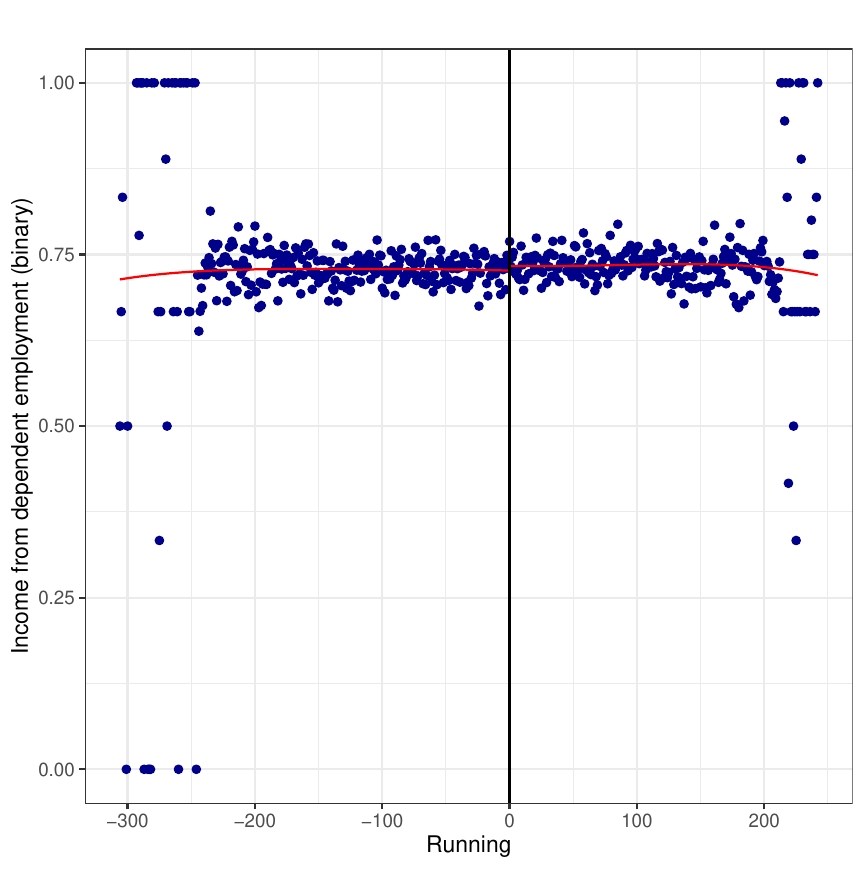}
       }
        \subfigure[Income from dependent employment (in CHF)]{%
        \label{fig:icdepe}
      \includegraphics[width=0.3\textwidth]{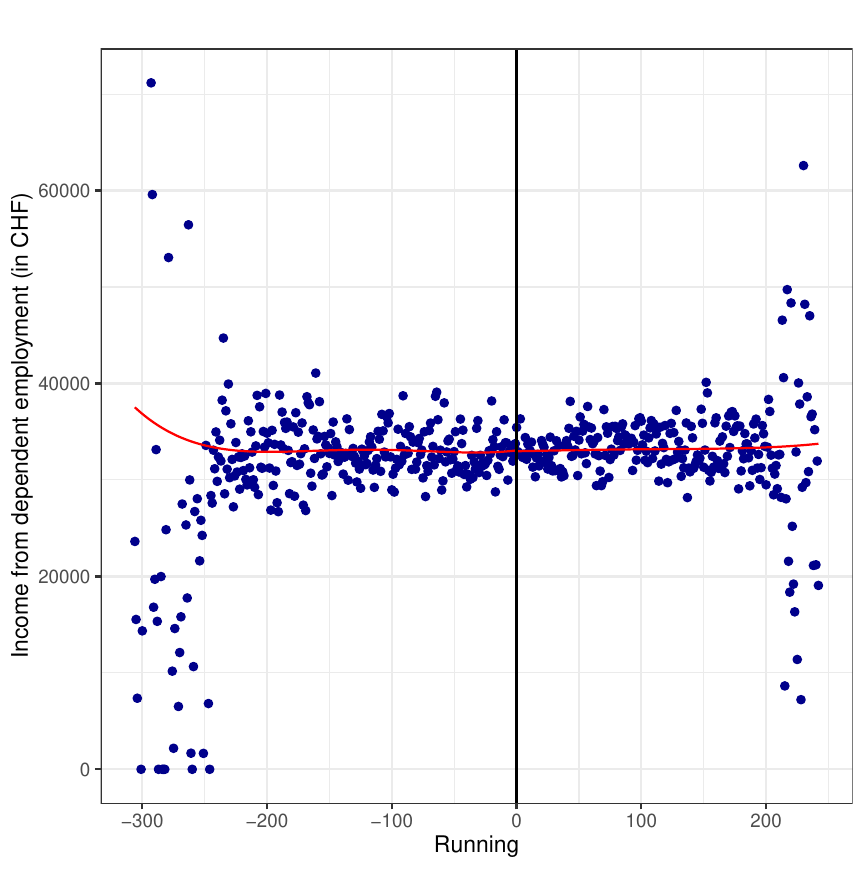}
       }
     \begin{tablenotes}[flushleft]
\footnotesize
\item Note: The figure provides plots for pooled outcome periods. The dots represent the conditional means of the respective outcome within bins defined upon values of the running variable and the solid line shows the quartic global polynomial fit when regressing the respective outcome on the running variable above and below the cut-off, respectively. The bin width is chosen according to the default option in the rdplot command of the rdrobust package. The following control variables are included: Number of children, born in Switzerland, resident permit, age, relationship status, labour market characteristics of the mother and the father of the four-year old child, cantonal characteristics and dummies. Data stems from STATPOP (2010 - 2017) and OASI (2010 - 2017), calculations are done by ourselves. Income is deflated, the base year is 2011. The official currency in Switzerland is the Swiss Franc (CHF), which had an average exchange rate of 1.04 USD/CHF in the last decade.
\end{tablenotes}
\end{threeparttable}
       \end{figure}

\newpage

\begin{landscape}
\begin{table}[!h]
\center
\footnotesize
\caption{RDD: Empirical results without covariates}
\label{result1oKov}
\centering
\begin{threeparttable}
 \renewcommand\TPTminimum{\textwidth} 
\begin{tabular}{lcccccc}
 \hline\hline
 & Sample mean & Coefficient & Standard error & P-value & Bandwidth  & Number obs within bw \\
  \hline
Out of labour force (binary) & 0.23 & -0.00 & 0.01 & 0.81 & 72.21 & 297,628\\ 
  Employed (binary) & 0.76 & 0.00 & 0.01 & 0.72 & 72.52 & 297,628\\ 
  Unemployed (binary) & 0.01 & -0.00 & 0.00 & 0.13 & 53.04 & 219,055\\ 
  Annual work income (in CHF) & 34,860.92 & -340.36 & 1,092.91 & 0.76 & 72.78 & 297,628\\ 
  Income dependent employment (binary) & 0.73 & 0.01 & 0.01 & 0.41 & 65.49 & 269,236\\ 
  Annual income from dependent employment (in CHF) & 33,075.03 & -169.70 & 1,025.17 & 0.87 & 73.76 & 301,689\\

  \hline
\end{tabular}
\begin{tablenotes}[flushleft]
\footnotesize
\item  This table reports the local linear estimates of  equation \ref{eqn2} but without covariates. Bandwidth shows the MSE-optimal bandwidth chosen by \citep{Calonico2021}. The standard errors are clustered at the individual level. Data stems from STATPOP (2010 - 2017) and OASI (2010 - 2017), calculations are done by ourselves.  Income is deflated, the base year is 2011. The official currency in Switzerland is the Swiss Franc (CHF), which had an average exchange rate of 1.04 USD/CHF in the last decade. Number of observations in total: 735,520. \\
\end{tablenotes}
\end{threeparttable}
\end{table}

\end{landscape}

\begin{figure}[h!]
    \centering
    \caption{RDD: Effects over years without covariates}
\label{fig:over_years_o_cov}
\begin{threeparttable}

        \subfigure[Out of labour force (binary)]{%

      \includegraphics[width=0.3\textwidth]{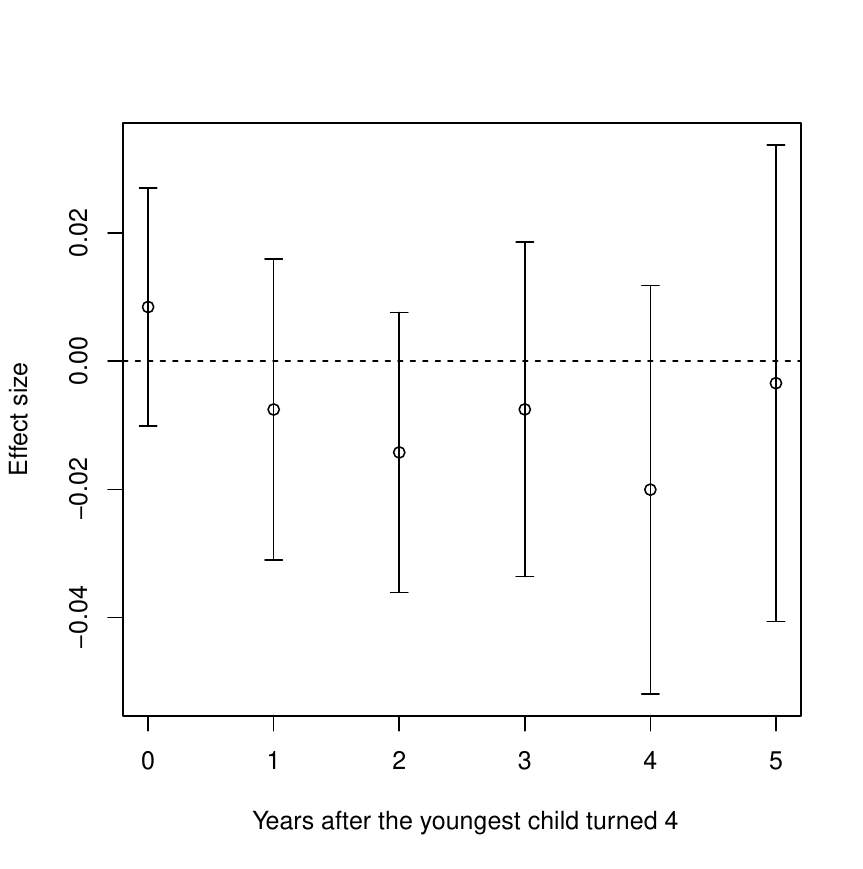}
      \label{fig:first} }
          \quad
        \subfigure[Employed (binary)]{%

      \includegraphics[width=0.3\textwidth]{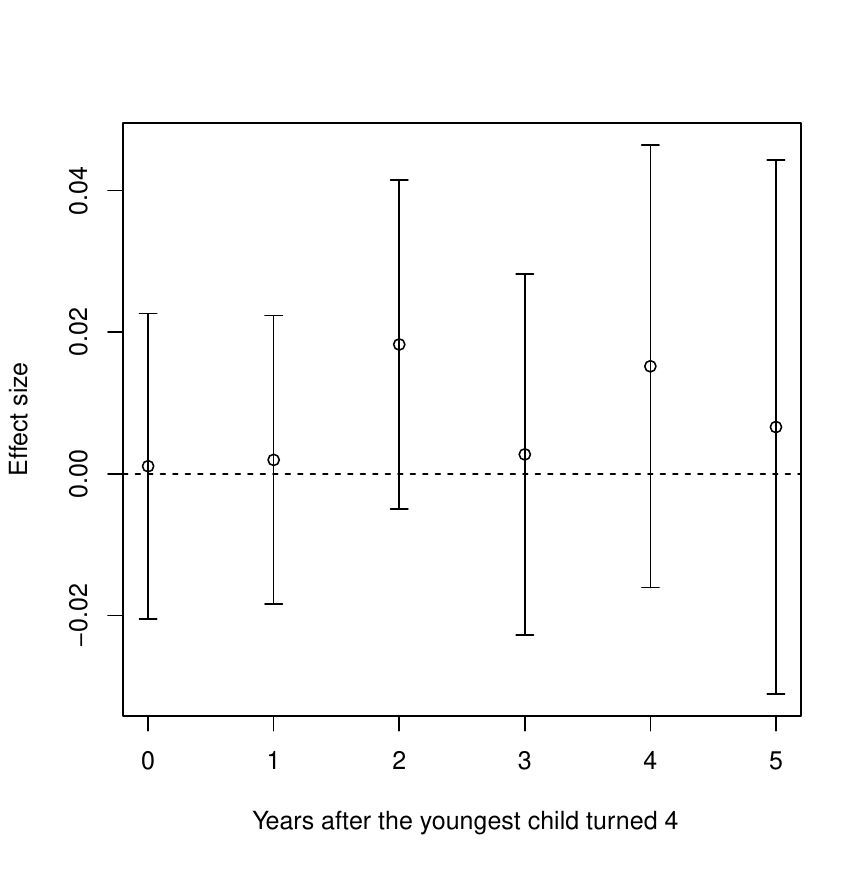}
      \label{fig:second}}
                \subfigure[Unemployed (binary)]{%
            \label{fig:third}
      \includegraphics[width=0.3\textwidth]{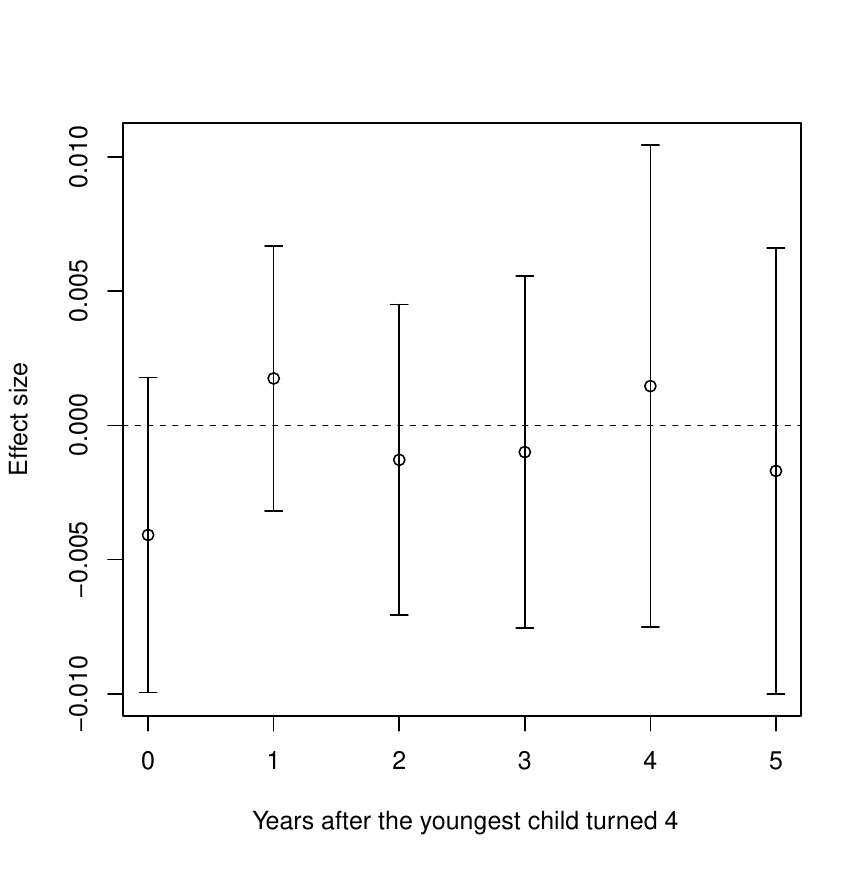}}
         \quad
        \subfigure[Total annual work income (in CHF)]{%
            \label{fig:first}
      \includegraphics[width=0.3\textwidth]{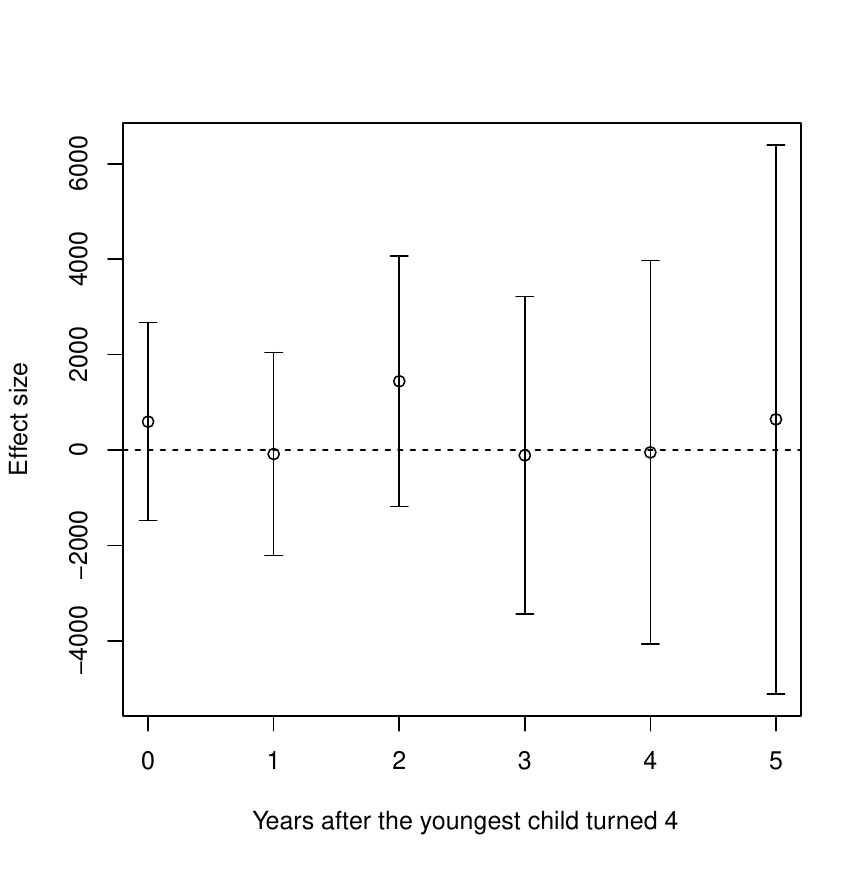}
       }
       \subfigure[Income dependent employment (binary)]{%
            \label{fig:third}
      \includegraphics[width=0.3\textwidth]{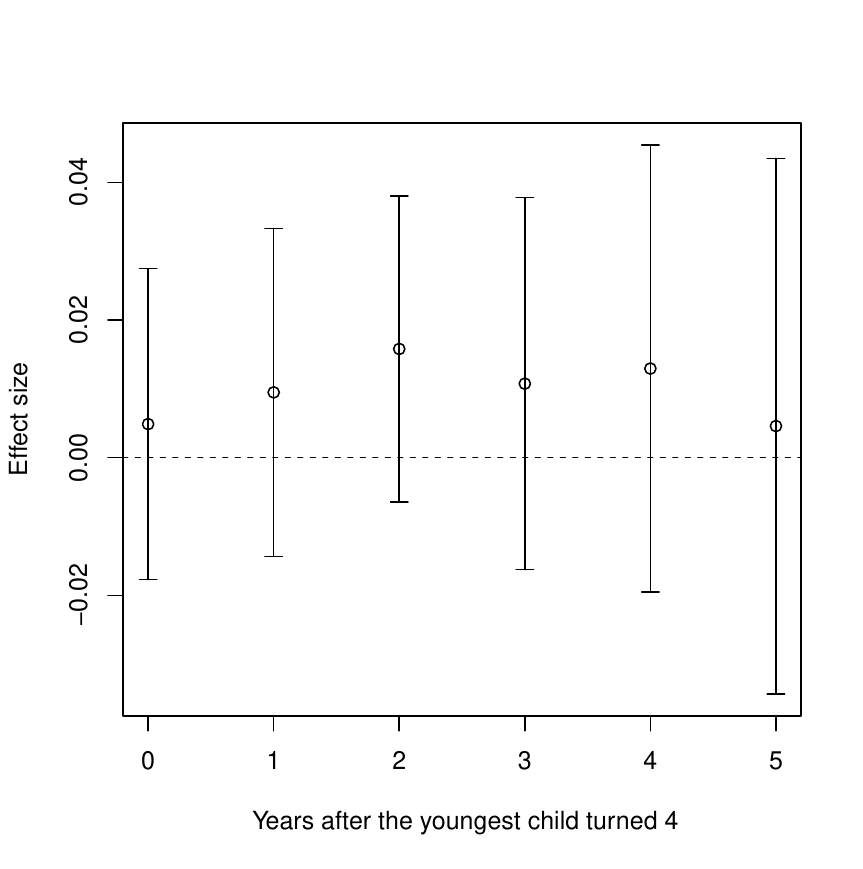}
       }
        \subfigure[Income dependent employment (in CHF)]{%
            \label{fig:second}
      \includegraphics[width=0.3\textwidth]{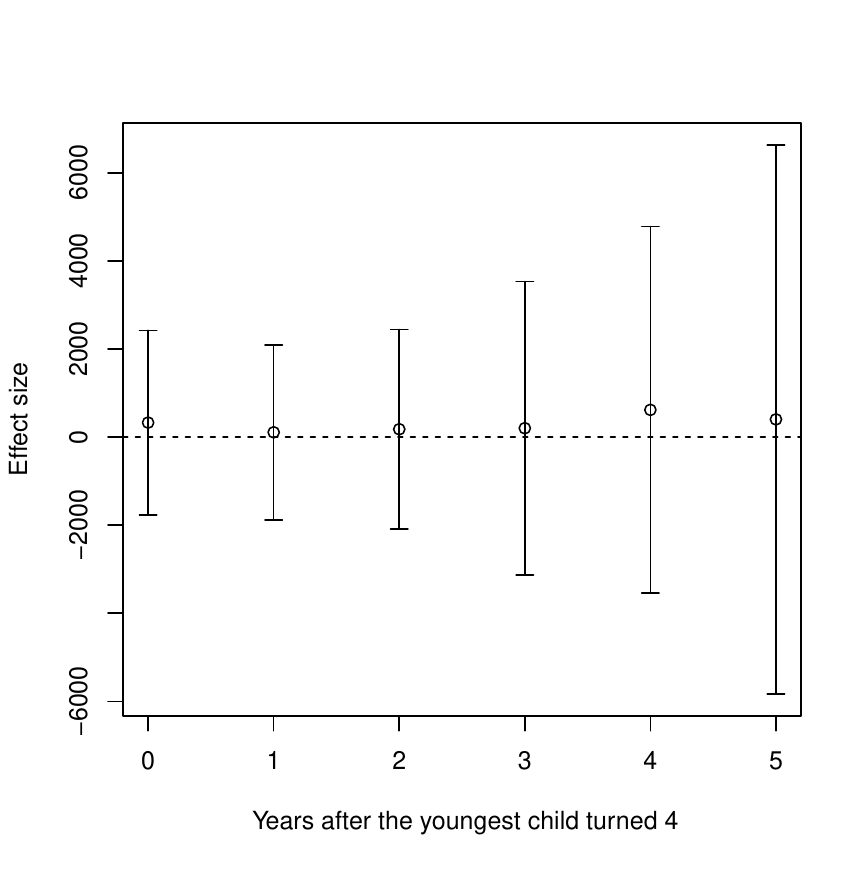}
       }
    \caption{RDD: Effects over years without covariates}\label{fig:over_years_o_cov}
\begin{tablenotes}[flushleft]
\footnotesize
\item  Note: This figure shows the effect of mandatory kindergarten on each outcome variable separately and over time. Dots represent ITTs at the threshold, bands correspond to 95\% confidence intervals. Data stems from STATPOP (2010 - 2017) and OASI (2010 - 2017), calculations are done by ourselves.  Income is deflated, the base year is 2011. The official currency in Switzerland is the Swiss Franc (CHF), which had an average exchange rate of 1.04 USD/CHF in the last decade.
\end{tablenotes}
\end{threeparttable}

\end{figure}

	\clearpage

\begin{landscape}
\begin{table}[!h]
\center
\footnotesize
\caption{RDD: Robustness: Bandwidth * 1.5}
\label{robust1}
\centering
\begin{threeparttable}
 \renewcommand\TPTminimum{\textwidth} 
\begin{tabular}{lcccccp{2cm}}
 \hline\hline

 & Sample mean & Coefficient & Standard error & P-value & Bandwidth   & Number of observations within bandwidth \\
  \hline
Out of labour force (binary) & 0.23 & -0.01 & 0.01 & 0.31 & 108.16  & 443,181 \\ 
  Employed (binary) & 0.76 & 0.01 & 0.01 & 0.12 & 105.78  & 431,395 \\ 
  Unemployed (binary) & 0.01 & -0.00 & 0.00 & 0.20 & 79.48  & 325,660 \\ 
  Annual work income (in CHF) & 34,860.92 & 590.74 & 611.71 & 0.33 & 78.17 & 321,448 \\ 
  Income dependent employment (binary) & 0.73 & 0.01 & 0.01 & 0.09 & 100.90  & 410,857 \\ 
  Annual income from dependent employment (in CHF) & 33,075.03 & 232.38 & 478.16 & 0.63 & 104.74 &  427,429 \\

 \hline
  \end{tabular}
\begin{tablenotes}[flushleft]
\footnotesize
\item  Note: This table reports the local linear estimates of  equation \ref{eqn2}. The optimal bandwidth is multiplied by 1.5. Bandwidth shows the MSE-optimal bandwidth chosen by \citep{Calonico2021}. The standard errors are clustered at the individual level. The following control variables are included: Number of children, born in Switzerland, resident permit, age, relationship status, labour market characteristics of the mother and the father of the four-year old child, cantonal characteristics and dummies. Data stems from STATPOP (2010 - 2017) and OASI (2010 - 2017), calculations are done by ourselves.  Income is deflated, the base year is 2011. The official currency in Switzerland is the Swiss Franc (CHF), which had an average exchange rate of 1.04 USD/CHF in the last decade.  Number of observations in total: 735,520.
 \end{tablenotes}
\end{threeparttable}
\end{table}

\end{landscape}

\begin{landscape}
\begin{table}[!h]
\center
\footnotesize
\caption{RDD: Robustness: Bandwidth * 2/3}
\label{robust2}
\centering
\begin{threeparttable}

 \renewcommand\TPTminimum{\textwidth} 
\begin{tabular}{lcccccp{2cm}}
 \hline\hline

 & Sample mean & Coefficient & Standard error & P-value & Bandwidth   & Number of observations within bandwidth \\
  \hline
Out of labour force (binary) & 0.23 & -0.01 & 0.01 & 0.57 & 48.07  & 198,603 \\ 
  Employed (binary) & 0.76 & 0.01 & 0.01 & 0.53 & 47.01  & 194,817 \\ 
  Unemployed (binary) & 0.01 & -0.00 & 0.00 & 0.23 & 35.33 &  145,453\\ 
  Annual work income (in CHF) & 34,860.92 & 1,046.00 & 957.31 & 0.27 & 34.74  & 141,416 \\ 
  Income dependent employment (binary) & 0.73 & 0.01 & 0.01 & 0.46 & 44.84 & 182,448\\ 
  Annual income from dependent employment (in CHF) & 33,075.03 & 157.24 & 745.27 & 0.83 & 46.55 &  190,654 \\

 \hline
  \end{tabular}
 \begin{tablenotes}[flushleft]
\footnotesize
\item Note: This table reports the local linear estimates of  equation \ref{eqn2}. The optimal bandwidth is multiplied by 3/2. Bandwidth shows the MSE-optimal bandwidth chosen by \citep{Calonico2021}. The standard errors are clustered at the individual level. The following control variables are included: Number of children, born in Switzerland, resident permit, age, relationship status, labour market characteristics of the mother and the father of the four-year old child, cantonal characteristics and dummies. Data stems from STATPOP (2010 - 2017) and OASI (2010 - 2017), calculations are done by ourselves.  Income is deflated, the base year is 2011. The official currency in Switzerland is the Swiss Franc (CHF), which had an average exchange rate of 1.04 USD/CHF in the last decade. Number of observations in total: 735,520.
\end{tablenotes}
\end{threeparttable}
\end{table}

\end{landscape}

\clearpage

\begin{table}[!h]
\center
\footnotesize
\caption{Subsample Employed -  Balance check: Covariates}
\label{descrRDDE}
\centering
\begin{threeparttable}

 \resizebox{\linewidth}{!}{

\begin{tabular}{lccccp{1.5cm}}
\hline\hline
 & Sample mean & Coefficient &Standard error & P-value & Relative number of missings in \% \\
  \hline
\textit{Mother's characteristics} & & & & &\\
[0.15cm]
Number Children & 1.82 & 0.02 & 0.02 & 0.29 & 0.00 \\ 
[0.15cm]
  Born in Switzerland & 0.56 & 0.00 & 0.01 & 0.83 & 2.47 \\ 
  [0.15cm]
  \textit{Nationality} & & & & & \\
  Resident permit B & 0.08 & -0.01 & 0.01 & 0.55 & 0.35 \\ 
  Resident permit C & 0.26 & 0.00 & 0.01 & 0.78 & 0.35 \\ 
  Other resident & 0.66 & 0.00 & 0.01 & 0.75 & 0.35 \\ 
  [0.15cm]
  \textit{Age} & & & & &\\
  Age & 35.63 & -0.00 & 0.17 & 0.98 & 0.35 \\ 
  [0.15cm]
  \textit{Relationship} & & & & &\\
  In Partnership & 0.80 & -0.02 & 0.01 & 0.10 & 0.35 \\ 
   Not in Partnership & 0.15 & 0.02 & 0.01 & 0.09 & 0.35 \\ 
   Terminated Partnership & 0.05 & 0.00 & 0.01 & 0.84 & 0.35 \\ 
   [0.15cm]
\textit{Mother's labour market characteristics} & & & & &\\
 Total income from work (in CHF) & 44,230.64 & -1,280.05 & 1,151.39 & 0.27 & 8.57 \\ 
Income from dependent employment (binary) & 0.96 & -0.00 & 0.01 & 0.49 & 8.57 \\ 
Income from dependent employment (in CHF)& 31,204.52 & 447.98 & 1,067.00 & 0.68 & 0.00 \\
  [0.3cm]
  \textit{Father's labour market characteristics} & & & & &\\
  Out of labour force (binary)  & 0.07 & -0.00 & 0.01 & 0.89 & 0.00 \\ 
  Employed (binary)&  0.923 & -0.00 & 0.01 & 0.74 & 0.00 \\ 
  Total income from work (in CHF)   & 92,063.52 & -1,561.61 & 3,079.85 & 0.61 & 0.00 \\ 
  Unemployed (binary) & 0.01 & 0.00 & 0.00 & 0.68 & 0.00 \\ 
  
  Income from dependent employment (binary)  & 0.89 & 0.01 & 0.01 & 0.40 & 0.00 \\ 
  Income from dependent employment (in CHF)    & 87,250.96 & 1,836.94 & 2,259.62 & 0.42 & 0.00 \\ 
   [0.3cm]
\textit{Cantonal characteristics\textit} & & & & &\\
 Harmos member & 0.95 & -0.00 & 0.01 & 0.99 & 0.00 \\ 
  Unemployment Rate & 3.56 & 0.05 & 0.04 & 0.18 & 0.00\\ 
   \hline
\end{tabular}
}
\begin{tablenotes}[flushleft]
\footnotesize
\item Note:  The table presents the balance check of the labour market relevant covariates as suggested in \cite{LEE2008}. The data comes from STATPOP (2010 - 2017) and OASI (2010 - 2017), the calculations are done by ourselves. Income is deflated, the base year is 2011. The official currency in Switzerland is the Swiss Franc (CHF), which had an average exchange rate of 1.04 USD/CHF in the last decade. Residence permit B allows foreign nationals in Switzerland for specific purposes with a five-year validity, while Residence permit C grants unrestricted residency after five or ten years of lawful stay, with specific agreements for citizens of certain EU/EFTA countries.
 \end{tablenotes}
\end{threeparttable}
\end{table}

\begin{table}[!h]
\center
\footnotesize
\caption{Subsample Non-employed - Balance check: Covariates}
\label{descrRDDN}
\centering
\begin{threeparttable}

 \resizebox{\linewidth}{!}{

\begin{tabular}{lccccp{1.5cm}}
\hline\hline
 & Sample mean & Coefficient &Standard error & P-value & Relative number of missings in \% \\
  \hline
\textit{Mother's characteristics} & & & & &\\
[0.15cm]
Number Children & 2.02 & -0.04 & 0.05 & 0.36 & 0.00 \\
[0.15cm] 
  Born in Switzerland & 0.41 & 0.00 & 0.03 & 0.87 & 2.96 \\ 
  [0.15cm]
  \textit{Nationality} & & & & & \\
  Resident permit B & 0.18 & -0.01 & 0.02 & 0.80 & 2.23 \\ 
  Resident permit C & 0.27 & 0.02 & 0.03 & 0.45 & 2.23 \\ 
  Other resident & 0.55 & -0.01 & 0.03 & 0.66 & 2.23 \\ 
    [0.15cm]
\textit{Age} & & & & &\\
  Age & 35.60 & -0.10 & 0.25 & 0.71 & 2.23 \\
  \textit{Relationship} & & & & &\\ 
  In Partnership & 0.86 & -0.00 & 0.02 & 0.91 & 2.23 \\ 
   Not in Partnership & 0.10 & 0.01 & 0.02 & 0.42 & 2.23 \\ 
   Terminated Partnership & 0.04 & -0.00 & 0.01 & 0.77 & 2.23 \\ 
       [0.15cm]
\textit{Mother's labour market characteristics} & & & & &\\
 Out of labour force (binary)  & 0.96 & -0.00 & 0.01 & 0.68 & 0.00 \\ 
 Unemployed (binary)  & 0.04 & 0.00 & 0.01 & 0.68 & 23.32 \\ 
    [0.3cm]
\textit{Father's labour market characteristics} & & & & &\\
Out of labour force (binary)  & 0.15 & 0.00 & 0.02 & 0.87 & 0.00 \\ 
   Employed (binary)&  0.84 & -0.00 & 0.02 & 0.97 & 0.00 \\ 
    Unemployed (binary)  & 0.01 & -0.01 & 0.01 & 0.38 & 0.00 \\ 
 Total income from work (in CHF)   & 96,264.15 & -5,398.21 & 6,000.93 & 0.37 & 0.00 \\ 
  Income from dependent employment (binary)  & 0.81 & 0.01 & 0.02 & 0.78 & 0.00 \\ 
  Income from dependent employment (in CHF)  & 9,3578.42 & -5,317.87 & 5,699.39 & 0.35 & 0.00 \\ 
  [0.3cm]
\textit{Cantonal characteristics\textit} & & & & &\\
  Harmos member & 0.94 & -0.01 & 0.01 & 0.32 & 0.00 \\ 
  Unemployment Rate & 3.56 & -0.06 & 0.06 & 0.33 & 0.00 \\ 
   \hline 
\end{tabular}
}
\begin{tablenotes}[flushleft]
\footnotesize
\item Note:  The table presents the balance check of the labour market relevant covariates as suggested in \cite{LEE2008}. The data comes from STATPOP (2010 - 2017) and OASI (2010 - 2017), the calculations are done by ourselves. Income is deflated, the base year is 2011. The official currency in Switzerland is the Swiss Franc (CHF), which had an average exchange rate of 1.04 USD/CHF in the last decade. Residence permit B allows foreign nationals in Switzerland for specific purposes with a five-year validity, while Residence permit C grants unrestricted residency after five or ten years of lawful stay, with specific agreements for citizens of certain EU/EFTA countries.
 \end{tablenotes}
\end{threeparttable}
\end{table}

\begin{table}[!h]
\center
\footnotesize
\caption{Subsample Employed Balance check: Cantonal dummies}
\label{descr2emp}
\centering
\begin{threeparttable}

\begin{tabular}{lcccc}
 \hline\hline
 & Sample mean & Coefficient & Standard error & P-value \\
  \hline
Basel-Stadt & 0.03 & -0.00 & 0.01& 0.10 \\ 
  St. Gallen & 0.09 & 0.00 & 0.01 & 0.69 \\ 
  Thurgau & 0.05 & 0.00 & 0.01 & 0.99 \\ 
  Zurich & 0.31 & 0.00 & 0.02 & 0.85 \\ 
  Fribourg & 0.06 & -0.00 & 0.01 & 0.82 \\ 
  Geneva & 0.10 & 0.01 & 0.01 & 0.32 \\ 
  Glarus & 0.08 & -0.00 & 0.00 & 0.34 \\ 
 Neuch\^atel  & 0.04 & 0.00& 0.01 & 0.74 \\ 
  Basel-Land & 0.05 & -0.02 & 0.01 & 0.02 \\ 
  Jura & 0.01 & -0.00 & 0.00 & 0.73 \\ 
  Solothurn & 0.04 & -0.00 & 0.01 & 0.85 \\ 
  Bern & 0.02 & -0.00 & 0.00 & 0.83 \\ 
  Vaud & 0.13 & -0.00 & 0.01& 0.70 \\ 
  Schaffhausen & 0.01 & 0.01 & 0.00 & 0.01 \\ 
  Ticino & 0.03 & 0.01 & 0.01 & 0.01 \\ 
  Upper Valais  & 0.01 & 0.00 & 0.00 & 0.89 \\ 
  Central and Lower Valais  & 0.03 & -0.00 & 0.01 & 0.82 \\ 
   \hline
\end{tabular}

\begin{tablenotes}[flushleft]
\footnotesize
\item Note:  The table presents the balance check of the cantonal dummies as suggested in \cite{LEE2008}. The data comes from STATPOP (2010 - 2017) and OASI (2010 - 2017), the calculations are done by ourselves.
 \end{tablenotes}
\end{threeparttable}
\end{table}

\begin{table}[!h]
\center
\footnotesize
\caption{Subsample Nonemployed - Balance check: Cantonal dummies}
\label{descr2non}
\centering
\begin{threeparttable}

\begin{tabular}{lcccc}
 \hline\hline
 & Sample mean & Coefficient & Standard error & P-value \\
  \hline

Basel-Stadt & 0.04 & -0.01 & 0.01 & 0.70 \\ 
  St. Gallen & 0.11 & 0.00 & 0.01 & 0.93 \\ 
  Thurgau & 0.06 & 0.01 & 0.01 & 0.32 \\ 
  Zurich & 0.30& 0.00 & 0.03 & 0.88 \\ 
  Fribourg & 0.04& -0.01 & 0.01 & 0.50 \\ 
  Geneva & 0.12 & -0.00 & 0.02 & 0.93 \\ 
  Glarus & 0.01 & -0.00 & 0.00 & 0.90 \\ 
  Neuch\^atel  & 0.04& -0.01 & 0.01 & 0.68 \\ 
  Basel-Land & 0.05 & -0.02 & 0.01 & 0.14 \\ 
  Jura & 0.00 & -0.01 & 0.01 & 0.23 \\ 
  Solothurn & 0.05 & -0.01 & 0.01 & 0.46 \\ 
  Bern & 0.01 & 0.00 & 0.01 & 0.70 \\ 
  Vaud & 0.12 & 0.01& 0.02 & 0.51 \\ 
  Schaffhausen & 0.01 & -0.00 & 0.00 & 0.55 \\ 
  Ticino & 0.04 & 0.01 & 0.01 & 0.23 \\ 
  Upper Valais & 0.01 & -0.00 & 0.01 & 0.37\\ 
  Central and Lower Valais  & 0.02 & -0.02 & 0.01 & 0.04 \\ 
   \hline
\end{tabular}

\begin{tablenotes}[flushleft]
\footnotesize
\item Note:  The table presents the balance check of the cantonal dummies as suggested in \cite{LEE2008}. The data comes from STATPOP (2010 - 2017) and OASI (2010 - 2017), the calculations are done by ourselves.
 \end{tablenotes}
\end{threeparttable}
\end{table}

\vskip0.5cm
\begin{figure}[!h]
\centering
\caption{Subsample Employed: Density plot of running variable and Frandsen's manipulation test\\
}
\label{tab:Frandsenstestempl}
\begin{threeparttable}
{ \includegraphics[scale=0.9]{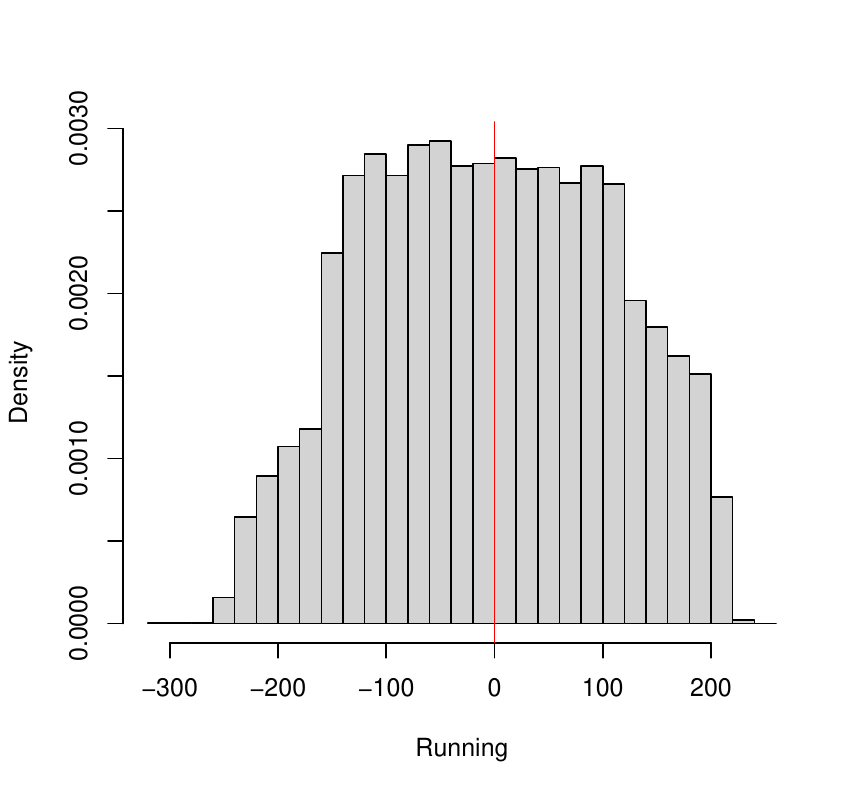}}
\centering

 Manipulation test (k = 0) p-value = 0.00\\
 Manipulation test (k = 0.1) p-value = 0.38\\
 Manipulation test (k = 0.2) p-value =  1.00 \\
 \vskip0.3cm
\begin{tablenotes}[flushleft]
\item Note:  The figure presents the density plot of the running variable and the Frandsen's test for the non-sorting assumption at the birthday cut-off. The running variable is defined as the difference between the birthday cut-off date and children's birth date. K corresponds to the self-defined maximum of the probability mass function (pmf) curve which is still considered as "non-sorting," see \cite{Frandsen2017}. Data stems from STATPOP (2010 - 2017) and OASI (2010 - 2017), calculations are done by ourselves.
 \end{tablenotes}
\end{threeparttable}
\end{figure}

\vskip0.5cm
\begin{figure}[!h]
\centering
\caption{Subsample non-employed: Density plot of running variable and Frandsen's manipulation test\\
}
\label{tab:Frandsenstestnempl}
\begin{threeparttable}
{ \includegraphics[scale=0.9]{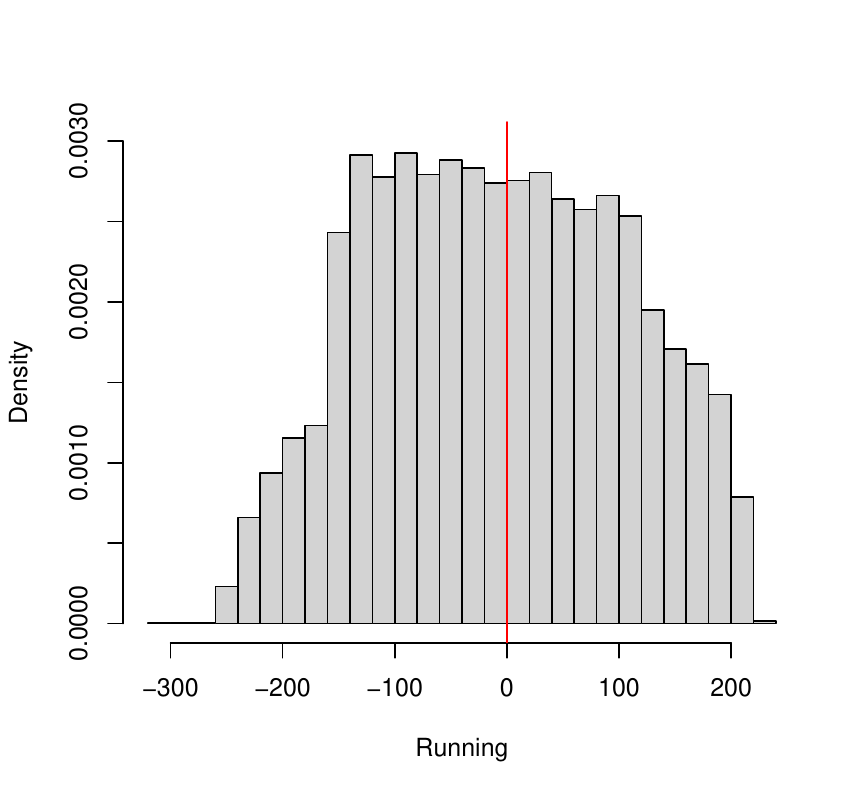}}
\centering

 Manipulation test (k = 0) p-value = 0.00\\
 Manipulation test (k = 0.1) p-value = 0.10\\
 Manipulation test (k = 0.2) p-value =  0.67 \\
 \vskip0.3cm
\begin{tablenotes}[flushleft]
\item Note:  The figure presents the density plot of the running variable and the Frandsen's test for the non-sorting assumption at the birthday cut-off. The running variable is defined as the difference between the birthday cut-off date and children's birth date. K corresponds to the self-defined maximum of the probability mass function (pmf) curve which is still considered as "non-sorting," see \cite{Frandsen2017}. Data stems from STATPOP (2010 - 2017) and OASI (2010 - 2017), calculations are done by ourselves.
 \end{tablenotes}
\end{threeparttable}
\end{figure}


\begin{landscape}
\begin{table}[!h]
\center
\footnotesize
\caption{RDD with Fake Cutoff: Heterogeneous Effects for Non-employed mothers}
\label{nempl_fake}
\centering
\begin{threeparttable}
 \renewcommand\TPTminimum{\textwidth} 
\begin{tabular}{lcccccp{2cm}}
 \hline\hline
  & Sample mean & Coefficient & Standard error & P-value & Bandwidth  & Observations within bandwidth \\
  \hline
Out of labour force (binary) & 0.70 & 0.02 & 0.02 & 0.38 & 40.89& 39,566 \\ 
  Employed (binary) & 0.30 & -0.02 & 0.02 & 0.48 & 41.07 & 40,607 \\ 
  Unemployed (binary) & 0.01 & -0.00 & 0.00 & 0.80 & 51.62 & 50,131 \\ 
  Annual work income (in CHF) & 5,582.01 & 715.98 & 980.61 & 0.47 & 44.61 & 43,274\\ 
  Income dependent employment (binary) & 0.28 & -0.01 & 0.02 & 0.65 & 40.92 & 39,566 \\ 
  Annual income from dependent employment (in CHF) & 4,789.60 & 585.49 & 938.06 & 0.53 & 43.24 & 42,317 \\ 
  \hline
\end{tabular}
\begin{tablenotes}[flushleft]
\footnotesize
\item Note: This table reports the local linear estimates of equation \ref{eqn2} only for previously non-employed mothers, using a fake cut-off set 30 days earlier than the actual threshold. Bandwidth shows the MSE-optimal bandwidth chosen by \citep{Calonico2021}. The following control variables are included: Number of children, born in Switzerland, resident permit, age, relationship status, labour market characteristics of the mother and the father of the four-year old child, cantonal characteristics and dummies. The standard errors are clustered at the individual level. Data stems from STATPOP (2010 - 2017) and OASI (2010 - 2017), calculations are done by ourselves. Income is deflated, the base year is 2011. The official currency in Switzerland is the Swiss Franc (CHF), which had an average exchange rate of 1.04 USD/CHF in the last decade.
 \end{tablenotes}
\end{threeparttable}
\end{table}

\end{landscape}

\begin{landscape}
\begin{table}[!h]
\center
\footnotesize
\caption{Donut-RDD: Heterogenous effects: Non-employed mothers}
\label{donut}
\centering
\begin{threeparttable}
 \renewcommand\TPTminimum{\textwidth} 
\begin{tabular}{lcccccp{2cm}}
 \hline\hline
  & Sample mean & Coefficient & Standard error & P-value & Bandwidth  & Observations within bandwidth \\
  \hline
 Out of labour force (binary)          & 0.70  & -0.03 & 0.02  & 0.12  & 52.07  & 52,578 \\ 
Employed (binary)                     & 0.30  & 0.03  & 0.02  & 0.11  & 49.44  & 49,589 \\ 
Unemployed (binary)                   & 0.01  & -0.00 & 0.00  & 0.59  & 58.32  & 58,731\\ 
Annual work income (in CHF)           & 5,573.86 & 733.75 & 678.40 & 0.28  & 45.67  & 45,805 \\ 
Income dependent employment (binary)  & 0.28  & 0.04  & 0.02  & 0.09  & 51.55  & 51,643 \\ 
Annual income from dependent employment (in CHF) & 4,783.63 & 481.32 & 623.40 & 0.44  & 44.99  & 44,855 \\ 

  \hline
\end{tabular}
\begin{tablenotes}[flushleft]
\footnotesize
\item Note: This table reports the local linear estimates of equation \ref{eqn2} only for previously non-employed mothers excluding observations with running variables close to the cut-off (values between -2 and 1 are omitted). Bandwidth shows the MSE-optimal bandwidth chosen by \citep{Calonico2021}. The following control variables are included: Number of children, born in Switzerland, resident permit, age, relationship status, labour market characteristics of the mother and the father of the four-year old child, cantonal characteristics and dummies. The standard errors are clustered at the individual level. Data stems from STATPOP (2010 - 2017) and OASI (2010 - 2017), calculations are done by ourselves.  Income is deflated, the base year is 2011. The official currency in Switzerland is the Swiss Franc (CHF), which had an average exchange rate of 1.04 USD/CHF in the last decade. Number of observations in total: 178,079.
 \end{tablenotes}
\end{threeparttable}
\end{table}

\end{landscape}

\newgeometry{top=3cm,bottom=3cm,right=1.5cm,left=1.5cm}

\subsection{Additional Difference-in-Differences Approach}
\label{s:app_comp}

\subsubsection{Institutional background: Kindergarten reform}
\label{s:background}

 This section provides an overview of the initial situation and the education policy reform in Switzerland. Switzerland regulates  education policy at the cantonal level, consequently, the quality and quantity of education varies from canton to canton. For example, the school curriculum, entrance age, and the number of school years differ across the 26 cantons \citep{Abstimmung2006}. A national referendum about an educational reform took place in 2006, which was accepted by 86\% of the voters. One year later, an inter-cantonal  "HarmoS" concordat was established, 15 cantons entered into the concordat, while seven cantons rejected it and four cantons are still indecisive.\footnote{as of 2019.} The main goal of the concordat is to harmonise the mandatory school education among the cantons. This agreement also includes an age decrease for mandatory school attendance. All children turning four before the first of August must enter kindergarten \citep{EDK2007}. But also in this case, neither the implementation year nor the shift of the birthday cut-off date is regulated on a central level, see Table \ref{tab: Cut-off dates} in the Online Appendix for an overview.

\begin{figure}[h!t]
	\caption{Expansion of the kindergarten reform}
	\label{fig: Expansion of the kindergarten reform}
	\centering
\begin{threeparttable}

   \includegraphics[width=0.475\textwidth]{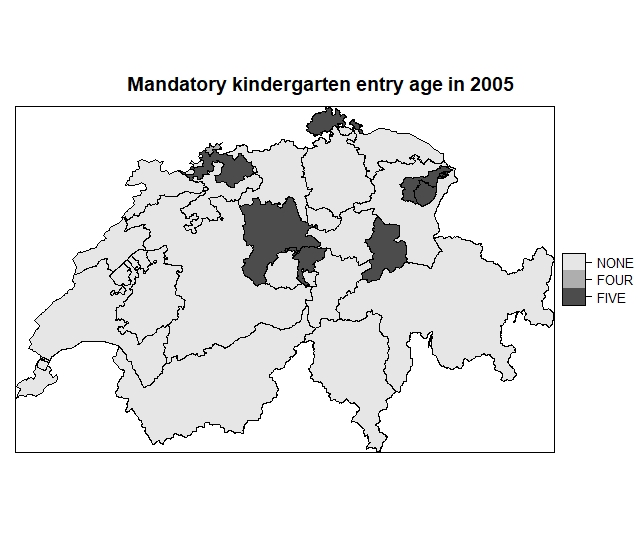}
   \hfill
   \includegraphics[width=0.475\textwidth]{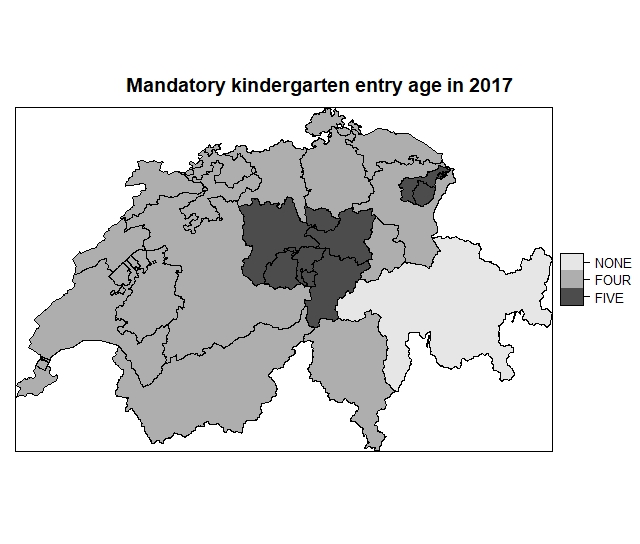}\\
\begin{tablenotes}[flushleft]
\footnotesize
\item Note: The figure compares the spread of mandatory kindergarten for four-year-olds in 2005 (left panel) and 2017 (right panel). The map of Switzerland and its subdivisions stems from the GADM,\footnote{https://gadm.org/maps/CHE.html} the information about the implementation of the mandatory kindergarten was collected by ourselves.
   \end{tablenotes}
\end{threeparttable}

\end{figure}

Even before this inter-cantonal reform, several cantons had started to implement mandatory kindergarten for four-year-olds.  Figure \ref{fig: Expansion of the kindergarten reform} gives an overview of the spread of this policy. Basel-Stadt was the first canton in which four-year-old children are supposed to enter kindergarten in 2005, the other cantons followed over time. In 2017, 17 of 26 cantons had implemented an obligation to attend kindergarten for four-year-olds.\footnote{Aargau, Basel-Land, Basel-Stadt, Bern, Fribourg, Geneva, Glarus, Jura, Neuch\^atel, St Gallen, Schaffhausen, Solothurn, Thurgau, Ticino, Vaud, Valais, and Zurich.} Eight cantons have implemented mandatory kindergarten for five-year-olds.\footnote{Appenzell Outer-Rhodes, Appenzell Inner-Rhodes, Lucerne, Nidwalden, Obwalden, Schwyz, Uri, and Zug.} In the canton Grisons, children are generally not obliged to attend kindergarten. Not directly observable from Figure \ref{fig: Expansion of the kindergarten reform} is the change in the policy in Basel-Land, Glarus, and Schaffhausen. These cantons have implemented a mandatory kindergarten for five-year-old children at first and expanded the policy to four-year-olds.

%

Table \ref{tab: Implementation dates} in the Online Appendix compares the implementation date of mandatory kindergarten (the so-called "Besuchsobligatorium," i.e. children must attend kindergarten) and the mandatory offer of kindergarten places (the so-called "Angebotsobligatorium," i.e. kindergarten places must be offered) for four-year-old children. The second column reveals the year in which each  canton has implemented a mandatory kindergarten for four-year-olds, if ever. The third column lists the year of the obligation to offer kindergarten.  In some cases, the exact implementation year is not known because of data limitation, yet we know that the obligation to offer kindergarten exist at least since 2007.

\newpage

\begin{landscape}
\begin{table}[!h]
\caption{Overview of the implementation dates}
\label{tab:  Implementation dates}
\centering
\begin{threeparttable}
\begin{tabular}{ |p{4.5cm}||p{4.5cm}|p{2cm}||p{4.5cm}| p{4.5cm}|}
\hline
    Canton  & Mandatory kindergarten for four-year-old children & Mandatory kindergarten extended& Obligation to offer kindergarten for four-year-old children &Mandatory offer extended  \\
    \hline
   Aargau & 2013 & no& 2013 & yes \\
   Appenzell Outer-Rhodes & - &no& at least since 2007&at least not since 2007\\
    Appenzell Inner-Rhodes & -& no& at least since 2007&at least not since 2007\\
   Basel-Land & 2012&yes& at least since 2007&at least not since 2007\\
   Basel-Stadt & 2005& no& at least since 2005&at least not since 2005\\
   Bern & 2013&no & 2013&yes\\
   Fribourg &Stepwise since 2009/10&no&2009&yes\\
   Geneva & 2011&no& at least since 2007&at least not since 2007\\
   Glarus &2011 &yes& at least since 2007&at least not since 2007\\
   Grisons& -&no&2013&yes\\
   Jura & 2012&no& at least since 2007&at least not since 2007\\
   Lucerne & -& no&2011&no\\
   Neuch\^{a}tel & 2011& no& at least since 2007&at least not since 2007\\
   Nidwalden & -&no& 2008&no\\
   Obwalden &-&no&-&no\\
   St Gallen &2008&no& at least since 2007&at least not since 2007\\
   Schaffhausen &2014&yes& at least since 2007&at least not since 2007\\
   Schwyz &-&no&2017&no\\
   Solothurn&2012&no& at least since 2007&at least not since 2007\\
   Ticino & 2015&no& at least since 2007&at least not since 2007\\
   Thurgau&2008&no& at least since 2007&at least not since 2007\\
   Uri&-&no&2016&no\\
   Vaud&2013&no& at least since 2007&at least not since 2007\\
   Valais&2015&no&2015&yes\\
   Zug&-&no&-&no\\
   Zurich&2008&no& at least since 2007&at least not since 2007\\

\hline
\end{tabular}

\begin{tablenotes}[flushleft]
\footnotesize
\item Sources: \cite{EDK2007-08}, \cite{EDK2008-09}, \cite{EDK2009-10}, \cite{EDK2010-11}, \cite{EDK2011-12},  \cite{EDK2012-13}, \cite{EDK2013-14},  \cite{EDK2014-15},  \cite {EDK2015-16},  \cite{EDK2016-17}, and \cite{SZ2012}.

\end{tablenotes}
\end{threeparttable}
\end{table}

\end{landscape}

\clearpage
 We can distinguish between the implementation behaviour of four groups: Cantons that have implemented mandatory kindergarten (Table \ref{tab: Mandatory kindergarten and Mandatory offer}) have either implemented the mandatory offer simultaneously or prior to the reform.\footnote{Since the canton Basel-Stadt has implemented the kindergarten policy already in 2005, we cannot ascertain whether the mandatory offer is implemented at the same time or before.} Cantons which have a voluntary kindergarten (Table \ref{tab: Voluntary kindergarten and Mandatory offer}) have either implemented the mandatory offer or not.

\begin{table}[h]
\caption{Mandatory kindergarten \& Mandatory offer}
\label{tab: Mandatory kindergarten and Mandatory offer}
\centering
\begin{threeparttable}
\footnotesize
\begin{tabular}{p{4cm}p{6cm}p{6cm}}
  \hline
\multicolumn{1}{p{4cm}}{ }    & \multicolumn{2}{c}{Mandatory offer}\\
& \multicolumn{1}{c}{Simultaneously} & \multicolumn{1}{c}{Prior}\\
\hline
     Mandatory kindergarten & Aargau, Bern, Fribourg, and Valais &  Basel-Land,   Geneva,   Glarus,   Jura, Neuch\^atel, St Gallen, Schaffhausen, Solothurn, Ticino, Thurgau,Vaud, and Zurich\\
\hline
\end{tabular}
\begin{tablenotes}[flushleft]
\footnotesize
\item Note:  The table shows that cantons with a mandatory kindergarten, either implemented a mandatory offer simultaneously or prior to the policy.
  \end{tablenotes}
\end{threeparttable}

\par

\caption{Voluntary kindergarten \& Mandatory offer}
\label{tab: Voluntary kindergarten and Mandatory offer}
\centering
\begin{threeparttable}

\footnotesize
\begin{tabular}{p{4cm}p{6cm}p{6cm}}
  \hline
\multicolumn{1}{p{4cm}}{ }  & \multicolumn{2}{c}{Mandatory offer} \\
& \multicolumn{1}{c}{Implemented} & \multicolumn{1}{c} {Not implemented}\\
\hline
 Voluntary kindergarten & Appenzell Outer-Rhodes,
    Appenzell Inner-Rhodes, Grisons, Lucerne, Nidwalden, and Uri & Obwalden, Schwyz, and Zug\\
  \\

\hline
\end{tabular}

\begin{tablenotes}[flushleft]
\footnotesize
\item Note:  The table reveals that cantons with a voluntary kindergarten have either implemented a mandatory offer or not.
\end{tablenotes}
\end{threeparttable}
\end{table}

Kindergarten is free of charge for both, the mandatory  and the voluntary kindergarten \citep{EDK2008}. Hence, the implementation of  the kindergarten reform is not a subsidy, merely an obligation. According to a cantonal survey by the Swiss Conference of Directors of Education, almost 77\% of four-year-old children attended kindergarten in the school year 2007/08 \citep{EDK2007-08}, although most cantons had not yet implemented the obligation. The aim of the kindergarten policy is twofold: Children's development should be promoted, but also  the compatibility of family and work should be encouraged \citep{EDK2014}. In this paper, we analyse the latter and examine whether the obligation to attend kindergarten at the age of four affects mothers' work behaviour.

Although data on kindergarten attendance rates are not available for all cantons, we know that in the cantons of Basel-Stadt, Thurgau, Neuchâtel, Aargau and Glarus on average 94\% of four-year-olds attended kindergarten before the introduction of the obligation.\footnote{Departments of Education of the respective cantons.}

Red-shirting, i.e. deferring a four-year-old child from kindergarten, is possible, but is associated with hurdles that vary from canton to canton. In some cantons, an application by the parents is sufficient, in other cantons a medical certificate is required which proves that the admission of a child to the kindergarten would not be reasonable even with other educational assistance.\footnote{https://www.swissmom.ch/de/kind/kindergarten-und-schule/mit-4-jahren-schon-in-den-kindergarten-19063.} 

\subsubsection{Data}\label{SHP}
For the Difference-in-Differences strategy, we link data from different sources. Firstly, we use survey data from the Swiss Household Panel (SHP). This panel starts in 1999 and has drawn households by NUTS II regions. The sample is representative with respect to social groups across Switzerland. In the years 2004 and 2013 refreshment samples are added. The main aim of the SHP is  to follow the participants over a longer period of time in order to study social change \citep{SHP2017}. For this reason, each person has a unique identifier (id). In our analysis, we use data about mothers' demographic variables and their labour market behaviour. This information comes from the individual dataset. The number of children in the household and the residential canton are provided by the household dataset. The birth month and year of birth of the children as well as the id number of the mother stem from the grid dataset.

Secondly, data concerning the mandatory kindergarten entry age, the year of policy implementation, and the birthday cut-off dates are collected from the cantonal departments of education, the cantonal laws, and the Swiss Conference of Cantonal department of education (EDK). Furthermore, the EDK conducts a yearly survey among the cantons. From these documents, we extract the information about the obligation to offer kindergarten. However, this information is only available for the time span between 2007 and 2017. The EDK provided also data about the member cantons of the HarmoS concordat. Thirdly, cantonal data about the unemployment rate in each year were provided by the Staatsekretariat f\"ur Wirtschaft (SECO). Information about the official languages in each canton stems from cantonal laws.

To get one dataset, we merge all waves of the individual SHP data and the unique grid dataset by the identification number of a person. We use the same procedure for all waves of household data. In this case, the unique identifier is the household number. Then we merge these data with the masterfile. Afterwards, we merge the masterfile with the other collected data by canton and year. This approach was not possible for the canton Valais because the birthday cut-off dates differ between the German-speaking part (Upper Valais) and the francophone part (Central and Lower Valais). Therefore, we request the community numbers from the SHP additionally. We matched these administrative numbers with the two parts of the canton Valais. Subsequently, we merge the data about the kindergarten policy with this file and append it to the master file.

\begin{figure}[!h]
\centering
\caption{Definition of the age span}
\label{fig: Definition of the age span}
\begin{threeparttable}
   \includegraphics[width=1\textwidth]{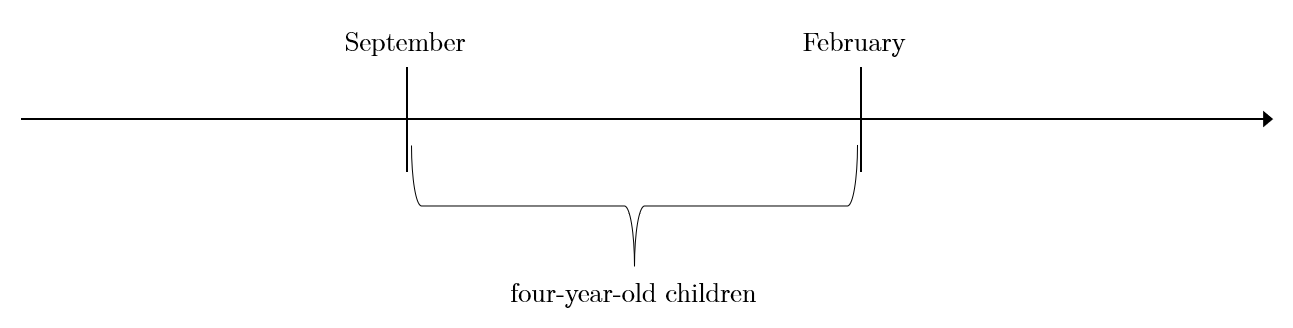}
\begin{tablenotes}[flushleft]
\footnotesize
\item Note:  The timeline presents the subsample restriction. Only four-year-old children born between September and February are included to obtain a constant treatment and control group because the birthday cut-offs differed across cantons and over years.
        \end{tablenotes}
\end{threeparttable}

\end{figure}

In a next step, we restrict the sample to  mothers whose youngest child is four years old between September and February, living in the same household for the years 2004 to 2017. The reasons for our approach are threefold: First, the empirical literature finds no effect for mothers with younger children in the same household. see, for example, \cite{Berlinski2007}, \cite{fitzpatrick2012}, and \cite{Gelbach2002}. Second, the birthday cut-off dates differ from canton to canton and from year to year. Table \ref{tab: Cut-off dates} in the Online Appendix shows, the earliest birthday cut-off date ever (28.02.) was implemented in the German-speaking part of the canton Valais  (Upper Valais). Whereas the latest birthday cut-off date ever (31.08.) was implemented in the French-speaking part of the canton Valais (Cenral and Lower Valais).  To get a constant treatment and control group,  we define the birthday cut-off as a time span from the latest ever implemented birthday cut-off to the earliest ever implemented birthday cut-off as in Figure \ref{fig: Definition of the age span}.

Since we use this time span, we exclude rather older children from our sample. Table \ref{tab: Treatment and control group} shows the treatment and control group: The age of the children is the same in both groups: They turn four before September and turn five after February. However, the children differ by their canton of residence. The treated children live in a canton in which the mandatory kindergarten for four-year-olds is implemented. Whereas the children in the control group live in a canton in which none or mandatory kindergarten for five-year-olds is implemented.

\begin{table}[!h]
\centering
\begin{threeparttable}
\caption{Treatment and control group}
\label{tab: Treatment and control group}
\begin{tabular}{cp{4.5cm}p{4.5cm}}
\hline
  & Treatment &  Control\\\hline
 Age \& birthday cut-off & Children turning four  & Children turning four \\
&before September & before September\\
 &and five after February&and five after February
  \\\hline
Canton & Mandatory kindergarten for four-year-old & None or mandatory kindergarten for five-year-old\\
& children  implemented&  children implemented
  \\\hline

\end{tabular}

\begin{tablenotes}[flushleft]
\footnotesize
\item Note:  The table shows that the treatment and control group consists of four-year-old children born between September and February. These children are assigned to the treatment group if mandatory kindergarten has been implemented in their canton of residence and to the control group otherwise.
        \end{tablenotes}
\end{threeparttable}
\end{table}

\hfill

Third, Basel-Stadt was the first canton that implemented mandatory kindergarten for four-year-olds. To have one pre-period, we use the time span from 2004 to 2017 (last year with available data). After this restriction procedure, we end up with 953 observations, out of which 394 observations are treated and 559 are part of the control group.

Table \ref{tab: Descriptive statistics} provides the descriptive statistics for the treatment and control group separately. The latter contains never treated observations as well as observations that are still in the pre-treatment period, yet treated later on. Since the treatment is implemented over time in the cantons, the number of observations in the treatment group is smaller than in the non-treated group. Consequently, the treated cases are observed later on average. Considering mothers' characteristics, the average age of a mother whose child is treated is 39 years old, whereas mothers of non-treated children are on average younger. Mothers in the treatment group are also higher educated than the control group. A value of 6 represents A-level as the highest degree.

\begin{table}[!h]
\center
\caption{SHP: Descriptive statistics}
\label{tab: Descriptive statistics}
\centering
\begin{threeparttable}
\footnotesize
\begin{tabular}{l|cc|cc| cc}
  \hline
   & \multicolumn{2}{c|}{Treated} & \multicolumn{2}{c|}{Non-treated} & & \\
& Mean & Std.dev & Mean & Std.dev & Mean difference & P-value\\
\hline
  & \multicolumn{4}{c|}{Time} \\
  \hline
Year & 2014.09 & 2.52 & 2008.11 & 3.52 & 5.98 & 0.00 \\
   \hline
  & \multicolumn{4}{c|}{Control variables} \\
\hline
\textit{Mothers' characteristics}  & & & & & & \\
Age & 39.17 & 4.28 & 38.12 & 3.86 & 1.05 & 0.00 \\
Highest education  & 6.53 & 2.84 & 5.91 & 2.72 & 0.62 & 0.00 \\
Married & 0.86 & 0.35 & 0.89 & 0.31 & -0.03 & 0.11 \\
Single & 0.10 & 0.30 & 0.04 & 0.19 & 0.06 & 0.00 \\
Foreign & 0.17 & 0.37 & 0.13 & 0.34 & 0.04 & 0.10 \\
Political attitude & 4.51 & 2.02 & 4.62 & 1.87 & -0.11 & 0.38 \\
Self-reported health status & 1.85 & 0.60 & 1.95 & 0.59 & -0.10 & 0.02 \\
Number of children  & 2.14 & 0.78 & 2.18 & 0.84 & -0.04 & 0.45 \\
\textit{Cantonal characteristics} & & & & & & \\
Unemploment rate & 3.31 & 0.92 & 3.26 & 1.31 & 0.05 & 0.48 \\
Mandatory kindergarten for 5-year-olds & 0.04 & 0.19 & 0.23 & 0.42 & -0.20 & 0.00 \\
HarmoS member & 0.89 & 0.31 & 0.69 & 0.46 & 0.20 & 0.00 \\
\hline
& \multicolumn{4}{c|}{Outcomes} \\
\hline
Out of labour force & 0.18 & 0.38 & 0.23 & 0.42 & -0.05 & 0.04 \\
Unemployed & 0.02 & 0.12 & 0.01 & 0.11 & 0.00 & 0.73 \\
Employed & 0.81 & 0.39 & 0.76 & 0.43 & 0.05 & 0.06 \\
Part-time employed & 0.77 & 0.42 & 0.70 & 0.46 & 0.07 & 0.02 \\
Full-time employed & 0.04 & 0.20 & 0.06 & 0.24 & -0.02 & 0.26 \\
Income from dependent employment & 0.78 & 0.42 & 0.72 & 0.45 & 0.06 & 0.04 \\
Number of observations & 394 &  & 559 &  &  &  \\
\hline
\end{tabular}

\begin{tablenotes}[flushleft]
\footnotesize
\item Note:  The table presents means and standard deviations separately for mothers with treated and non-treated children. P-values are derived from a t-test on the equality of means among mothers with a treated and a non-treated child. Data stems from the Swiss Household Panel (SHP), the calculations are done by ourselves.
 \end{tablenotes}
\end{threeparttable}
\end{table}

 Almost 90\% of mothers are married, yet mothers are more likely to be single in the treated group. Furthermore, the share of foreign mothers is higher in the treated group. The average political attitude of mothers is at the center of the political spectrum and does not differ significantly between the groups. The self-reported health status is good, yet better among the treated mothers. Two children live in an average household. The cantonal characteristics reveal an unemployment rate of 3.31\% on average in the treated observations with no statistically significant differences between the groups.  However,  the treated cantons are more likely to be a HarmoS member (89\%) and less likely to have implemented one kindergarten year (4\%) over time. The second part of  Table \ref{tab: Descriptive statistics} gives an overview of the descriptive statistics of the outcome variables. Mothers in the treatment group are more likely to take part in the labour force, have a higher probability of being employed, work more often in part-time work, and earn their income more frequently from dependent employment in comparison to the control group. There are no statistical differences in the probability of being unemployed or working full-time.

\subsubsection{The Difference-in-Differences Approach}
As preschool policies are decided on the cantonal level, we apply a difference-in-differences (DiD) strategy to exploit treatment variation across cantons and over time. As there has been a staggered adoption of mandatory kindergarten across cantons over time, we use a two way fixed effects model rather than the basic DiD setting with one pre- and one post-treatment period. The DiD estimation with different timings gives a weighted average of all DiD estimators across groups and times. We use ordinary least squares (OLS) to estimate the results.

Equation \eqref{did:model} shows the estimated model in our empirical analysis. The subscript i denotes the individual, s the canton, and t the time. Y$_{\text{ist}}$ is a vector of the following binary dependent variables:  "Out of labour force", "Unemployed", "Employed",  "Part-time employment", "Full-time employed",  "Income from dependent work."

\begin{equation}
Y_{ist}=\alpha+\gamma_{s}+\lambda_{t}+\delta D_{st}+\beta X'_{ist}+\epsilon_{ist}
\label{did:model}
\end{equation}

The right hand side of equation \eqref{did:model} presents the independent variables.  We include canton  ($\gamma_{s}$)  and time fixed effects ($\lambda_{t}$), the former controls for time-invariant differences between the cantons, the latter for differences across years which are common to all cantons. D$_{\text{st}}$ represents the treatment dummy indicating whether the child is eligible for the treatment, i.e. mandatory kindergarten in the respective canton and the current year. The parameter $\delta$ gives  the effect we are interested in, the average treatment effect on the treated (ATT). $\epsilon_{ist}$ corresponds to time-varying unobservables. We include a vector of  time-variant as well as time-invariant covariates
(X$_{\text{ist}}$)  to make the common trend assumption more plausible.

The common trend assumption is the key identifying assumption in a DiD approach. Intuitively speaking, the labour market outcomes of the treatment and control group would follow the same time trend in the absence of the treatment, i.e. the mandatory kindergarten.  Therefore, we need to control for those maternal and cantonal covariates which would lead to a different time trend. For instance, younger and less educated mothers are usually more affected by business cycles than older and better educated ones, such that the time trend differs among these groups. Another example represents mothers with a poorer  health status who might be laid off faster than healthy mothers in times of crisis. We must also control for cantonal covariates which might affect the employment trend of treated and control groups differently like the unemployment rate measure one year prior to the treatment year.  We provide placebo tests conditional on covariates in Table \ref{tab:placeboresultsnoweight} and \ref{tab:placeboresultsnoweight2} in Section~\ref{did}.

To check the robustness of the results, we use three specifications: First, we include socio-demographic characteristics of the mother which can affect the labour market situation. This subset of covariates is commonly used in the childcare literature. Like \cite{Felfe2016}, we include mothers' age, education, civil status, and the number of children in the household. In line with \cite{Ravazzini2018}, we also add mothers' citizenship as control. Since \cite{Cai2006} reports that a better health status leads to a rise in the likelihood of participating in the labour market, we control for mothers' self-reported health status additionally. Conservative mothers are more likely to stay out of the labour force and work less if they decide to participate in the labour market \citep{Stam2014}. For this reason, we control for mothers' political attitude. Second, we extend the former specification by adding cantonal characteristics to "control for forces that lead policies to change" \cite{besley2000}. Like \cite{Felfe2016} we use the cantonal unemployment rate as control. In the context of our study, we control for the implementation of a mandatory kindergarten for five-year-old children and  the membership in the HarmoS concordat. Third, we use double machine learning to pick the most important confounders, interactions, and higher order terms from the socio-demographic and cantonal characteristics in a data-driven way.


The latter approach prevents both overfitting (too many included covariates) and omitted variable bias (too few included variables). The variable selection is done by a least absolute shrinkage and selection operator (LASSO) of the R package "hdm", developed by \citep{Chernzhukov2016}.  Since we aim to do linear inference, we use the command "rlassoEffects" and estimate the results with the method "double selection", option "POST=TRUE".\footnote{The "rlassoEffects" command provides also the option "partialling-out". We have estimated the effects with this option too. The point estimates remain the same, in some cases the standard errors differ.} This procedure selects the relevant covariates first and does inference afterwards. Table \ref{tab: Picked covariates} in the Online Appendix gives an overview of the included covariates in each specification.

%
%
%
%
%

\subsubsection{Findings}\label{did}
Subsequently, we report the results of the DiD analysis.
Table \ref{DiD1} presents the main results of the kindergarten reform for mothers with four-year-old children. We display three specifications: Firstly, "socio-demographic X" controls only for socio-demographic characteristics, secondly, "all X" controls for all characteristics represented in Table~\ref{tab: Descriptive statistics} in Section~\ref{SHP}, and finally we let LASSO pick the controls "lasso picked-X". The standard errors are clustered to control for cantonal-specific effects in the first two specifications. However, clustering is not possible in the LASSO approach.\footnote{We estimated the first two specifications without clustered standard errors as well. The standard errors were not much different.} Therefore, we use heteroscedasticity robust standard errors in the last specification. The last column provides the number of observations "obs". The latter varies because of missing data in the outcome variables which leads to the exclusion of these observations.
\begin{table}[!h]
\center
\caption{Empirical results}
\label{DiD1}
\centering
\begin{threeparttable}
\footnotesize
\begin{tabular}{l|ccc|ccc|ccc|c}
  \hline\hline
   & \multicolumn{3}{c|}{socio-demographic $X$} & \multicolumn{3}{c|}{all $X$} & \multicolumn{3}{c|}{lasso-picked $X$} \\
 & effect & se & pval & effect & se & pval & effect & se & pval & obs \\
  \hline
Out of labour force & -0.09 & 0.04 & 0.04 & -0.10 & 0.04 & 0.02  & -0.09 & 0.05 & 0.05 & 953 \\
Unemployed & 0.02 & 0.02 & 0.28 & 0.02 & 0.02 & 0.28 & 0.02 & 0.02 & 0.22 & 953 \\
Employed & 0.07 & 0.05 & 0.14  & 0.08 & 0.05 & 0.10 & 0.07 & 0.05 & 0.14 & 953 \\
Part-time employed & 0.11 & 0.06 & 0.04 & 0.12 & 0.06 & 0.03 & 0.13 & 0.05 & 0.01 & 948 \\
Full-time employed & -0.04 & 0.02 & 0.13 & -0.04 & 0.02 & 0.11  & -0.04 & 0.03 & 0.15 & 948 \\
Income from dependent & & & & & & & & & &\\ employment & 0.06 & 0.05 & 0.26  & 0.07 & 0.05 & 0.13 & 0.09 & 0.05 & 0.09 & 952 \\
\hline
\end{tabular}

\begin{tablenotes}[flushleft]
\footnotesize
\item Note: This table reports the OLS estimates of  equation \ref{did:model} for three different model specification depending on the picked covariates (X). An overview of the picked covariates is available in \ref{tab: Picked covariates} in the Online Appendix. Standard errors in the model specifications "socio-demographic X" and "all X" account for clustering, the standard errors of lasso-based estimation do not. Data stems from Swiss Household Panel (SHP) 2004- 2017.

 \end{tablenotes}
\end{threeparttable}
\end{table}

The findings suggest a positive effect of the reform on the labour force supply of mothers: If the youngest child is supposed to enter kindergarten at the age of four, mothers' probability of being in the labour force increases by 9 percentage points. Therefore, the share of employed (unemployed) mothers increases by 7 (2) percentage points, yet not significantly. Results on the employment level reveal a significant effect on mothers' part-time work by 11 percentage points. In contrast, mothers' full-time rate decreases by 4 percentage points and is not significant. The obligation to attend kindergarten has no significant effect. It gets significant at the 10 \% level in the LASSO specification.\footnote{As the geographic distance between Swiss cantons is often small, we cannot rule out spillover effects, implying that control cantons not (yet) introducing mandatory kindergarten benefit to some extent from increased labour supply in treated cantons. If anything, this should bias estimated labour supply effects towards zero.} The Benjamini-Hochberg procedure suggests one (two) statistically significant effect at the 10\% level for the lasso (all X) specification, but none for the socio-demographic specification.\footnote{We note that if treatment effects are heterogeneous across time periods and cantons, linear models, as used for estimating the results presented in Table \ref{DiD1}, generally entail biased estimates of the overall Average Treatment Effect on the Treated (ATET), the causal parameter targeted by DiD approaches. It is theoretically possible that the ATET estimate in linear models does not correspond to any causal effect, not even a weighted average across cantons or time periods; see, e.g., the discussion in \cite{goodman2021difference}. As a robustness check that moves away from linear regression models, we also estimated the DiD specification when controlling for all covariates ($X$) and clustering at the cantonal level based on the semiparametric doubly robust estimator of \cite{CallawaySantAnna2018} as implemented in the R package "did" by \cite{didpackage} with its default options. The effects have the same sign as in the linear specifications in Table \ref{DiD1} (ATET on out of labor force: -0.14; unemployed: 0.02; employed: 0.12; part-time employed: 0.19; full-time employed: -0.07; income from dependent employment: 0.04), but are far from being statistically significant at any conventional level due to substantially larger standard errors.}

To investigate the plausibility of the parallel trend assumption, we conduct two placebo tests. In Table \ref{tab:placeboresultsnoweight}, we use a fake treatment group, namely mothers whose children are too young to be eligible for the reform. In the first specification, the youngest child in the household is between 0 and 1 years old. The other two columns use alternative specifications regarding the age of the children: 1 and 2 years, as well as 2 and 3 years. We control for the socio-demographic as well as the cantonal covariates in the placebo test. We do not find any statistically significant effect of the reform on mothers' whose children are younger than the eligibility age. In Table \ref{tab:placeboresultsnoweight2}, we test whether there was an anticipatory effect in the periods prior to the very first implemented reform (1999 - 2004). For this analysis, we split the pre-treatment periods in a fake pre-treatment period (1999 - 2001) and a fake post-treatment period (2002 - 2004). Since the p-values are too large to be significant, we cannot reject the null hypothesis and find no support for an anticipatory effect.

\begin{table}[!h]
\caption{Placebo tests with unaffected age groups}  \label{tab:placeboresultsnoweight}
\centering
\begin{threeparttable}
\footnotesize
\begin{tabular}{l|cccc|cccc|cccc}
  \hline\hline
   & \multicolumn{4}{c|}{0 and 1 years old} & \multicolumn{4}{c|}{1 and 2 years old}& \multicolumn{4}{c}{2 and 3 years old}  \\
 & est & se & pval & obs & est & se & pval & obs & est & se & pval & obs \\
  \hline
Out of labour force  & 0.04 & 0.05 & 0.42 & 440 & 0.02 & 0.04 & 0.68 & 680 & -0.00 & 0.05 & 0.95 & 732 \\
Unemployed & 0.02 & 0.04 & 0.70 & 440 & -0.00 & 0.03 & 0.90 & 680 & -0.00 & 0.02 & 0.86 & 732 \\
Employed & -0.06 & 0.05 & 0.28 & 440 & -0.01 & 0.05 & 0.77 & 680 & 0.01 & 0.05 & 0.89 & 732 \\
Part-time employed  & -0.11 & 0.09 & 0.19 & 439 & -0.02 & 0.06 & 0.76 & 676  & -0.01 & 0.05 & 0.86 & 728 \\
Full-time employed & 0.05 & 0.05 & 0.28 & 439 & 0.00 & 0.02 & 0.93 & 676 & 0.02 & 0.02 & 0.31 & 728 \\
Income from dependent & & & & & & & & & &\\ employment & -0.05 & 0.04 & 0.25 & 440 & -0.04 & 0.04 & 0.34 & 680 & -0.00 & 0.06 & 0.96 & 732 \\
\hline
\end{tabular}

\begin{tablenotes}[flushleft]
\footnotesize
\item Note:  This table displays the estimates of  equation \ref{did:model} for unaffected age groups, namely younger children.  It controls for all $X$, and the standard errors account for clustering. Data stems from Swiss Household Panel (SHP) 2004- 2017.

 \end{tablenotes}
\end{threeparttable}
\end{table}

\vskip3cm

\begin{table}[!h]
\captionsetup{justification=centering}
\caption{DiD: Placebo tests with unaffected periods \\Pre-treatment period: 1999-2001; fake post-treatment period: 2002-04}\label{tab:placeboresultsnoweight2}
\centering
\begin{threeparttable}
		\footnotesize
			\begin{tabular}{l|ccc|ccc|ccc|c}
				\hline\hline
				& \multicolumn{3}{c|}{treated: intro after 2008} & \multicolumn{3}{c|}{treated: intro after 2010}& \multicolumn{3}{c|}{treated: intro after 2012}  \\
				& est & se & pval & est & se & pval  & est & se & pval & obs \\
				\hline
Out of labour force & 0.06 & 0.06 & 0.34   & 0.04 & 0.06 & 0.52 & -0.01 & 0.07 & 0.90 & 490 \\
Unemployed & -0.01 & 0.03 & 0.62  & -0.01 & 0.03 & 0.74  & 0.01 & 0.02 & 0.81 & 490 \\
Employed & -0.05 & 0.07 & 0.51 & -0.03 & 0.07 & 0.65 & 0.00 & 0.07 & 0.95 & 490 \\
Part-time employed & -0.03 & 0.08 & 0.75 & -0.03 & 0.08 & 0.66 & 0.00 & 0.08 & 0.99 & 485 \\
Full-time employed & -0.02 & 0.04 & 0.69  & 0.01 & 0.04 & 0.88  & 0.02 & 0.02 & 0.53 & 485 \\
\hline
\end{tabular}

\begin{tablenotes}[flushleft]
\footnotesize
\item Note:  This table displays the estimates of  equation \ref{did:model} for unaffected periods, i.e. before the policy was implemented.  It controls for all $X$, and the standard errors account for clustering. Data stems from Swiss Household Panel (SHP) 2004- 2017.
 \end{tablenotes}
\end{threeparttable}
\end{table}

To test the robustness of the results, we run several checks. At first, we report the results with a changed control group. In Table \ref{tab:mainresultsnoweightnofirst}, we exclude mothers whose children have to attend kindergarten at the age of five. In other words, only mothers with children living in a canton without the kindergarten obligation are in the control group. The effect on the labour force participation gets more precise and bigger in absolute terms. The effect on being unemployed stays insignificant and is close to zero, whereas the effect on being employed increases slightly in comparison to the main results. Even in this robustness check, the effect on part-time rate is highly significant. The effect on the full-time rate stays insignificant. The effect on the income from dependent work gets highly significant.

\begin{table}[!h]
\caption{Empirical results with changed control group \\ (only eventually treated)}  \label{tab:mainresultsnoweightnofirst}
\centering
\begin{threeparttable}

\footnotesize
\begin{tabular}{r|ccc|ccc|ccc|c}
  \hline\hline
   & \multicolumn{3}{c|}{socio-demographic $X$} & \multicolumn{3}{c|}{all $X$} & \multicolumn{3}{c|}{lasso-picked $X$} \\
& est & se & pval & est & se & pval & est & se & pval & obs \\
  \hline
Out of labour force & -0.11 & 0.05 & 0.03 & -0.12 & 0.05 & 0.01 & -0.10 & 0.05 & 0.03 & 825 \\
Unemployed  & 0.02 & 0.02 & 0.34 & 0.02 & 0.02 & 0.26 & 0.02 & 0.02 & 0.31 & 825 \\
Employed  & 0.09 & 0.06 & 0.12  & 0.10 & 0.05 & 0.07 & 0.08 & 0.05 & 0.10 & 825 \\
Part-time employed & 0.12 & 0.07 & 0.08  & 0.12 & 0.06 & 0.04 & 0.13 & 0.06 & 0.02 & 821 \\
Full-time employed & -0.02 & 0.03 & 0.37 & -0.02 & 0.03 & 0.39 & -0.03 & 0.03 & 0.39 & 821 \\
Income from dependent & & & & & & & & & & \\
 employment & 0.11 & 0.04 & 0.01 & 0.10 & 0.05 & 0.03 & 0.11 & 0.05 & 0.04 & 824 \\
  \hline
\end{tabular}

\begin{tablenotes}[flushleft]
\footnotesize
\item Note:  This table displays the estimates of  equation \ref{did:model} with a changed control group, only eventually treated,  for three different model specification depending on the picked covariates (X). Standard errors in the model specifications "socio-demographic X" and "all X" account for clustering, the standard errors of lasso-based estimation do not. Data stems from Swiss Household Panel (SHP) 2004- 2017.
 \end{tablenotes}
\end{threeparttable}
\end{table}

For another robustness check, we exclude cases in which the information about the implementation year of a mandatory kindergarten was uncertain. This was the case for the cantons Glarus (2001-2010), Jura (1999-2011), and Ticino (1999-2014). The results are provided in Table \ref{tab: Empirical results dropping ambiguous kindergarten entry} in the Online Appendix. The effect on the labour force remains positive and significant. Whereas the impact of the kindergarten policy on the part-time employment is only significant in the lasso-picked specification. The effect on unemployment, and full-time employment and the income from dependent employment is insignificant.

In order to make the sample representative, we use individual cross-sectional weights which keep the sample size of the current year in a next step. The weights are provided by the SHP and refer to the population of the whole sample. Therefore, we have to adapt the weights, because we use a subsample of mothers with children of a specific age. We followed the approach of \cite{Antal2016} to adapt the weights. Furthermore, there are no weights available for only the second sample. For this reason, we use combined weights for the first and second sample and exclude observations that are not part of the survey from the beginning. Table \ref{tab:resultsweight} in the Online Appendix shows that the effect on part-time employment (controlling for all X) remains significant and the negative impact of the reform on full-time employment gets highly significant.

Since mothers with different educational levels could react differently to the implementation of mandatory kindergarten, we split the sample to analyse effect heterogeneity. However, we could not find any evidence for a different effect among the groups. This could also be the case because of the limited sample size.




All in all, the DiD results suggest a positive effect of mandatory kindergarten on mothers' employment, compared to rather insignificant effects in the RDD approach. However, we need to put our findings from these two different evaluation strategies into perspective: Firstly, the results from the RDD approach are based on administrative data, such that the number of observations is larger and the quality of data is better than the survey data used in the DiD part.  Due to the smaller sample size, the findings of the DiD approach are imprecisely estimated. Secondly, the RDD identifies in general the ITT at the threshold, whereas the DiD identifies the effect for the sub-population of treated individuals (ATT), such that these effects might not be directly comparable. Finally, we analyse the effect in both evaluation strategies on rather different outcomes.

\renewcommand\thetable{3.\thesection.2.\arabic{table}}
\setcounter{table}{0}

\newgeometry{top=3cm,bottom=3cm,right=2.5cm,left=2.5cm}

\begin{landscape}
\begin{table}[!h]
\caption{Picked covariates}
\label{tab: Picked covariates}
\centering
\begin{threeparttable}
\begin{tabular}{ |p{7cm}||p{7cm}||p{7cm}|  }
\hline
    socio-demographic $X$ & all $X$ & lasso-picked $X$ \\
    \hline
  \textit{Mothers' characteristics} & \textit{Mothers' characteristics}&\textit{Mothers' characteristics}\\
    age  & age&age\\
    age\textsuperscript{2}& age\textsuperscript{2}&\\
    education & education &education\\
    citizenship & citizenship & citizenship\\
     civil status & civil status&\\
     political attitude & political attitude & \\
      political attitude\textsuperscript{2}&and political attitude\textsuperscript{2}&\\
      self-reported health&self-reported health&self-reported health\\
      \hline
      \textit{Household characteristics}& \textit{Household characteristics}&\textit{Household characteristics}\\
       number of children in the household & number of children in the household & number of children in the household\\
       \hline
       &\textit{Cantonal characteristics}&\textit{Cantonal characteristics}\\
       & unemployment rate& unemployment rate\\
       &unemployment rate\textsuperscript{2}&\\
     & mandatory kindergarten for five-year-old children& mandatory kindergarten for five-year-old children\\
      & canton is member of Harmos&canton is member of Harmos\\

\hline

\end{tabular}

\begin{tablenotes}[flushleft]
\footnotesize
\item Note:  This table reports the included covariates in the estimations of each model specification.
   \end{tablenotes}
\end{threeparttable}
\end{table}

\end{landscape}

\begin{table}[!h]
\caption{Empirical results dropping ambiguous kindergarten entry}  \label{tab: Empirical results dropping ambiguous kindergarten entry}
\centering
\footnotesize
\begin{threeparttable}
\begin{tabular}{r|ccc|ccc|ccc|c}
  \hline\hline
   & \multicolumn{3}{c|}{socio-demographic $X$} & \multicolumn{3}{c|}{all $X$} & \multicolumn{3}{c|}{lasso-picked $X$} \\
 & effect & se & pval & effect & se & pval & effect & se & pval & obs \\
  \hline
Out of labour force & -0.07 & 0.04 & 0.10  & -0.08 & 0.05 & 0.08 & -0.08 & 0.04 & 0.08 & 915 \\
Unemployed & 0.02 & 0.02 & 0.14 & 0.02 & 0.02 & 0.14 & 0.02 & 0.02 & 0.19 & 915 \\
Employed & 0.05 & 0.05 & 0.31 & 0.06 & 0.05 & 0.26 & 0.06 & 0.05 & 0.22 & 915 \\
Part-time employed  & 0.09 & 0.06 & 0.11 & 0.10 & 0.06 & 0.09  & 0.11 & 0.05 & 0.03 & 910 \\
Full-time employed  & -0.04 & 0.03 & 0.19 & -0.04 & 0.03 & 0.18 & -0.04 & 0.03 & 0.16 & 910 \\
Income from dependent employment & 0.05 & 0.06 & 0.39 & 0.06 & 0.05 & 0.20  & 0.05 & 0.05 & 0.32 & 914 \\
  \hline
\end{tabular}

\begin{tablenotes}[flushleft]
\footnotesize
\item Note:  This table reports the OLS estimates of  equation \ref{did:model} and excludes cases with uncertain information about the implementation year of a mandatory kindergarten. An overview of the picked covariates is available in \ref{tab: Picked covariates}. Standard errors in the model specifications "socio-demographic X" and "all X" account for clustering, the standard errors of lasso-based estimation do not. Data stems from Swiss Household Panel (SHP) 2004- 2017.
 \end{tablenotes}
\end{threeparttable}
\end{table}

\begin{table}[!h]
\caption{Empirical results with weights}  \label{tab:resultsweight}
\centering
\begin{threeparttable}
\footnotesize
\begin{tabular}{r|ccc|ccc|c}
  \hline\hline
   & \multicolumn{3}{c|}{socio-demographic $X$} & \multicolumn{3}{c|}{all $X$}  \\
   & est & se & pval & est & se & pval  & obs \\
   \hline
 Out of labour force & -0.05 & 0.05 & 0.37 & -0.07 & 0.05 & 0.13 & 946 \\
Unemployed & 0.02 & 0.01 & 0.14 & 0.02 & 0.01 & 0.07 & 946 \\
Employed & 0.03 & 0.06 & 0.57 & 0.05 & 0.05 & 0.32 & 946 \\
Part-time employed & 0.10 & 0.07 & 0.13  & 0.13 & 0.06 & 0.04 & 941 \\
Full-time employed & -0.07 & 0.03 & 0.03 & -0.07 & 0.03 & 0.02 & 941 \\
Income from dependent employment & 0.02 & 0.06 & 0.69 & 0.04 & 0.06 & 0.54 & 945 \\
\hline
\end{tabular}

\begin{tablenotes}[flushleft]
\footnotesize
\item Note:  This table reports the OLS estimates of  equation \ref{did:model}, in which observation weights from the Swiss Household Panel (SHP)  are used. An overview of the picked covariates is available in \ref{tab: Picked covariates}. Standard errors in the model specifications "socio-demographic X" and "all X" account for clustering, the standard errors of lasso-based estimation do not. Data stems from Swiss Household Panel (SHP) 2004- 2017.
 \end{tablenotes}
\end{threeparttable}
\end{table}

\clearpage

\begin{table}[!h]
\caption{Restriction of the sample size}
\label{tab: Restriction of the sample size}
\centering
\begin{threeparttable}
\begin{tabular}{ |p{6cm}|p{7cm}|}
\hline
Description & Sample size\\
\hline
    All mothers with children in the SHP dataset  & 35.337 observations  \\
    \hline
    children turning 4 before September and 5 after February & 3.510 observations\\
    \hline
    Selected children must be youngest child in household & 2.006 observations\\
    \hline
    Periods from the year 2004 onwards & 1.472 observations\\
    \hline
    Exclude observations with missings in the outcomes or
controls & 953 observations in evaluation sample\\

    \hline

\end{tabular}

\begin{tablenotes}[flushleft]
\footnotesize
\item Note: This table reports the restriction of the sample size based on the Swiss Household Panel (SHP) 2004- 2017.
\end{tablenotes}
\end{threeparttable}

\end{table}

\begin{table}[!h]
\caption{Variable description}
\label{tab: Variable description}
\centering
\begin{threeparttable}
\begin{tabular}{ |p{4cm}|p{9cm}|}
\hline
Outcome variable & Description\\
\hline
    Out of labour force  & Dummy working status "not in labour force", constructed from wstat==3 \\
    \hline
    Unemployed & Dummy working status "unemployed", constructed from wstat==2\\
    \hline
    Employed & Dummy  working status "active occupied", constructed from wstat==1\\
    \hline
    Part-time employed  & Dummy working in part-time, constructed from pw39==1\\
    & \textit{"Currently, in your main job, do you work parttime or 100\% ?"}\\

    \hline
    Full-time employed & Dummy working in full-time, constructed from pw39==2\\
    & \textit{"Currently, in your main job, do you work parttime or 100\% ?"}\\

    \hline

\end{tabular}

\begin{tablenotes}[flushleft]
\footnotesize
\item Note: This table describes the outcome variables used in the estimation and based on the Swiss Household Panel (SHP) 2004 - 2017.
\end{tablenotes}
\end{threeparttable}

\end{table}

\clearpage
\newpage

\subsection{Acknowledgement}
The DiD part of the study has been realised using the data collected by the Swiss Household Panel (SHP), which is based at the Swiss Centre of Expertise in the Social Sciences (FORS). The project is financed by the Swiss National Science Foundation.
\end{appendices}
\newpage

\begingroup
    \setlength{\bibsep}{10pt}
    \setstretch{1}
   \bibliography{DissResearch}{}

\begin{thebibliography}{85}
\expandafter\ifx\csname natexlab\endcsname\relax\def\natexlab#1{#1}\fi
\providecommand{\url}[1]{\texttt{#1}}
\providecommand{\href}[2]{#2}
\providecommand{\path}[1]{#1}
\providecommand{\DOIprefix}{doi:}
\providecommand{\ArXivprefix}{arXiv:}
\providecommand{\URLprefix}{URL: }
\providecommand{\Pubmedprefix}{pmid:}
\providecommand{\doi}[1]{\href{http://dx.doi.org/#1}{\path{#1}}}
\providecommand{\Pubmed}[1]{\href{pmid:#1}{\path{#1}}}
\providecommand{\bibinfo}[2]{#2}
\ifx\xfnm\relax \def\xfnm[#1]{\unskip,\space#1}\fi
\bibitem[{{1815.ch}(2014)}]{Wallis2014}
\bibinfo{author}{{1815.ch}}, \bibinfo{year}{2014}.
\newblock \bibinfo{title}{Einf\"uhrung des neuen {P}rimarschulgesetzes
  {E}inschulungsalter von 4 {J}ahren ab n\"achstem {S}ommer}.
\newblock
  \bibinfo{howpublished}{\url{https://1815.ch/news/wallis/aktuell/einschulungsalter-von-4-jahren-ab-naechstem-sommer-20140903111319/}}.
\newblock \bibinfo{note}{Accessed: 2021-11-13}.
\bibitem[{Antal(2016)}]{Antal2016}
\bibinfo{author}{Antal, E.}, \bibinfo{year}{2016}.
\newblock \bibinfo{title}{Some remarks on the use of weights}.
\newblock \URLprefix
  \url{https://forscenter.ch/wp-content/uploads/2018/07/some-remarkes-on-the-use-of-weights.pdf}.
\bibitem[{Barua(2014)}]{BARUA2014}
\bibinfo{author}{Barua, R.}, \bibinfo{year}{2014}.
\newblock \bibinfo{title}{Intertemporal substitution in maternal labor supply:
  Evidence using state school entrance age laws}.
\newblock \bibinfo{journal}{Labour Economics} \bibinfo{volume}{31},
  \bibinfo{pages}{129 -- 140}.
\newblock \URLprefix
  \url{http://www.sciencedirect.com/science/article/pii/S0927537114000852},
  \DOIprefix\doi{https://doi.org/10.1016/j.labeco.2014.07.002}.
\bibitem[{Bauernschuster and Schlotter(2015)}]{BAUERNSCHUSTER2015}
\bibinfo{author}{Bauernschuster, S.}, \bibinfo{author}{Schlotter, M.},
  \bibinfo{year}{2015}.
\newblock \bibinfo{title}{Public child care and mothers' labor
  supply—evidence from two quasi-experiments}.
\newblock \bibinfo{journal}{Journal of Public Economics} \bibinfo{volume}{123},
  \bibinfo{pages}{1--16}.
\newblock \URLprefix
  \url{https://www.sciencedirect.com/science/article/pii/S004727271500002X},
  \DOIprefix\doi{https://doi.org/10.1016/j.jpubeco.2014.12.013}.
\bibitem[{Becker(1965)}]{Becker1965}
\bibinfo{author}{Becker, G.S.}, \bibinfo{year}{1965}.
\newblock \bibinfo{title}{A theory of the allocation of time}.
\newblock \bibinfo{journal}{The Economic Journal} \bibinfo{volume}{75},
  \bibinfo{pages}{493--517}.
\newblock \URLprefix \url{http://www.jstor.org/stable/2228949}.
\bibitem[{Berlinski and Galiani(2007)}]{Berlinski2007}
\bibinfo{author}{Berlinski, S.}, \bibinfo{author}{Galiani, S.},
  \bibinfo{year}{2007}.
\newblock \bibinfo{title}{The effect of a large expansion of pre-primary school
  facilities on preschool attendance and maternal employment}.
\newblock \bibinfo{journal}{Labour Economics} \bibinfo{volume}{14},
  \bibinfo{pages}{665--680}.
\bibitem[{Besley and Case(2000)}]{besley2000}
\bibinfo{author}{Besley, T.}, \bibinfo{author}{Case, A.}, \bibinfo{year}{2000}.
\newblock \bibinfo{title}{Unnatural experiments? estimating the incidence of
  endogenous policies}.
\newblock \bibinfo{journal}{The Economic Journal} \bibinfo{volume}{110},
  \bibinfo{pages}{672--694}.
\bibitem[{Bettendorf et~al.(2015)Bettendorf, Jongen and
  Muller}]{Bettendorf2015}
\bibinfo{author}{Bettendorf, L.J.}, \bibinfo{author}{Jongen, E.L.},
  \bibinfo{author}{Muller, P.}, \bibinfo{year}{2015}.
\newblock \bibinfo{title}{Childcare subsidies and labour supply — evidence
  from a large dutch reform}.
\newblock \bibinfo{journal}{Labour Economics} \bibinfo{volume}{36},
  \bibinfo{pages}{112 -- 123}.
\bibitem[{{Bundesamt f{\"u}r
  Sozialversicherungen}(2006)}]{BundesamtfurSozialversicherungen}
\bibinfo{author}{{Bundesamt f{\"u}r Sozialversicherungen}},
  \bibinfo{year}{2006}.
\newblock \bibinfo{title}{{Botschaft zum Bundesbeschluss {\"u}ber Finanzhilfen
  für familienerg{\"a}nzende Kinderbetreuung}}.
\newblock \URLprefix \url{https://www.fedlex.admin.ch/eli/fga/2006/357/de}.
\bibitem[{Cai and Kalb(2006)}]{Cai2006}
\bibinfo{author}{Cai, L.}, \bibinfo{author}{Kalb, G.}, \bibinfo{year}{2006}.
\newblock \bibinfo{title}{Health status and labour force participation:
  evidence from australia}.
\newblock \bibinfo{journal}{Health Economics} \bibinfo{volume}{15},
  \bibinfo{pages}{241--261}.
\bibitem[{Callaway and Sant'Anna(2021a)}]{didpackage}
\bibinfo{author}{Callaway, B.}, \bibinfo{author}{Sant'Anna, P.H.},
  \bibinfo{year}{2021}a.
\newblock \bibinfo{title}{did: Difference in differences}.
\newblock \URLprefix \url{https://bcallaway11.github.io/did/}. \bibinfo{note}{r
  package version 2.1.2}.
\bibitem[{Callaway and Sant'Anna(2021b)}]{CallawaySantAnna2018}
\bibinfo{author}{Callaway, B.}, \bibinfo{author}{Sant'Anna, P.H.},
  \bibinfo{year}{2021}b.
\newblock \bibinfo{title}{Difference-in-differences with multiple time
  periods}.
\newblock \bibinfo{journal}{Journal of Econometrics} \bibinfo{volume}{225},
  \bibinfo{pages}{200--230}.
\bibitem[{Calonico et~al.(2021)Calonico, D., Farrell and
  Titiunik}]{Calonico2021}
\bibinfo{author}{Calonico, S.}, \bibinfo{author}{D., C.M.},
  \bibinfo{author}{Farrell, M.H.}, \bibinfo{author}{Titiunik, R.},
  \bibinfo{year}{2021}.
\newblock \bibinfo{title}{Robust Data-Driven Statistical Inference in
  Regression-Discontinuity Designs}.
\newblock \URLprefix
  \url{https://www.google.com/url?sa=t&rct=j&q=&esrc=s&source=web&cd=&cad=rja&uact=8&ved=2ahUKEwiywffQiqvyAhWG_rsIHRdnA60QFnoECAgQAQ&url=https%3A%2F%2Fcran.r-project.org%2Fweb%2Fpackages%2Frdrobust%2Frdrobust.pdf&usg=AOvVaw1D9Rf0BUIsPFdj2akTOvar}.
\bibitem[{Carta and Rizzica(2018)}]{CARTA2018}
\bibinfo{author}{Carta, F.}, \bibinfo{author}{Rizzica, L.},
  \bibinfo{year}{2018}.
\newblock \bibinfo{title}{Early kindergarten, maternal labor supply and
  children's outcomes: Evidence from italy}.
\newblock \bibinfo{journal}{Journal of Public Economics} \bibinfo{volume}{158},
  \bibinfo{pages}{79 -- 102}.
\newblock \URLprefix
  \url{http://www.sciencedirect.com/science/article/pii/S0047272717302141},
  \DOIprefix\doi{https://doi.org/10.1016/j.jpubeco.2017.12.012}.
\bibitem[{Chernozhukov et~al.(2016)Chernozhukov, Hansen and
  Spindler}]{Chernzhukov2016}
\bibinfo{author}{Chernozhukov, V.}, \bibinfo{author}{Hansen, C.},
  \bibinfo{author}{Spindler, M.}, \bibinfo{year}{2016}.
\newblock \bibinfo{title}{{hdm}: High-dimensional metrics}.
\newblock \bibinfo{journal}{R Journal} \bibinfo{volume}{8},
  \bibinfo{pages}{185--199}.
\newblock \URLprefix
  \url{https://journal.r-project.org/archive/2016/RJ-2016-040/index.html}.
\bibitem[{{Eckhoff Andresen} and Havnes(2019)}]{ECKHOFFANDRESEN2019}
\bibinfo{author}{{Eckhoff Andresen}, M.}, \bibinfo{author}{Havnes, T.},
  \bibinfo{year}{2019}.
\newblock \bibinfo{title}{Child care, parental labor supply and tax revenue}.
\newblock \bibinfo{journal}{Labour Economics} \bibinfo{volume}{61},
  \bibinfo{pages}{101762}.
\newblock \URLprefix
  \url{https://www.sciencedirect.com/science/article/pii/S0927537119300880},
  \DOIprefix\doi{https://doi.org/10.1016/j.labeco.2019.101762}.
\bibitem[{{Erziehungsdepartement des Kantons
  Basel-Stadt}(2013)}]{Basel-Stadt2013}
\bibinfo{author}{{Erziehungsdepartement des Kantons Basel-Stadt}},
  \bibinfo{year}{2013}.
\newblock \bibinfo{title}{{Volksschulen: Handreichung Schullaufbahn, Mappe B -
  Primarstufe 2013}}.
\bibitem[{{European Commission}(2023)}]{EuropeanComm2023}
\bibinfo{author}{{European Commission}}, \bibinfo{year}{2023}.
\newblock \bibinfo{title}{Commission report finds labour and skills shortages
  persist and looks at possible ways to tackle them}.
\newblock
  \bibinfo{howpublished}{\url{https://ec.europa.eu/social/main.jsp?langId=en\&catId=89\&furtherNews=yes\&newsId=10619\#navItem-3}}.
\newblock \bibinfo{note}{Accessed: 2023-10-05}.
\bibitem[{{Eurydice}(2016/17)}]{EuropeanComm16/17}
\bibinfo{author}{{Eurydice}}, \bibinfo{year}{2016/17}.
\newblock \bibinfo{title}{Compulsory education in europe – 2016/17. eurydice
  facts and figures.}
\newblock
  \bibinfo{howpublished}{\url{https://publications.europa.eu/en/publication-detail/-/publication/2f15cd79-9a83-11e6-9bca-01aa75ed71a1/language-en}}.
\newblock \bibinfo{note}{Accessed: 2019-08-13}.
\bibitem[{{Federal Statistical Office}(2022)}]{BFS2022}
\bibinfo{author}{{Federal Statistical Office}}, \bibinfo{year}{2022}.
\newblock \bibinfo{title}{Umzugsquote der ständigen wohnbevölkerung nach
  kanton}.
\newblock
  \bibinfo{howpublished}{\url{https://www.bfs.admin.ch/bfs/en/home/statistics/construction-housing/dwellings/home-moves.html},
  note={Accessed: 2024-02-14}}.
\bibitem[{Felfe et~al.(2016)Felfe, Lechner and Thiemann}]{Felfe2016}
\bibinfo{author}{Felfe, C.}, \bibinfo{author}{Lechner, M.},
  \bibinfo{author}{Thiemann, P.}, \bibinfo{year}{2016}.
\newblock \bibinfo{title}{After-school care and parents' labor supply}.
\newblock \bibinfo{journal}{Labour Economics} \bibinfo{volume}{42},
  \bibinfo{pages}{64 -- 75}.
\newblock \URLprefix
  \url{http://www.sciencedirect.com/science/article/pii/S0927537116300616}.
\bibitem[{Finseraas et~al.(2017)Finseraas, Hardoy and
  Sch{\o}ne}]{Finseraas2017}
\bibinfo{author}{Finseraas, H.}, \bibinfo{author}{Hardoy, I.},
  \bibinfo{author}{Sch{\o}ne, P.}, \bibinfo{year}{2017}.
\newblock \bibinfo{title}{School enrolment and mothers' labor supply: evidence
  from a regression discontinuity approach}.
\newblock \bibinfo{journal}{Review of Economics of the Household}
  \bibinfo{volume}{15}, \bibinfo{pages}{621--638}.
\bibitem[{Fitzpatrick(2012)}]{fitzpatrick2012}
\bibinfo{author}{Fitzpatrick, M.D.}, \bibinfo{year}{2012}.
\newblock \bibinfo{title}{Revising our thinking about the relationship between
  maternal labor supply and preschool}.
\newblock \bibinfo{journal}{Journal of Human Resources} \bibinfo{volume}{47},
  \bibinfo{pages}{583--612}.
\bibitem[{Frandsen(2017)}]{Frandsen2017}
\bibinfo{author}{Frandsen, B.}, \bibinfo{year}{2017}.
\newblock \bibinfo{title}{Party bias in union representation elections: Testing
  for manipulation in the regression discontinuity design when the running
  variable is discrete}.
\newblock \bibinfo{journal}{Advances in Econometrics} \bibinfo{volume}{38},
  \bibinfo{pages}{281--315}.
\newblock \URLprefix \url{https://doi.org/10.1108/S0731-905320170000038012}.
\bibitem[{Gelbach(2002)}]{Gelbach2002}
\bibinfo{author}{Gelbach, J.B.}, \bibinfo{year}{2002}.
\newblock \bibinfo{title}{Public schooling for young children and maternal
  labor supply}.
\newblock \bibinfo{journal}{The American Economic Review} \bibinfo{volume}{92},
  \bibinfo{pages}{307--322}.
\newblock \URLprefix \url{http://www.jstor.org/stable/3083335}.
\bibitem[{Gelman and Imbens(2019)}]{GelmanandImbens2019}
\bibinfo{author}{Gelman, A.}, \bibinfo{author}{Imbens, G.},
  \bibinfo{year}{2019}.
\newblock \bibinfo{title}{Why high-order polynomials should not be used in
  regression discontinuity designs}.
\newblock \bibinfo{journal}{Journal of Business \& Economic Statistics}
  \bibinfo{volume}{37}, \bibinfo{pages}{447--456}.
\newblock \URLprefix \url{https://doi.org/10.1080/07350015.2017.1366909},
  \DOIprefix\doi{10.1080/07350015.2017.1366909},
  \href{http://arxiv.org/abs/https://doi.org/10.1080/07350015.2017.1366909}{{\tt
  arXiv:https://doi.org/10.1080/07350015.2017.1366909}}.
\bibitem[{Geyer et~al.(2015)Geyer, Haan and Wrohlich}]{GEYER2015}
\bibinfo{author}{Geyer, J.}, \bibinfo{author}{Haan, P.},
  \bibinfo{author}{Wrohlich, K.}, \bibinfo{year}{2015}.
\newblock \bibinfo{title}{The effects of family policy on maternal labor
  supply: Combining evidence from a structural model and a quasi-experimental
  approach}.
\newblock \bibinfo{journal}{Labour Economics} \bibinfo{volume}{36},
  \bibinfo{pages}{84 -- 98}.
\newblock \URLprefix
  \url{http://www.sciencedirect.com/science/article/pii/S0927537115000755},
  \DOIprefix\doi{https://doi.org/10.1016/j.labeco.2015.07.001}.
\bibitem[{Goux and Maurin(2010)}]{GOUX}
\bibinfo{author}{Goux, D.}, \bibinfo{author}{Maurin, E.}, \bibinfo{year}{2010}.
\newblock \bibinfo{title}{Public school availability for two-year olds and
  mothers' labour supply}.
\newblock \bibinfo{journal}{Labour Economics} \bibinfo{volume}{17},
  \bibinfo{pages}{951 -- 962}.
\newblock \URLprefix
  \url{http://www.sciencedirect.com/science/article/pii/S0927537110000576},
  \DOIprefix\doi{https://doi.org/10.1016/j.labeco.2010.04.012}.
\bibitem[{{Gran Consiglio Repubblica e Cantone Ticino}(2011)}]{Tessin2011}
\bibinfo{author}{{Gran Consiglio Repubblica e Cantone Ticino}},
  \bibinfo{year}{2011}.
\newblock \bibinfo{title}{{Legge della scuola, 400.100}}.
\bibitem[{{Grosser Rat des Kantons Basel-Stadt}(2010)}]{Basel-Stadt2016}
\bibinfo{author}{{Grosser Rat des Kantons Basel-Stadt}}, \bibinfo{year}{2010}.
\newblock \bibinfo{title}{{ 410.100 Schulgesetz in der Fassung des GRB vom 19.
  5. 2010}}.
\newblock \URLprefix
  \url{https://www.gesetzessammlung.bs.ch/app/de/texts_of_law/410.100}.
\bibitem[{{Grosser Rat des Kantons Freiburg}(2008)}]{Fribourg2008}
\bibinfo{author}{{Grosser Rat des Kantons Freiburg}}, \bibinfo{year}{2008}.
\newblock \bibinfo{title}{{Gesetz vom 5. September 2008 zur Änderung des
  Schulgesetzes (Kindergarten)}}.
\bibitem[{{Grosser Rat des Kantons Schaffhausen}(2014)}]{Schaffhausen2014}
\bibinfo{author}{{Grosser Rat des Kantons Schaffhausen}}, \bibinfo{year}{2014}.
\newblock \bibinfo{title}{{410.100 Schulgesetz}}.
\bibitem[{{Grosser Rat des Kantons St.Gallen}(2007)}]{StGallen2007}
\bibinfo{author}{{Grosser Rat des Kantons St.Gallen}}, \bibinfo{year}{2007}.
\newblock \bibinfo{title}{{sGS 213.1 - Volksschulgesetz (VSG)}}.
\bibitem[{Hahn et~al.(2001)Hahn, Todd and der Klaauw}]{Hahn2001}
\bibinfo{author}{Hahn, J.}, \bibinfo{author}{Todd, P.}, \bibinfo{author}{der
  Klaauw, W.V.}, \bibinfo{year}{2001}.
\newblock \bibinfo{title}{Identification and estimation of treatment effects
  with a regression-discontinuity design}.
\newblock \bibinfo{journal}{Econometrica} \bibinfo{volume}{69},
  \bibinfo{pages}{201--209}.
\newblock \URLprefix \url{http://www.jstor.org/stable/2692190}.
\bibitem[{Hardoy and Sch{\o}ne(2015)}]{Hardoy2015}
\bibinfo{author}{Hardoy, I.}, \bibinfo{author}{Sch{\o}ne, P.},
  \bibinfo{year}{2015}.
\newblock \bibinfo{title}{Enticing even higher female labor supply: the impact
  of cheaper day care}.
\newblock \bibinfo{journal}{Review of Economics of the Household}
  \bibinfo{volume}{13}, \bibinfo{pages}{815--836}.
\bibitem[{Havnes and Mogstad(2011)}]{HAVNES2011}
\bibinfo{author}{Havnes, T.}, \bibinfo{author}{Mogstad, M.},
  \bibinfo{year}{2011}.
\newblock \bibinfo{title}{Money for nothing? universal child care and maternal
  employment}.
\newblock \bibinfo{journal}{Journal of Public Economics} \bibinfo{volume}{95},
  \bibinfo{pages}{1455--1465}.
\newblock \URLprefix
  \url{https://www.sciencedirect.com/science/article/pii/S0047272711000880},
  \DOIprefix\doi{https://doi.org/10.1016/j.jpubeco.2011.05.016}.
  \bibinfo{note}{special Issue: International Seminar for Public Economics on
  Normative Tax Theory}.
\bibitem[{Herrmann and Murier(2016)}]{MutterArbeitsm}
\bibinfo{author}{Herrmann, A.B.}, \bibinfo{author}{Murier, T.},
  \bibinfo{year}{2016}.
\newblock \bibinfo{title}{Schweizerische Arbeitskr\"afteerhebung - M\"utter auf
  dem Arbeitsmarkt}.
\newblock \bibinfo{publisher}{Bundesamt f\"ur Statistik},
  \bibinfo{address}{Neuch\^atel}.
\newblock \URLprefix
  \url{https://www.bfs.admin.ch/bfs/de/home/statistiken/kataloge-datenbanken/publikationen.assetdetail.1061095.html}.
\bibitem[{{Jacobs Foundation}(2018)}]{JacobsFoundation2018}
\bibinfo{author}{{Jacobs Foundation}}, \bibinfo{year}{2018}.
\newblock \bibinfo{title}{Whitepaper zur Vereinbarkeit von Familie und Beruf:
  Zwischen Wunsch und Realität}.
\newblock \bibinfo{type}{Technical Report}. Jacobs Foundation.
\bibitem[{{Kanton Aargau - Departement Bildung, Kultur und Sport - Abteilung
  Volksschule}(2010)}]{Aargau2010}
\bibinfo{author}{{Kanton Aargau - Departement Bildung, Kultur und Sport -
  Abteilung Volksschule}}, \bibinfo{year}{2010}.
\newblock \bibinfo{title}{{Teilrevision der Kantonsverfassung und des
  Schulgesetzes betreffend St\"arkung der Volksschule Aargau}}.
\bibitem[{{Kanton Solothurn - Amt für Volksschule und
  Kindergarten}(2012)}]{Solothurn2012}
\bibinfo{author}{{Kanton Solothurn - Amt für Volksschule und Kindergarten}},
  \bibinfo{year}{2012}.
\newblock \bibinfo{title}{{HarmoS: der Kindergarten ist die erste Stufe der
  Volksschule Umsetzung auf das Schuljahr 2012/2013}}.
\bibitem[{{Kanton Solothurn - Amt für Volksschule und
  Kindergarten}(2013)}]{Bern2013}
\bibinfo{author}{{Kanton Solothurn - Amt für Volksschule und Kindergarten}},
  \bibinfo{year}{2013}.
\newblock \bibinfo{title}{{Kurznachrichten aus dem Gemeinderat}}.
\newblock \URLprefix
  \url{https://www.bern.ch/mediencenter/medienmitteilungen/aktuell_ptk/2013-04-kurznach,
  last downloaded 2021/11/02.}
\bibitem[{{Kantonsrat Z\"urich}(2007)}]{Kantonsrat}
\bibinfo{author}{{Kantonsrat Z\"urich}}, \bibinfo{year}{2007}.
\newblock \bibinfo{title}{{412.100 Volksschulgesetz (VSG) (vom 7. Februar
  2005)}}.
\bibitem[{Kleven et~al.(2019)Kleven, Landais, Posch, Steinhauer and
  Zweim\"uller}]{Kleven2019}
\bibinfo{author}{Kleven, H.}, \bibinfo{author}{Landais, C.},
  \bibinfo{author}{Posch, J.}, \bibinfo{author}{Steinhauer, A.},
  \bibinfo{author}{Zweim\"uller, J.}, \bibinfo{year}{2019}.
\newblock \bibinfo{title}{Child penalties across countries: Evidence and
  explanations}.
\newblock \bibinfo{journal}{AEA Papers and Proceedings} \bibinfo{volume}{109},
  \bibinfo{pages}{122--26}.
\newblock \URLprefix
  \url{https://www.aeaweb.org/articles?id=10.1257/pandp.20191078},
  \DOIprefix\doi{10.1257/pandp.20191078}.
\bibitem[{Kleven et~al.(2020)Kleven, Landais, Posch, Steinhauer and
  Zweimüller}]{Kleven2020}
\bibinfo{author}{Kleven, H.}, \bibinfo{author}{Landais, C.},
  \bibinfo{author}{Posch, J.}, \bibinfo{author}{Steinhauer, A.},
  \bibinfo{author}{Zweimüller, J.}, \bibinfo{year}{2020}.
\newblock \bibinfo{title}{Do Family Policies Reduce Gender Inequality? Evidence
  from 60 Years of Policy Experimentation}.
\newblock \bibinfo{type}{Working Paper} \bibinfo{number}{28082}. National
  Bureau of Economic Research.
\newblock \URLprefix \url{http://www.nber.org/papers/w28082},
  \DOIprefix\doi{10.3386/w28082}.
\bibitem[{Krapf et~al.(2020)Krapf, Roth and Slotwinski}]{Krapf2020}
\bibinfo{author}{Krapf, M.}, \bibinfo{author}{Roth, A.},
  \bibinfo{author}{Slotwinski, M.}, \bibinfo{year}{2020}.
\newblock \bibinfo{title}{The effect of childcare on parental earnings
  trajectories}.
\newblock \URLprefix
  \url{https://www.cesifo.org/en/publikationen/2020/working-paper/effect-childcare-parental-earnings-trajectories}.
  \bibinfo{note}{cESifo Working Papers}.
\bibitem[{Kunze and Liu(2019)}]{Kunze19}
\bibinfo{author}{Kunze, A.}, \bibinfo{author}{Liu, X.}, \bibinfo{year}{2019}.
\newblock \bibinfo{title}{Universal Childcare for the Youngest and the Maternal
  Labour Supply}.
\newblock \bibinfo{type}{Technical Report}.
\newblock \bibinfo{note}{CESifo Working Paper No. 7509}.
\bibitem[{Landers{\o} et~al.(2020)Landers{\o}, Nielsen and Simonsen}]{Landerso}
\bibinfo{author}{Landers{\o}, R.K.}, \bibinfo{author}{Nielsen, H.S.},
  \bibinfo{author}{Simonsen, M.}, \bibinfo{year}{2020}.
\newblock \bibinfo{title}{Effects of school starting age on the family}.
\newblock \bibinfo{journal}{Journal of Human Resources} \bibinfo{volume}{55},
  \bibinfo{pages}{1258--1286}.
\newblock \URLprefix \url{https://jhr.uwpress.org/content/55/4/1258},
  \DOIprefix\doi{10.3368/jhr.55.4.1117-9174R1},
  \href{http://arxiv.org/abs/https://jhr.uwpress.org/content/55/4/1258.full.pdf}{{\tt
  arXiv:https://jhr.uwpress.org/content/55/4/1258.full.pdf}}.
\bibitem[{{Landsgemeinde Glarus}(2009)}]{Glarus2009}
\bibinfo{author}{{Landsgemeinde Glarus}}, \bibinfo{year}{2009}.
\newblock \bibinfo{title}{{Gesetz über Schule und Bildung(Bildungsgesetz) Vom
  6. Mai 2001 (Stand 1. August 2017)}}.
\bibitem[{{Le Grand Conseil de la République et Canton de
  Neuch\^{a}tel}(2011)}]{Neuchatel2011}
\bibinfo{author}{{Le Grand Conseil de la République et Canton de
  Neuch\^{a}tel}}, \bibinfo{year}{2011}.
\newblock \bibinfo{title}{{410.10 Loi sur l'organisation scolaire (LOS)}}.
\bibitem[{Lee(2008)}]{LEE2008}
\bibinfo{author}{Lee, D.S.}, \bibinfo{year}{2008}.
\newblock \bibinfo{title}{Randomized experiments from non-random selection in
  u.s. house elections}.
\newblock \bibinfo{journal}{Journal of Econometrics} \bibinfo{volume}{142},
  \bibinfo{pages}{675--697}.
\newblock \URLprefix
  \url{https://www.sciencedirect.com/science/article/pii/S0304407607001121},
  \DOIprefix\doi{https://doi.org/10.1016/j.jeconom.2007.05.004}.
  \bibinfo{note}{the regression discontinuity design: Theory and applications}.
\bibitem[{Lee and Lemieux(2010)}]{LeeandLemieux2010}
\bibinfo{author}{Lee, D.S.}, \bibinfo{author}{Lemieux, T.},
  \bibinfo{year}{2010}.
\newblock \bibinfo{title}{Regression discontinuity designs in economics}.
\newblock \bibinfo{journal}{Journal of Economic Literature}
  \bibinfo{volume}{48}, \bibinfo{pages}{281--355}.
\newblock \URLprefix
  \url{https://www.aeaweb.org/articles?id=10.1257/jel.48.2.281},
  \DOIprefix\doi{10.1257/jel.48.2.281}.
\bibitem[{Lefebvre and Merrigan(2008)}]{Lefebvre2008}
\bibinfo{author}{Lefebvre, P.}, \bibinfo{author}{Merrigan, P.},
  \bibinfo{year}{2008}.
\newblock \bibinfo{title}{Child‐care policy and the labor supply of mothers
  with young children: A natural experiment from canada}.
\newblock \bibinfo{journal}{Journal of Labor Economics} \bibinfo{volume}{26},
  \bibinfo{pages}{519--548}.
\bibitem[{Lundin et~al.(2008)Lundin, M\"ork and \"Ockert}]{Lundin2008}
\bibinfo{author}{Lundin, D.}, \bibinfo{author}{M\"ork, E.},
  \bibinfo{author}{\"Ockert, B.}, \bibinfo{year}{2008}.
\newblock \bibinfo{title}{How far can reduced childcare prices push female
  labour supply?}
\newblock \bibinfo{journal}{Labour Economics} \bibinfo{volume}{15},
  \bibinfo{pages}{647--659}.
\bibitem[{Nollenberger and Rodr\'{i}guez-Planas(2015)}]{NOLLENBERGER2015}
\bibinfo{author}{Nollenberger, N.}, \bibinfo{author}{Rodr\'{i}guez-Planas, N.},
  \bibinfo{year}{2015}.
\newblock \bibinfo{title}{Full-time universal childcare in a context of low
  maternal employment: Quasi-experimental evidence from spain}.
\newblock \bibinfo{journal}{Labour Economics} \bibinfo{volume}{36},
  \bibinfo{pages}{124 -- 136}.
\newblock \URLprefix
  \url{http://www.sciencedirect.com/science/article/pii/S0927537115000238},
  \DOIprefix\doi{https://doi.org/10.1016/j.labeco.2015.02.008}.
\bibitem[{OECD(2015)}]{OECD2015}
\bibinfo{author}{OECD}, \bibinfo{year}{2015}.
\newblock \bibinfo{title}{Isced 2011 level 0: Early childhood education, in
  isced 2011 operational manual: Guidelines for classifying national education
  programmes and related qualifications}.
\newblock
  \bibinfo{howpublished}{\url{https://www.oecd-ilibrary.org/docserver/9789264228368-4-en.pdf?expires=1565703850&id=id&accname=ocid56025002&checksum=FABD3CBC97D44F307409C640E5CCAC30}}.
\newblock \bibinfo{note}{Accessed: 2019-08-13}.
\bibitem[{Oosterbeek et~al.(2021)Oosterbeek, {ter Meulen} and {van der
  Klaauw}}]{OOSTERBEEK2021}
\bibinfo{author}{Oosterbeek, H.}, \bibinfo{author}{{ter Meulen}, S.},
  \bibinfo{author}{{van der Klaauw}, B.}, \bibinfo{year}{2021}.
\newblock \bibinfo{title}{Long-term effects of school-starting-age rules}.
\newblock \bibinfo{journal}{Economics of Education Review}
  \bibinfo{volume}{84}, \bibinfo{pages}{102144}.
\newblock \URLprefix
  \url{https://www.sciencedirect.com/science/article/pii/S0272775721000637},
  \DOIprefix\doi{https://doi.org/10.1016/j.econedurev.2021.102144}.
\bibitem[{{Parlement de la République et Canton du Jura}(2011)}]{Jura2011}
\bibinfo{author}{{Parlement de la République et Canton du Jura}},
  \bibinfo{year}{2011}.
\newblock \bibinfo{title}{{Loi sur l' \'{e}cole enfantine, l'\'{e}cole primaire
  et l'\'{e}cole secondaire (Loi scolaire, 410.11)}}.
\bibitem[{Ravazzini(2018)}]{Ravazzini2018}
\bibinfo{author}{Ravazzini, L.}, \bibinfo{year}{2018}.
\newblock \bibinfo{title}{Childcare and maternal part-time employment: a
  natural experiment using swiss cantons}.
\newblock \bibinfo{journal}{Swiss Journal of Economics and Statistics}
  \bibinfo{volume}{154}, \bibinfo{pages}{15}.
\newblock \URLprefix \url{https://doi.org/10.1186/s41937-017-0003-x}.
\bibitem[{{Regierungsrat Basel-Stadt}(2010)}]{Basel-Stadt2010}
\bibinfo{author}{{Regierungsrat Basel-Stadt}}, \bibinfo{year}{2010}.
\newblock \bibinfo{title}{{SG 410.101 - Regierungsratsbeschluss betreffend
  Stichtag für die Einschulung für die Schuljahre 2011/12 bis 2015/16}}.
\newblock \bibinfo{howpublished}{§ 56 Abs. 1 Schulgesetz}.
\newblock \URLprefix
  \url{https://www.gesetzessammlung.bs.ch/app/de/texts_of_law/410.101}.
  \bibinfo{note}{{Accessed: 2019-10-07}}.
\bibitem[{{Regierungsrat des Kantons Basel-Landschaft}(2011)}]{BaselLand2011}
\bibinfo{author}{{Regierungsrat des Kantons Basel-Landschaft}},
  \bibinfo{year}{2011}.
\newblock \bibinfo{title}{{641.11 Verordnung für den Kindergarten und die
  Primarschule}}.
\bibitem[{{Regierungsrat Thurgau}(2007)}]{Thurgau2007}
\bibinfo{author}{{Regierungsrat Thurgau}}, \bibinfo{year}{2007}.
\newblock \bibinfo{title}{{411.11 Gesetz \"uber die Volksschule (VG) vom 29.
  August 2007 (Stand 1. Januar 2014)}}.
\bibitem[{{Schule K\"usnacht}({no date})}]{Kuesnacht}
\bibinfo{author}{{Schule K\"usnacht}}, \bibinfo{year}{{no date}}.
\newblock \bibinfo{title}{{Kindergartenstufe}}.
\newblock \URLprefix
  \url{https://www.schule-kuesnacht.ch/allgemeines/kindergartenstufe/}.
  \bibinfo{note}{{Accessed: 2019/10/07}}.
\bibitem[{{Schweizerische Konferenz der kantonalen
  Erziehungsdirektoren}(2007)}]{EDK2007}
\bibinfo{author}{{Schweizerische Konferenz der kantonalen
  Erziehungsdirektoren}}, \bibinfo{year}{2007}.
\newblock \bibinfo{title}{{Interkantonale Vereinbarung \"uber die
  Harmonisierung der obligatorischen Schule (HarmoS-Konkordat)}} \URLprefix
  \url{https://edudoc.ch/record/24711/files/HarmoS_d.pdf}.
  \bibinfo{note}{{Accessed: 2019-01-30}}.
\bibitem[{{Schweizerische Konferenz der kantonalen
  Erziehungsdirektoren}(2008a)}]{EDK2008}
\bibinfo{author}{{Schweizerische Konferenz der kantonalen
  Erziehungsdirektoren}}, \bibinfo{year}{2008}a.
\newblock \bibinfo{title}{{Faktenblatt: Kindergarten-Obligatorium und fr\"uhere
  Einschulung}}.
\bibitem[{{Schweizerische Konferenz der kantonalen
  Erziehungsdirektoren}(2008b)}]{EDK2007-08}
\bibinfo{author}{{Schweizerische Konferenz der kantonalen
  Erziehungsdirektoren}}, \bibinfo{year}{2008}b.
\newblock \bibinfo{title}{Kantonsumfrage 2007/08} \bibinfo{note}{{Accessed:
  2019-01-30}}.
\bibitem[{{Schweizerische Konferenz der kantonalen
  Erziehungsdirektoren}(2009)}]{EDK2008-09}
\bibinfo{author}{{Schweizerische Konferenz der kantonalen
  Erziehungsdirektoren}}, \bibinfo{year}{2009}.
\newblock \bibinfo{title}{Kantonsumfrage 2008/09}.
\newblock
  \bibinfo{howpublished}{\url{https://edudoc.ch/record/38708/files/Kantonsumfrage_d.pdf}}.
\newblock \bibinfo{note}{{Accessed: 2019-01-30}}.
\bibitem[{{Schweizerische Konferenz der kantonalen
  Erziehungsdirektoren}(2010)}]{EDK2009-10}
\bibinfo{author}{{Schweizerische Konferenz der kantonalen
  Erziehungsdirektoren}}, \bibinfo{year}{2010}.
\newblock \bibinfo{title}{Kantonsumfrage 2009/10}.
\newblock
  \bibinfo{howpublished}{\url{https://edudoc.ch/record/98068/files/KU_09_10_d.pdf}}.
\newblock \bibinfo{note}{{Accessed: 2019-01-30}}.
\bibitem[{{Schweizerische Konferenz der kantonalen
  Erziehungsdirektoren}(2011)}]{EDK2010-11}
\bibinfo{author}{{Schweizerische Konferenz der kantonalen
  Erziehungsdirektoren}}, \bibinfo{year}{2011}.
\newblock \bibinfo{title}{Kantonsumfrage 2010/11}.
\newblock
  \bibinfo{howpublished}{\url{https://edudoc.ch/record/106651/files/KU_10_11_d.pdf}}.
\newblock \bibinfo{note}{{Accessed: 2019-01-30}}.
\bibitem[{{Schweizerische Konferenz der kantonalen
  Erziehungsdirektoren}(2012)}]{EDK2011-12}
\bibinfo{author}{{Schweizerische Konferenz der kantonalen
  Erziehungsdirektoren}}, \bibinfo{year}{2012}.
\newblock \bibinfo{title}{Kantonsumfrage 2011/12}.
\newblock
  \bibinfo{howpublished}{\url{https://edudoc.ch/record/115193/files/d_Kantonsumfrage_2011_2012.pdf}}.
\newblock \bibinfo{note}{{Accessed: 2019-01-30}}.
\bibitem[{{Schweizerische Konferenz der kantonalen
  Erziehungsdirektoren}(2013)}]{EDK2012-13}
\bibinfo{author}{{Schweizerische Konferenz der kantonalen
  Erziehungsdirektoren}}, \bibinfo{year}{2013}.
\newblock \bibinfo{title}{Kantonsumfrage 2012/13}.
\newblock
  \bibinfo{howpublished}{\url{https://edudoc.ch/record/115194/files/KU_12_13_d.pdf}}.
\newblock \bibinfo{note}{{Accessed: 2019-01-30}}.
\bibitem[{{Schweizerische Konferenz der kantonalen
  Erziehungsdirektoren}(2014a)}]{EDK2013-14}
\bibinfo{author}{{Schweizerische Konferenz der kantonalen
  Erziehungsdirektoren}}, \bibinfo{year}{2014}a.
\newblock \bibinfo{title}{Kantonsumfrage 2013/14}.
\newblock
  \bibinfo{howpublished}{\url{https://edudoc.ch/record/122836/files/Archiv_KU_13_14_d.pdf}}.
\newblock \bibinfo{note}{{Accessed: 2019-01-30}}.
\bibitem[{{Schweizerische Konferenz der kantonalen
  Erziehungsdirektoren}(2014b)}]{EDK2014}
\bibinfo{author}{{Schweizerische Konferenz der kantonalen
  Erziehungsdirektoren}}, \bibinfo{year}{2014}b.
\newblock \bibinfo{title}{Obligatorische schule: Schuleintritt und erste
  jahre}.
\newblock
  \bibinfo{howpublished}{\url{https://edudoc.ch/record/111988/files/schuleintritt_d.pdf}}.
\newblock \bibinfo{note}{Accessed: 2019-01-30}.
\bibitem[{{Schweizerische Konferenz der kantonalen
  Erziehungsdirektoren}(2015)}]{EDK2014-15}
\bibinfo{author}{{Schweizerische Konferenz der kantonalen
  Erziehungsdirektoren}}, \bibinfo{year}{2015}.
\newblock \bibinfo{title}{Kantonsumfrage 2014/15}.
\newblock
  \bibinfo{howpublished}{\url{https://edudoc.ch/record/122866/files/Archiv_KU_14_15_d.pdf}}.
\newblock \bibinfo{note}{{Accessed: 2019-01-30}}.
\bibitem[{{Schweizerische Konferenz der kantonalen
  Erziehungsdirektoren}(2016a)}]{EDK2015-16}
\bibinfo{author}{{Schweizerische Konferenz der kantonalen
  Erziehungsdirektoren}}, \bibinfo{year}{2016}a.
\newblock \bibinfo{title}{Kantonsumfrage 2015/16}.
\newblock
  \bibinfo{howpublished}{\url{https://edudoc.ch/record/129856/files/kantonsu_15_16_d_archiv.pdf?version=1}}.
\newblock \bibinfo{note}{{Accessed: 2019-01-30}}.
\bibitem[{{Schweizerische Konferenz der kantonalen
  Erziehungsdirektoren}(2016b)}]{EDK2016}
\bibinfo{author}{{Schweizerische Konferenz der kantonalen
  Erziehungsdirektoren}}, \bibinfo{year}{2016}b.
\newblock \bibinfo{title}{Kurz-info: Obligatorische schule: Schulstufen,
  zählweise der schuljahre}.
\newblock
  \bibinfo{howpublished}{\url{https://www.edudoc.ch/static/web/arbeiten/sprach_unterr/kurzinfo_zaehlweise_d.pdf}}.
\newblock \bibinfo{note}{Accessed: 2019-01-30}.
\bibitem[{{Schweizerische Konferenz der kantonalen
  Erziehungsdirektoren}(2017)}]{EDK2016-17}
\bibinfo{author}{{Schweizerische Konferenz der kantonalen
  Erziehungsdirektoren}}, \bibinfo{year}{2017}.
\newblock \bibinfo{title}{Kantonsumfrage 2016/17}.
\newblock
  \bibinfo{howpublished}{\url{https://edudoc.ch/record/133445/files/Archiv_16_17_d.pdf}}.
\newblock \bibinfo{note}{{Accessed: 2019-01-30}}.
\bibitem[{Schwyz(2012)}]{SZ2012}
\bibinfo{author}{Schwyz, K.}, \bibinfo{year}{2012}.
\newblock \bibinfo{title}{Regierungsrat des kantons schwyz; beschluss nr.
  383/2012}.
\newblock
  \bibinfo{howpublished}{\url{https:https://www.sz.ch/public/upload/assets/2080/rrb_383_2012.pdf}}.
\newblock \bibinfo{note}{Accessed: 2019-03-03}.
\bibitem[{{Secrétariat du Grand Conseil Genève}(2010)}]{Genf2010}
\bibinfo{author}{{Secrétariat du Grand Conseil Genève}},
  \bibinfo{year}{2010}.
\newblock \bibinfo{title}{{Projet de loi modifiant la loi sur l'instruction
  publique (HarmoS) (C 1 10)}}.
\bibitem[{Stam et~al.(2014)Stam, Verbakel and de~Graaf}]{Stam2014}
\bibinfo{author}{Stam, K.}, \bibinfo{author}{Verbakel, E.},
  \bibinfo{author}{de~Graaf, P.M.}, \bibinfo{year}{2014}.
\newblock \bibinfo{title}{Do values matter? the impact of work ethic and
  traditional gender role values on female labour market supply}.
\newblock \bibinfo{journal}{Social Indicators Research} \bibinfo{volume}{116},
  \bibinfo{pages}{593--610}.
\bibitem[{Stern and Felfe(2015)}]{SternFelfe2015}
\bibinfo{author}{Stern, S.}, \bibinfo{author}{Felfe, C.}, \bibinfo{year}{2015}.
\newblock \bibinfo{title}{Krippenkosten und -finanzierung im internationalen
  Vergleich}.
\newblock \bibinfo{type}{Technical Report}. Soziale Sicherheit.
\bibitem[{Swart et~al.(2019)Swart, van~den Berge and van~der
  Wiel}]{SwartBergeWiel2017}
\bibinfo{author}{Swart, L.}, \bibinfo{author}{van~den Berge, W.},
  \bibinfo{author}{van~der Wiel, K.}, \bibinfo{year}{2019}.
\newblock \bibinfo{title}{Do parents work more when children start school?
  Evidence from the Netherlands}.
\newblock \bibinfo{type}{Technical Report}.
\newblock \bibinfo{note}{CPB Discussion Paper}.
\bibitem[{SWI(2006)}]{Abstimmung2006}
\bibinfo{author}{SWI}, \bibinfo{year}{2006}.
\newblock \bibinfo{title}{{Wuchtiges Ja f\"ur die Bildungsverfassung}}.
\newblock
  \bibinfo{howpublished}{\url{https://www.swissinfo.ch/ger/wuchtiges-ja-fuer-die-bildungsverfassung/5206776}}.
\newblock \bibinfo{note}{Accessed: 2018-12-01}.
\bibitem[{{United Nations}(2022)}]{UN2022}
\bibinfo{author}{{United Nations}}, \bibinfo{year}{2022}.
\newblock \bibinfo{title}{Old-age poverty has a woman’s face}.
\newblock
  \bibinfo{howpublished}{\url{https://www.un.org/development/desa/dspd/2022/11/old-age-poverty/}}.
\newblock \bibinfo{note}{Accessed: 2023-10-06}.
\bibitem[{Voorpostel et~al.(2017)Voorpostel, Tillmann, Lebert, Kuhn, Lipps,
  Ryser, Antal, Monsch, Dasoki and Wernli}]{SHP2017}
\bibinfo{author}{Voorpostel, M.}, \bibinfo{author}{Tillmann, R.},
  \bibinfo{author}{Lebert, F.}, \bibinfo{author}{Kuhn, U.},
  \bibinfo{author}{Lipps, O.}, \bibinfo{author}{Ryser, V.A.},
  \bibinfo{author}{Antal, E.}, \bibinfo{author}{Monsch, G.A.},
  \bibinfo{author}{Dasoki, N.}, \bibinfo{author}{Wernli, B.},
  \bibinfo{year}{2017}.
\newblock \bibinfo{title}{Swiss household panel, user guide (1999 - 2016)}.
\bibitem[{Zhu and Bradbury(2015)}]{Zhu2015}
\bibinfo{author}{Zhu, A.}, \bibinfo{author}{Bradbury, B.},
  \bibinfo{year}{2015}.
\newblock \bibinfo{title}{Delaying school entry: Short- and longer-term effects
  on mothers' employment}.
\newblock \bibinfo{journal}{Economic Record} \bibinfo{volume}{91},
  \bibinfo{pages}{233--246}.

\end{thebibliography}
\endgroup

\end{document}